\documentclass{egpubl}
\usepackage{preprint}
\SpecialIssueSubmission
\usepackage[T1]{fontenc}
\usepackage{dfadobe}  

\usepackage{cite}
\BibtexOrBiblatex

\electronicVersion
\PrintedOrElectronic
\ifpdf \usepackage[pdftex]{graphicx} \pdfcompresslevel=9
\else \usepackage[dvips]{graphicx} \fi

\usepackage{egweblnk}

\usepackage{amssymb}
\usepackage{relsize}
\usepackage{booktabs}
\usepackage{tabu}
\usepackage{multirow}
\usepackage[capitalise]{cleveref}
\usepackage{tikz}
\usepackage{listings}
\usepackage{stackengine}
\usepackage{algorithm}
\usepackage{algpseudocode}
\usepackage{makecell}

\definecolor{ForestGreen}{HTML}{009B55}

\usepackage{listings}
\newif\ifsubmission

\title{Beyond ExaBricks: GPU Volume Path Tracing of AMR Data}
\ifsubmission
\author[Submission ID 1136]{Submission ID: 1136}
\else
\author[S.~Zellmann et al.]
{\parbox{\textwidth}{\vspace{-4em}\centering Stefan~Zellmann$^{1}$\orcid{0000-0003-2880-9090},
    Qi~Wu$^{2}$\orcid{0000-0003-0342-9366},
    Alper~Sahistan$^{3}$\orcid{0000-0002-3480-7713},
    Kwan-Liu Ma$^{2}$\orcid{0000-0001-8086-0366},
    and Ingo~Wald$^{4}$\orcid{0000-0003-0046-713X} 
        }
        \\
{\parbox{\textwidth}{\vspace{-2em}\centering $^1$University of Cologne~$^2$University of California - Davis~$^3$University of Utah~$^4$NVIDIA
       }
}
}
\fi

\definecolor{darkgreen}{rgb}{0,0.5,0}
\definecolor{midgreen}{rgb}{0,0.6,0}
\definecolor{lightgreen}{rgb}{0,0.8,0}
\definecolor{darkred}{rgb}{0.6,0,0}

\def\exabrick{\emph{ExaBrick}}

\newif\ifdiff

\ifdiff
\usepackage{soul}
\def\added#1{{\color{midgreen}#1}}
\def\removed#1{\textrm{\color{red}\st{#1}}}
\else
\def\added#1{#1}
\def\removed#1{}
\fi

\teaser{
  \vspace{-3em}
 \includegraphics[width=\linewidth]{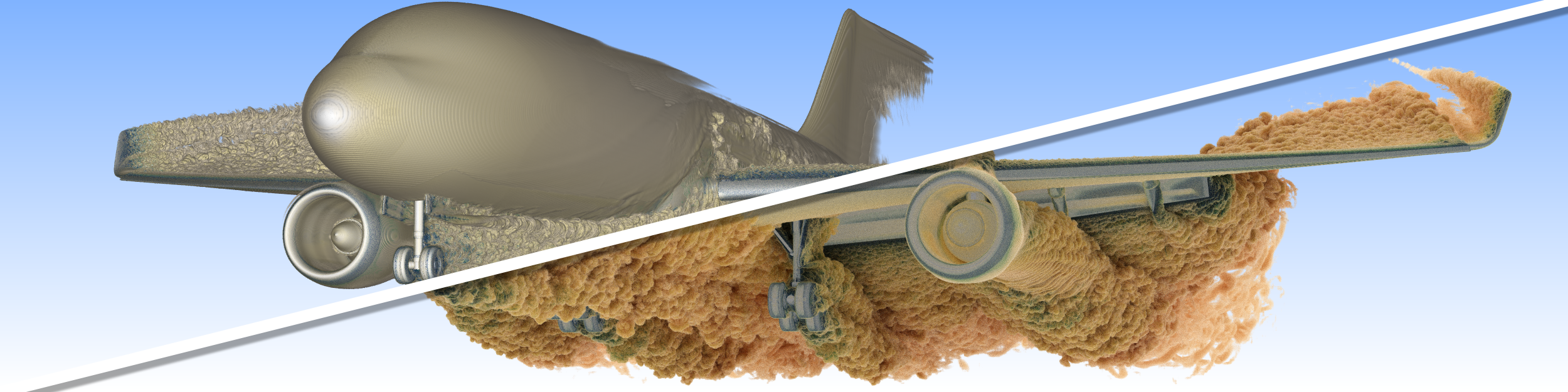}
 \centering
 \caption{\label{fig:teaser}%
 AMR rendering comparison using \exabrick{}~\cite{wald2021exabrick} vs.\ our
 extended framework. Top-left: original ``sci-vis style'' rendering (old) with
 ray marching, local shading with on-the-fly gradients, and a delta light
 source. Bottom-right: volumetric path tracing (new) with multi scattering,
 isotropic phase function and ambient lighting. The original software used two
 RTX~8000 GPUs to render a convergence frame with all quality settings set to
 maximum at 4~frames/sec.; our framework, with the best combination of
 optimizations discussed in this paper, renders path-traced convergence frames
 with global illumination and significantly better image quality at
 \added{16~frames/sec.\ on a single GeForce RTX~4090
 GPU.}\removed{6.7~frames/sec.}
 }
}

\begin{document}

\maketitle
\begin{abstract}
\added{Adaptive Mesh Refinement (AMR) is becoming a prevalent data
representation for HPC, and thus also for scientific visualization. AMR data is
usually cell centric (which imposes numerous challenges), complex, and generally
hard to render. Recent work on GPU-accelerated AMR rendering has made much
progress towards real-time volume and isosurface rendering of such data, but so
far this work has focused exclusively on ray marching, with simple lighting
models and without scattering events or global illumination. True high-quality
rendering requires a modified approach that is able to trace arbitrary
incoherent paths; but this may not be a perfect fit for the types of data
structures recently developed for ray marching. In this paper, we describe a
novel approach to high-quality path tracing of complex AMR data, with a
specific focus on analyzing and comparing different data structures and
algorithms to achieve this goal.}
\removed{Adaptive Mesh Refinement (AMR) is becoming a prevalent data representation for
scientific visualization. AMR data is usually cell centric, imposing a number
of challenges for high quality reconstruction at sample positions. While recent
work has concentrated on real-time volume and isosurface rendering on GPUs, the
rendering methods used still focus on ray marching with simple lighting models
and without scattering events or global illumination. As in other areas of
rendering, acceleration data structures are key to real-time performance; it is
not clear though if the acceleration structures originally used for ray
marching also work well for volumetric path tracing. In this work we analyze
the major bottlenecks of data structures that were originally optimized for
primary ray marching when used with the incoherent ray tracing workload of a
volumetric path tracer, and propose strategies to overcome the challenges
coming with this.}
\end{abstract}

\section{Introduction}
\label{sec:intro}

Adaptive Mesh Refinement (AMR) data is currently emerging as one of
the most prevalent and important types of data that any
visualization-focused renderer needs to be able to handle.
Unfortunately, from a rendering point of view AMR data comes with several
challenges: first, AMR data can come in many forms, from
octree-AMR~\cite{p4est} to various forms of block-structured AMR~\cite{chombo}.
Second, for rendering AMR data one requires a sample reconstruction method to
be defined over the AMR cells' scalar values; and how this is defined can have
a huge influence on the images that the renderer produces. Third,
most AMR data is highly non-uniform in nature, requiring the
renderer to be able to concentrate its sampling effort where it matters most.
Fourth, rendering AMR data is expensive 
partly because of the aforementioned
issue of non-uniformity of the signal, and partly because of the hierarchical
nature of the data. This means that each sample typically requires some form of
hierarchical data structure traversal. Fifth, the recent advances in real-time
ray- and path-tracing have raised the stakes 
for sci-vis rendering, too. While
arguably, effects such as indirect illumination or soft shadows and ambient lighting
are most important for photorealism, demand for high-quality Monte
Carlo rendering is gradually becoming the de facto for sci-vis rendering as
well.
Particularly, this change in paradigm has been initiated by
industry-quality tools like OSPRay~\cite{wald2017ospray} or
OpenVKL~\cite{Knoll2021}, which are readily available to sci-vis users through
ParaView or VisIt plugins~\cite{Wu_VisItOSPRay_2018}.


We explore the problem of high-quality path traced rendering of non-trivial AMR
data sets (cf. \cref{fig:teaser}). For that we start with the \exabrick{}
AMR rendering framework by Wald et al.~\cite{wald2021exabrick}, and look at
what is required to extend this to volumetric \emph{path} tracing. \exabrick{}
is both a framework and an acceleration data structure, similar
to what kd-trees and BVHs are for surface ray tracing, and while
\emph{fundamentally}, there is no reason this acceleration data
structure should not work with volumetric path tracing, it was
not originally optimized for that. 
It is an open research question if this
data structure optimized for sci-vis style volume ray marching can be
easily adapted to support the fundamentally different operation of computing
transmittance estimates as performed by a volumetric path tracer.

\removed{In this paper we focus on path tracing with \emph{Woodcock
tracking} [], which requires majorant extinctions $\bar \mu$
to be defined for the sampling regions. Those are used to determine if a volume
interaction is a scattering event or a \emph{null collision}, in which case we
continue sampling the free-flight distance. The more exact the majorants match the
local extinction, the fewer volume/extinction samples must be taken. We propose
different options to implement efficient volumetric path tracing for AMR data
that comprise different alternative traversal data structures to obtain
spatially varying local majorants. We evaluate the quality of the majorants
obtained, and their influence on overall sample count and rendering
performance.}
\added{In contrast to ray marching, the dominating operation in volume path tracing
is computing free flight distances, which require an upper bound estimate for the
extinction of the volume density, called \emph{majorant}. Though the method is
correct irrespective of how much that upper bound over-shoots the real value,
too high a difference between conservative majorant and actual
extinction that varies in space causes unnecessary samples to be taken, so that
key to accelerating volume path tracing is finding majorants that vary in space
themselves and then form an acceleration structure.}

In sci-vis, however--and in particular, for the type of data we are looking at
in this paper--computing good majorants is hard: one factor is that actual
extinction at any point in space depends on a \emph{transfer function} that can
map any data value to any extinction value, and which can---and typically
will---undergo radical changes as the user explores the volume. A second factor
is that the original AMR structure wants to adapt to the \emph{data}, so that
even if we were hypothetically allowed to relax the real-time requirement,
finding good \emph{alpha-mapped} majorants is a hard problem in itself.


Building upon these challenges in AMR visualization, the key contributions
of this paper are the following:
\begin{itemize}
  \item
We extend the AMR framework \exabrick{} by replacing alpha-composited ray
marching with volumetric path tracing, giving rise to significantly
improved time to image for sci-vis rendering, plus the possibility to
interactively create images with much better visual fidelity by using
full global illumination.

  \item
We investigate the question how efficient the existing \exabrick{} data structure
is by systematically identifying and implementing all of the most obvious
alternative variations of that approach.

  \item
We thoroughly analyze and compare the different alternatives and discuss their
technical implications. This includes a comparison to a hypothetical (and due
to its construction times impractical) reference data structure that is allowed
to perform near unlimited preprocessing.
\end{itemize}
\vspace{-1em}




\section{Related Work} \label{sec:related}
Volume rendering is important for both scientific visualization (sci-vis) and
production rendering, but opposing goals regarding visual fidelity on the one
hand, and interactivity on the other hand, have led the two fields to take
largely orthogonal approaches. Even early production rendering systems used
scattering and virtual point lights~\cite{laine:2007} and have since
transitioned to full global illumination with multi scattering and hundreds of
bounces~\cite{fascione:2018}. Scientific visualization has traditionally
focused on interactive exploration~\cite{engel:2006} based on ray
marching~\cite{levoy:1988} with absorption and emission~\cite{max:1995}.
Rendering with this lighting model can still be found in common scientific
visualization systems like VisIt~\cite{childs2011visit} or
ParaView~\cite{ahrens2005paraview}.

Exposure renderer~\cite{kroes2012exposure} is one of the first examples to
use volumetric path tracing for sci-vis. It
supports multi-scattering and can produce highly realistic
images~\cite{magnus2017interactive, igouchkine2017multi}. However, volumetric
path tracing only started to gain more attention from the community recently when
optimized software and hardware ray tracing frameworks became widely
available~\cite{wald2017ospray, iglesias:2022}.
Sci-vis
renderers use simple isotropic phase functions covering the whole
medium, RGB albedo from color lookup tables as opposed to full spectral
rendering, and simple lighting models.
\added{Common techniques to compute free-flight distance estimates---the stochastic
distance a photon travels without colliding with a particle from
the density---are \emph{unbiased}, which is attractive also for scientific
visualization~\cite{Morrical2022QuickClusters}}.
\removed{Woodcock tracking and null collision
techniques [] used for Monte Carlo path tracing have the
property of being unbiased, which has been identified as being attractive for
scientific visualization as well.}

\added{A number of techniques exist to compute free flight distances and
transmittance estimates using stochastic sampling.  Woodcock (or delta-)
tracking~\cite{woodcock:1965} is one of the oldest
methods and virtually ``homogenizes'' the volume by introducing fictitious
particles that produce null collisions where the ray is extended without
a change in direction. Other estimators include decomposition
tracking~\cite{Kutz2017SDT}, residual ratio tracking~\cite{novak:2014}, or
unbiased ray marching using power series expansion~\cite{kettunen:2021}.}

Combined with spatially varying local majorants~\cite{novak:2014}, unbiased
transmittance and radiance estimates can be found much faster than with
quadrature techniques, which is attractive for interactive exploration. In this sense, local
majorants allow for the equivalent of empty space skipping and adaptive
sampling, which are popular acceleration techniques in scientific volume
rendering~\cite{zellmann:2019b,Zellmann2021,wald:owlDVRSpaceSkipping}. In
contrast to approaches that approximate the local frequency using the number of
local cells and their sizes~\cite{morrical2019spaceskip}, majorants directly
consider the actual frequency of the, which is represented by the local
extinction.

Commonly used data structures that allow for traversal with local majorants are
kd-trees~\cite{Yue:2010} as well as uniform macrocell grids that are
traversed using the 3D digital differential analyzer (DDA)
algorithm~\cite{szirmaykalos2011freeps}. \removed{Bounding volume hierarchies are less
useful to this end as they in general do not provide a \emph{spatial} subdivision;
Morrical et al.~[] have explored this option to
make use of hardware ray tracing, but found that grids are generally superior.
If the domain boxes of the BVH do not overlap, as is sometimes the case with
structured or semi-structured data, BVHs can still be used, but then are
semantically equivalent to spatial data structures}\added{Bounding
volume hierarchies (BVH) have received less attention as direct volume rendering accelerators,
as the predominant techniques in sci-vis traditionally required ordered traversal.
Morrical et~al.~\cite{Morrical2022QuickClusters} used BVHs to accelerate an interactive
volume path tracer, but concluded that a simple DDA grid implementation was more efficient.
Approaches using BVHs as empty space skipping accelerators for volume rendering with
front-to-back alpha compositing
required the bounding boxes at the leaf nodes to not overlap~\cite{zellmann:2019}.}
The OpenVDB data structure that is popular in production rendering uses a
hierarchical grid and DDA traversal with a fixed hierarchy size~\cite{openvdb}.
Hofmann et~al.~\cite{hofmann:2021} also provide a recent summarization of multi-level DDA traversal for volumetric path tracing on GPUs.


The term \emph{adaptive mesh refinement} (AMR) was originally coined by Berger
and Collela~\cite{berger:1984,berger:1989} at NASA. It refers to simulation
codes that hierarchically subdivide a (often uniform) grid, both in space and
in time, in regions where the simulation domain is more interesting than in
others. Subdivision schemes include octrees~\cite{p4est}, or block-structured
AMR~\cite{chombo} where the rectangular regions are allowed to be irregular.

Virtually all existing AMR simulation codes (e.g., FLASH~\cite{Dubey2008},
Lava~\cite{lava}, etc.) output data that is cell-centric and results in \mbox{\emph{T-junctions}}
at odd level boundaries. Work by Wald et al.~\cite{wald:17:AMR}
and by Wang et al.~\cite{wang:18:iso-amr} has proposed interpolators that
are continuous even at level boundaries, which is important for
artifact-free visualization. These interpolators use sampling
routines that perform neighbor queries by traversing tree data structures.
If not carefully designed they require traversing those data structures per sample
taken, resulting in non-trivial reconstruction costs infeasible for GPUs.
\exabrick{} by Wald et al.~\cite{wald2021exabrick}, which is explained in more
detail in \cref{sec:exabrick}, provides an acceleration structure to accomplish
fast AMR cell location on GPUs.

For rendering octree-based AMR specifically, 
Wang et al.~\cite{wang2020cpu} developed high-quality reconstruction filters for direct ray tracing.
Other authors have also focused on large-scale out-of-core AMR rendering, such as the interactive streaming and caching framework proposed by Wu et al.~\cite{wu2022pgv} targeted at CPU rendering,
or the framework by Zellmann et al.~\cite{zellmann2022exajet-animated}, who proposed a similar method for AMR streaming and rendering on GPUs, but focused on time-varying data.

\added{GPU path tracing native AMR data, to our knowledge, has not been proposed by
the literature yet. Closest to this are OpenVKL~\cite{Knoll2021} that can path
trace on the CPU but does not use any of the optimizations necessary for
GPUs, and the framework proposed by Zellmann et al.~\cite{stitcher} that
renders the AMR data by converting it to an unstructured mesh with voxels.}

\section{Background and Method Overview} \label{sec:method-overview}
\added{In this section we discuss how to integrate volumetric
path tracing (VPT) into a framework like Wald et al.'s
\exabrick{}~\cite{wald2021exabrick}, which can render AMR data natively. A key
requirement of our method is that VPT is compatible with RGB$\alpha$ transfer
functions that contribute albedo and extinction coefficients.}

\subsection{\added{Scientific Visualization and Volumetric Path Tracing}}
\algdef{SE}[DOWHILE]{Do}{DoWhile}{\algorithmicdo}[1]{\algorithmicwhile\ #1}%
\newcommand{\Break}{\State \textbf{break} }
\begin{algorithm}[t]
\begin{algorithmic}[1]
\Function{DeltaTracking}{$o$, $\omega$, $t_{min}$, $t_{max}$}
    \State $t = t_{min}$
    \For{$all~segments~\langle t_{0i},t_{1i},\bar \mu_{i} \rangle \in range\left(t_{min},t_{max}\right)$}
        \Do
        \State $t_{curr} = max({t, t_{0i}})$
            \State $\zeta =$ \Call{rand()}{}
            \State $t_{curr} = t_{curr} - \frac{\log{(1-\zeta)}}{\bar \mu_{i}}$
            \If{$t_{curr} \ge t_{1i}$}
                \Break
            \EndIf{}
            \State $\xi =$ \Call{rand()}{}
        \DoWhile{$\xi > \frac{\mu(o+t_{curr}*\omega)}{\bar \mu_{i}}$}
        \If{$t_{curr} < t_{1i}$}
            \State $t = min({t, t_{curr}})$
        \EndIf{}
    \EndFor
    \State \Return{t}
\EndFunction
\end{algorithmic}
\caption{\label{alg:woodcock}%
Delta tracking over the range $\left(t_{min},t_{max}\right)$, which is subdivided into
segments $\langle t_{0i},t_{1i},\bar \mu_{i} \rangle$ that have local majorants
$\bar \mu_{i}$.}
\end{algorithm}
\added{In scientific visualization (sci-vis), volume renderers traditionally
use the absorption plus emission model~\cite{max:1995}, often implemented with
ray marching~\cite{levoy:1988}, where for each screen space sample the volume
is sampled at positions with uniform step size $dt$. At each position an
indirect RGB$\alpha$ transfer function lookup provides color and opacity. Final
colors for the screen space sample are obtained via alpha
compositing~\cite{Porter84}}.

\added{Rendering with volumetric path tracing (VPT) is fundamentally different
in that opacity sampling and compositing are replaced with stochastically
sampling the transmittance of light along a ray segment. The light then
possibly (but not necessarily) bounces. Final screen space samples are averaged
in the accumulation buffer instead of using alpha compositing. Despite (or
probably even because of) those differences, the approach has merits also for
sci-vis where it so far has not gained much attention yet.}

\added{The most notable advantage is that transmittance
estimates are computed with much fewer volume samples than with ray marching.
This comes at the cost of Monte Carlo noise, which however tends to converge
over a few samples. Where the fast transmittance estimates pay off
particularly is shadow computation: with a ray marcher, marching \emph{another
volume ray} towards the light source at \emph{each} sample position is
infeasible. With VPT a shadow ray is traced only when a \emph{collision}
occurs. This happens when (stochastically) the light is fully extinct and the
\emph{free flight distance} reached. Another advantage is that VPT is
unbiased, which can give better correctness guarantees. Finally, it is
perfectly viable to implement the absorption plus emission model based on the
transmittance estimate routine used in VPT, and by that improving the time to
(first) image, yet at the cost of introducing some noise.}


\added{We add VPT to the \exabrick{} framework~\cite{wald2021exabrick}. A key
change necessary is replacing the ray marcher used by \exabrick{} with a
tracking estimator that computes stochastic \emph{free fight distances} that
light passes through the volume without being blocked. Considering the Woodcock
(also delta-) tracking algorithm~\cite{woodcock:1965} (cf. \cref{alg:woodcock})
as an example, this estimator employs two stochastic processes--rejection and
inversion sampling; rejection sampling depends on the so called \emph{majorant
extinction coefficient} for a region of space (in the following called $\bar
\mu$, or $\bar \mu_{i}$ if multiple majorants are encountered along a ray
segment).}

\added{In \cref{alg:woodcock} a ray, given by its origin $o$, direction
$\omega$, and range $[t_{min},t_{max}]$, is traversed through the volume.
Rejection sampling decides if a \emph{null collision} occurred; in that case,
if the ray is still inside the tested segment, traversal continues in the same
direction and another sample must be taken. This decision is made in line $12$
of \cref{alg:woodcock}. The closer the majorant $\bar \mu_{i}$ matches the
\emph{actual} density $\mu(x)$ at the sample position $x \in \mathbb{R}^3$, the
less often the \textbf{do...while} loop is executed, and the less often the
volume is sampled.}

\added{This gives rise to acceleration data structures over regions of
space---or groups of objects---to find majorants that tightly bound the local
density. This is indicated in \cref{alg:woodcock} by the outer \textbf{for}
loop over all segments $\langle t_{0i},t_{1i}, \bar \mu_{i} \rangle$ that the
range $[t_{min},t_{max}]$ is divided into; in practice, those segments are
provided by a spatial accelerator like a grid or kd-tree, or by a bounding
volume hierarchy (BVH). Below we explain in more detail how to implement such
acceleration structures.}

\added{A requirement of sci-vis is that the volumes are color-mapped using
RGB$\alpha$ transfer functions; then $\alpha$ becomes the extinction
coefficient $\mu(x)$ from \cref{alg:woodcock}. RGB is interpreted as albedo,
which gets evaluated during shading. Since $\mu(x)$ now depends on the alpha
transfer function, indirectly, so do the majorants $\bar \mu_{i}$. When
spatially varying majorants are used as an acceleration data structure, hence,
it is required that majorants can be recomputed interactively when the
RGB$\alpha$ map changes. However, na\"ively iterating over every cell of the volume to
recompute the majorants is infeasible for typical sci-vis data set
sizes.}

\added{Instead, we pre-process the data storing, the \mbox{min/max} ranges with
the accelerator leaves, as suggested by Knoll et~al.~\cite{knoll-min-max}. When
the RGB$\alpha$ map changes, this allows us to iterate over the accelerator
leaf nodes and use the \mbox{min/max} ranges as indices into the RGB$\alpha$
map. The majorant for the accelerator leaf is the maximum $\alpha$ value the
RGB$\alpha$ map takes on inside that range. In this paper, we assume that
majorants are computed indirectly to maintain interactivity when updating
transfer functions, but also seek to answer the question how much worse this is
compared to a hypothetical pre-classified data structure without this
restriction.}


\vspace{-0.5em}
\subsection{Extending the \exabrick{} Framework} \label{sec:exabrick}
The \exabrick{} framework and data structure have been explained in detail by
prior
works~\cite{wald2021exabrick,zellmann2022cise,zellmann2022exajet-animated}. We
refer the reader to these for details. \exabrick{} is a \emph{native} AMR data
structure, i.e., it does not resample or transform the AMR subgrids to
unstructured elements, but samples the given cells directly. In its original
form, \exabrick{} was optimized to perform ray marching on GPUs. We extend
this data structure to support volumetric path tracing. To our knowledge,
volumetric path tracing using a native AMR sampling structure has not yet been
proposed by the literature.

\exabrick{} first generates bricks of same-level cells in the spirit of
K{\"a}hler et al.~\cite{kaehler:2003} so the builtin ray marcher can traverse
relatively large coherent regions of space without requiring hierarchy
traversals.\removed{(cf. Fig~2a).} Those bricks are much bigger in size than
the original octree leaf nodes or block-structured AMR subgrids.

At the level boundaries, the bricks overlap by half a cell, to support
reconstruction with tent-shaped basis functions\removed{ (cf. Fig~2b)}. In the
resulting \emph{overlap regions}, filtered data values are obtained as a linear
combination of the overlapping cell values. To make the non-convex overlap
regions more rendering-friendly, Wald et al.\ subdivide them using a kd-tree.
The domain boxes of the kd-tree leaves form what the authors call the
\emph{active brick regions} (ABRs).

\added{An OptiX BVH is used to traverse the ABRs in hardware. Empty ABRs are
culled when the alpha transfer function changes; adaptive sampling is achieved
by adjusting the marcher's step size to the size of the finest cell covered by
the ABR.}\removed{Finally, Wald et al. build an OptiX
BVH over the ABRs to traverse them in hardware. The BVH is also used for empty
space skipping, by pre-classifying the ABRs using the \mbox{min/max} values of
the cells they cover when the alpha transfer function changes; empty ABRs are
culled completely by the hardware during traversal. The OptiX BVH is also used
for adaptive sampling by setting the ray marching step size to the finest cell
size of the bricks covered by the ABR.  Note that the \exabrick{} data
structure solves the cell location and adaptive sampling problems at once; the
ABR BVH allows to quickly find a spatial partition and associated sampling
rate, while taking samples and locating the constituent cells is efficiently
implemented by iterating over and accumulating the contributions of the bricks
that the ABR covers; for that, each ABR stores a list of brick IDs.}
\added{We extend the data structure to support VPT. For that we need to be able
to support ray segment traversal as in \cref{alg:woodcock}, to obtain local
majorants $\bar \mu_{i}$ that tightly bound the density along those segments,
and support for point location queries to compute the local density and
extinction coefficients $\mu(x)$ using an AMR reconstruction filter.}

\begin{figure}[tb]
 \begin{tikzpicture}
 \node[anchor=south west,inner sep=0] (image) at (0,0) {\includegraphics[width=0.49\columnwidth]{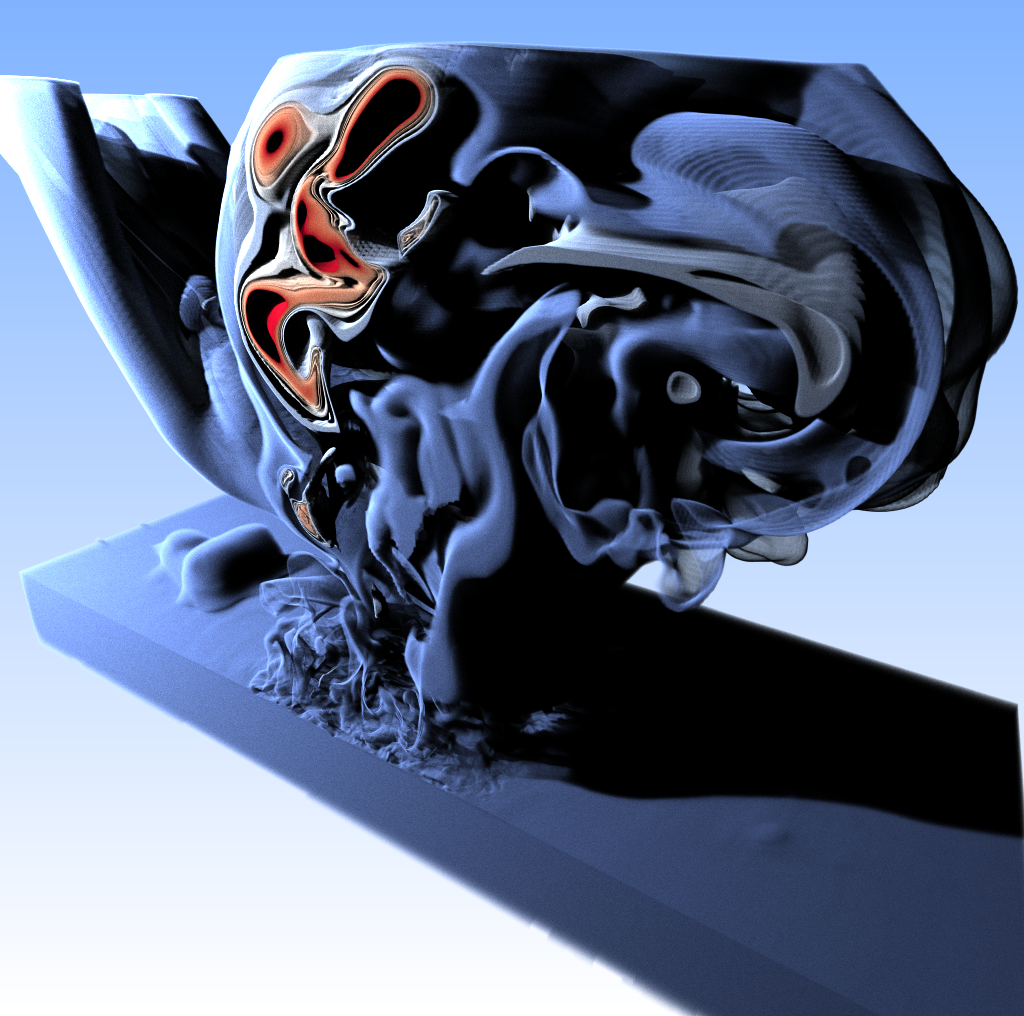}};
 \begin{scope}[x={(image.south east)},y={(image.north west)}]
 \draw[white,ultra thick,rounded corners] (0.575,0.000) rectangle (0.745,0.170);
 \end{scope}
 \begin{scope}[x={(image.south east)},y={(image.north west)}]
 \draw[white,ultra thick,rounded corners] (0.169,0.251) rectangle (0.332,0.406);
 \end{scope}
 \end{tikzpicture}
 \begin{tikzpicture}
 \node[anchor=south west,inner sep=0] (image) at (0,0) {\includegraphics[width=0.49\columnwidth]{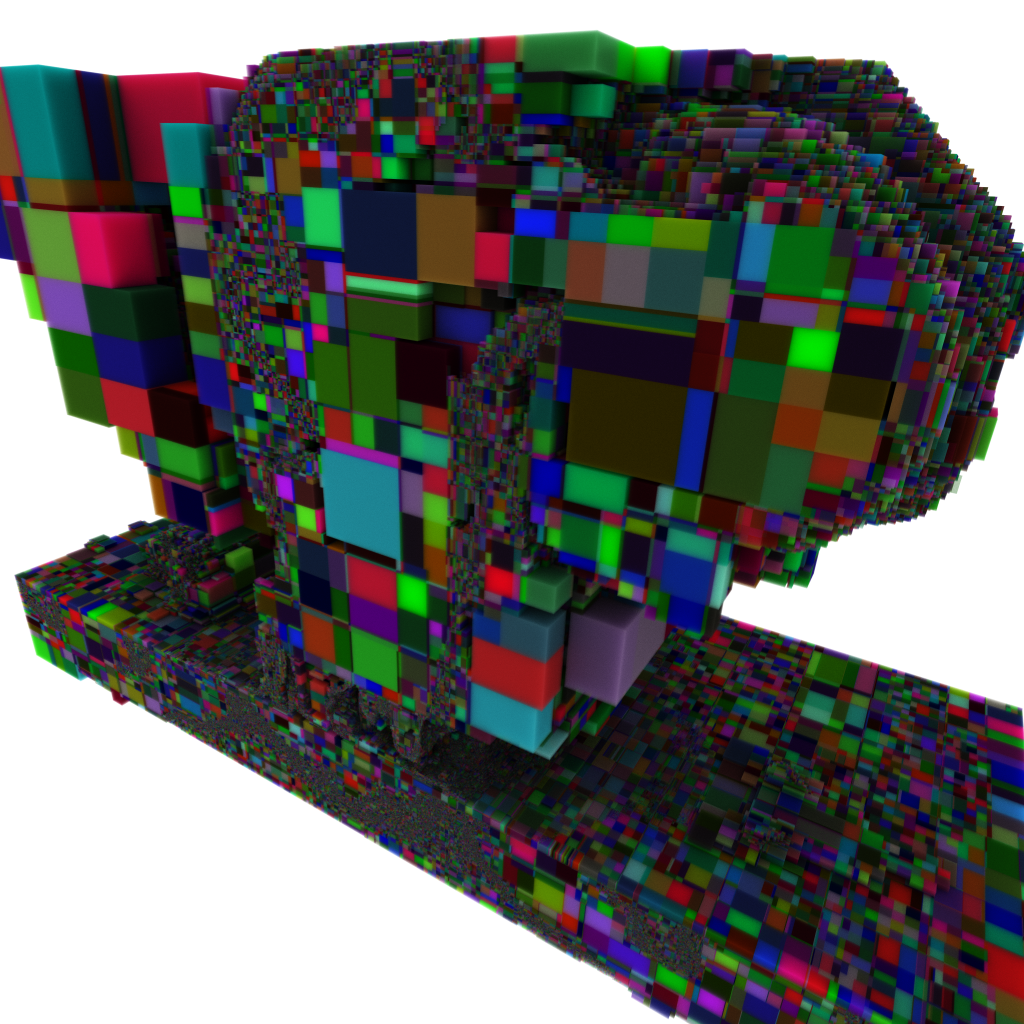}};
 \begin{scope}[x={(image.south east)},y={(image.north west)}]
 \draw[red,ultra thick,rounded corners] (0.575,0.000) rectangle (0.745,0.170);
 \end{scope}
 \begin{scope}[x={(image.south east)},y={(image.north west)}]
 \draw[red,ultra thick,rounded corners] (0.169,0.251) rectangle (0.332,0.406);
 \end{scope}
 \end{tikzpicture}
 \\
 \includegraphics[width=0.49\columnwidth]{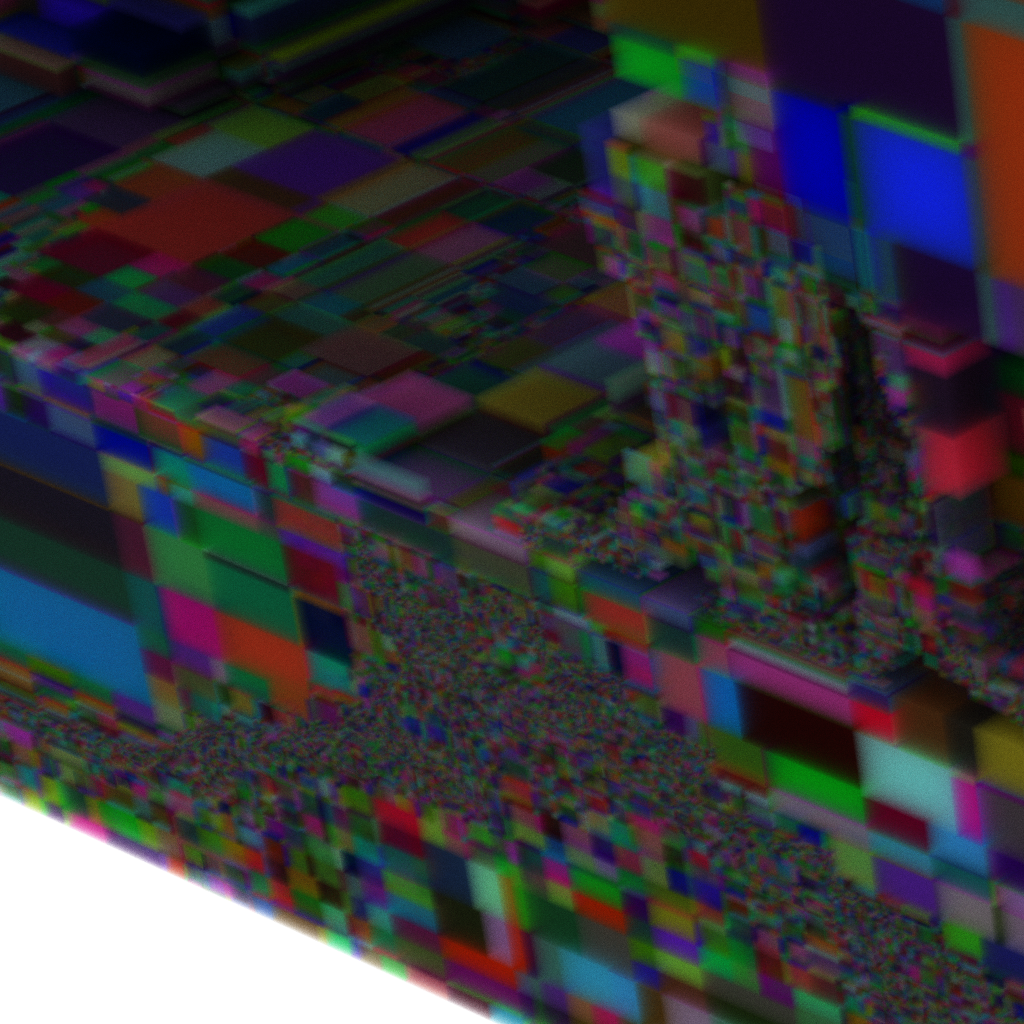}
 \includegraphics[width=0.49\columnwidth]{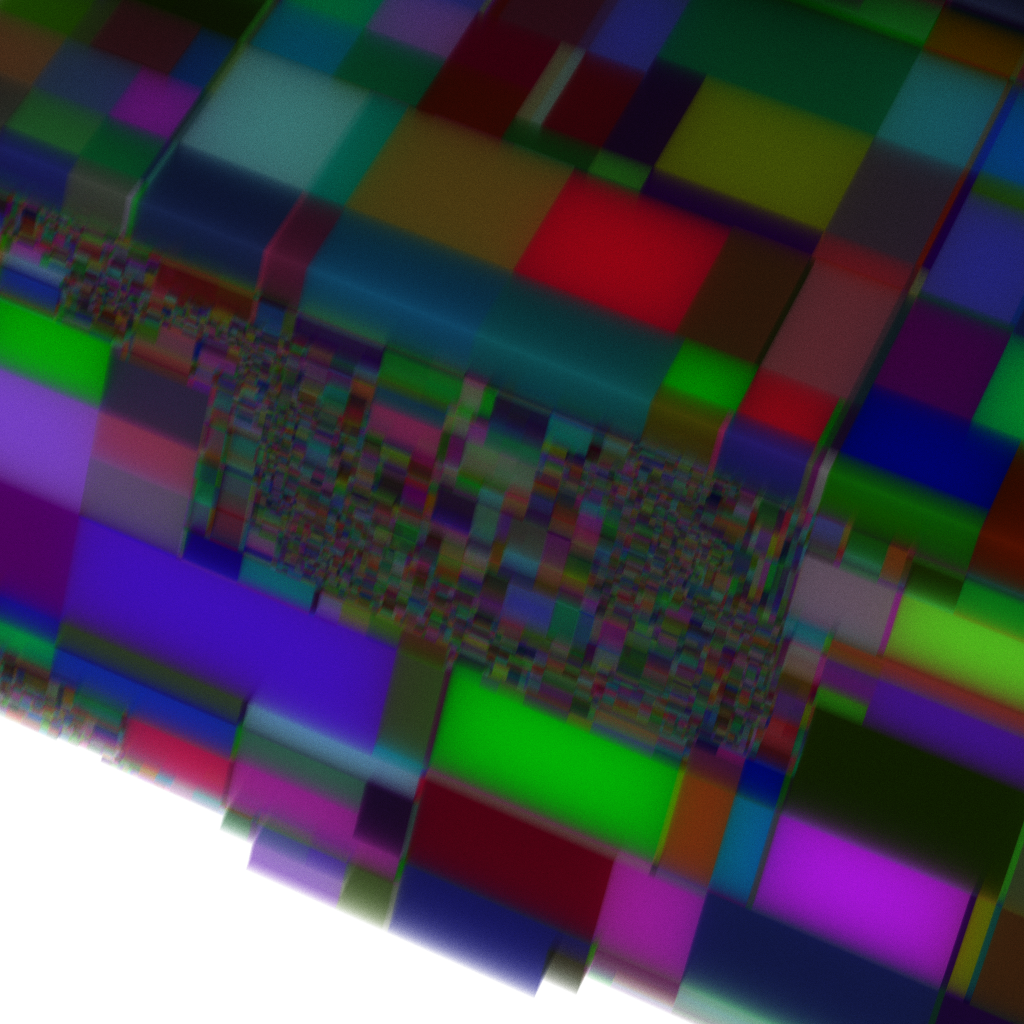}
\vspace{-2pt}
\caption{\label{fig:abrs}%
Visualization of \emph{active brick regions} (ABRs) for the LANL impact data
set. Top left: volume rendering. Top right: the extinction of the
ABRs is set to their majorant extinction, the RGB albedo is derived from the ABR
ID. Bottom: the two insets from above; ABRs vary in size and shape, and so do
their majorants.
\vspace{-2.0em}
}
\end{figure}

\begin{figure}[tb]
\centering
\begin{tikzpicture}
\node[anchor=south west,inner sep=0] (image) at (0,0) {\includegraphics[width=0.99\columnwidth]{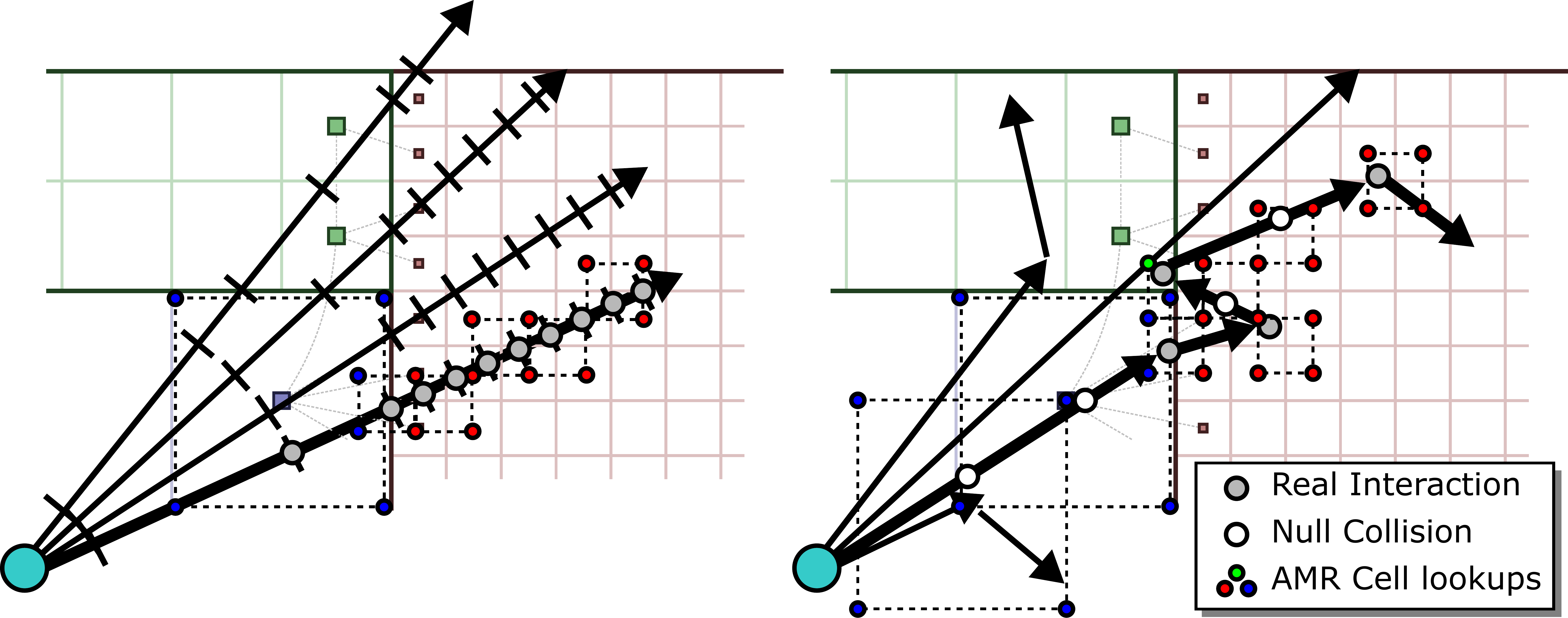}};
\node[anchor=south west,inner sep=0] at (1.6,-0.4) {(a)};
\node[anchor=south west,inner sep=0] at (6.1,-0.4) {(b)};
\end{tikzpicture}
\vspace{-3pt}
\caption{\label{fig:marcher-vs-pt}%
Traversal and cell location with ray marching and with Woodcock tracking.
(a) A marcher amortizes traversal costs when all rays in a warp take
multiple samples from the same ABR or brick.
(b) A path tracer often traverses the ABRs or bricks, only takes a single
sample, and then leaves the region again due to scattering, or due to
non-uniform steps taken due to null collisions.
}
\end{figure}

\added{The differences between a volume ray marcher and volumetric path tracer,
and the challenges regarding memory access patterns are illustrated in
\cref{fig:marcher-vs-pt}. In the paper we explore different acceleration
structures, including directly using the ABRs for segment traversal and sample
reconstruction, but also evaluate alternative strategies. As in surface ray
tracing, the quality of the acceleration structure has a large influence on
performance, and it is unclear if the ABRs serve that purpose well. An example
visualization of majorants obtained from ABRs is shown in \cref{fig:abrs}.}

\added{We use CUDA and OptiX~7 to implement our extensions as this is the
software framework also used by ExaBrick. We try to make use of hardware ray
tracing where possible to accelerate segment traversal and sampling the AMR
data structure.}

%

\removed{\section{Problem Statement}} \label{sec:problem}
\begin{figure}[tb]
\centering
\includegraphics[width=0.32\columnwidth]{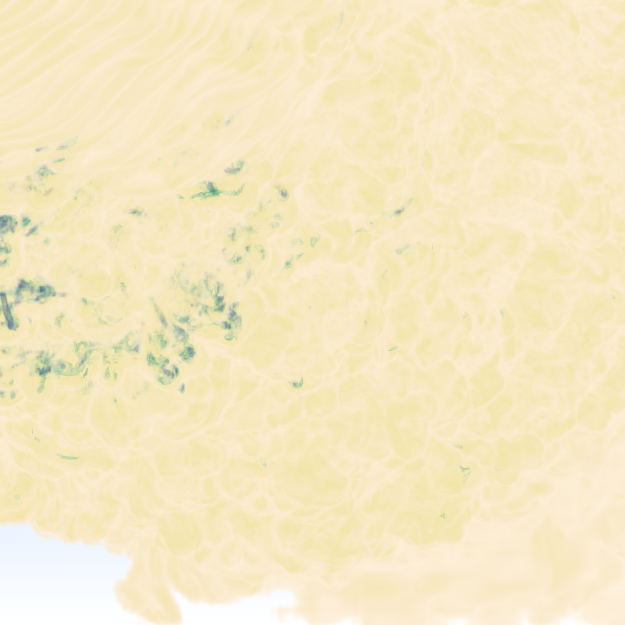}
\includegraphics[width=0.32\columnwidth]{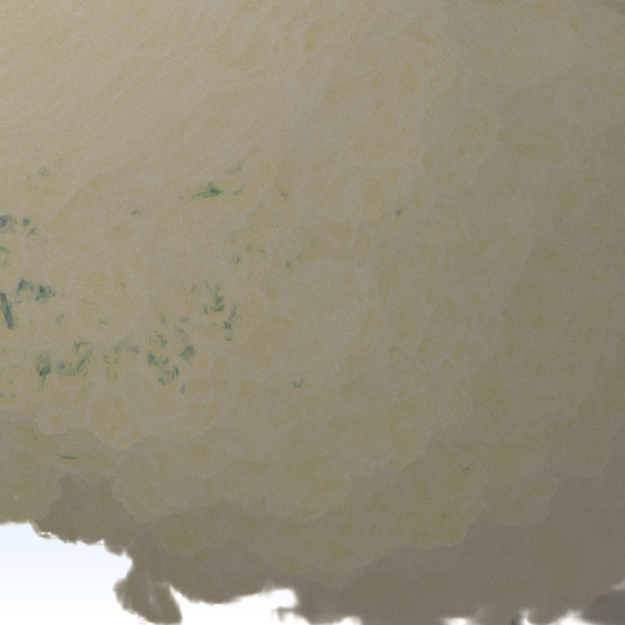}
\includegraphics[width=0.32\columnwidth]{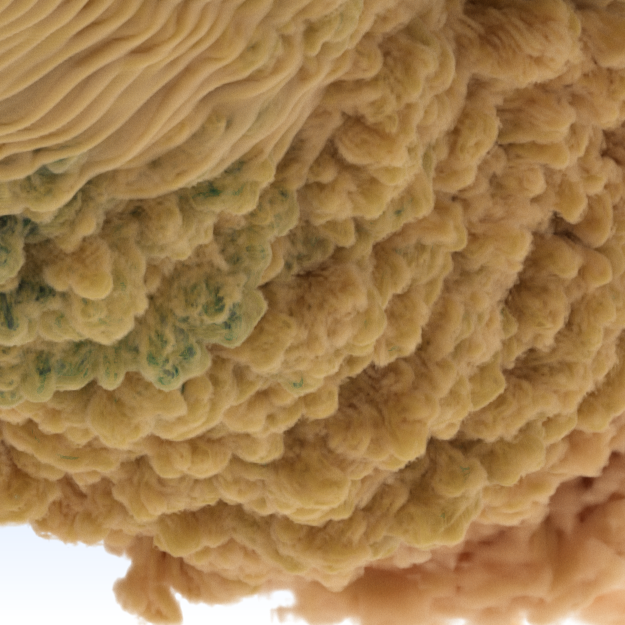}\\
\includegraphics[width=0.32\columnwidth]{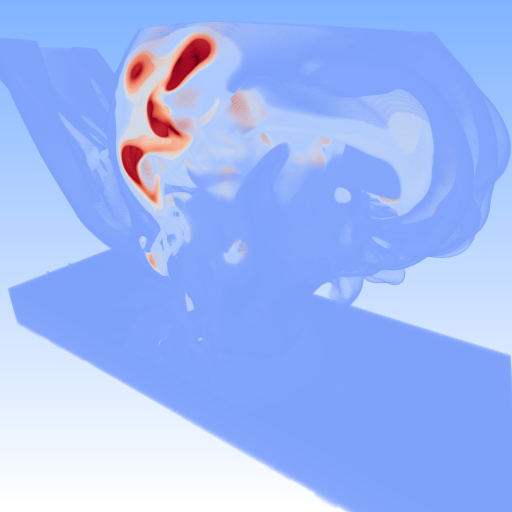}
\includegraphics[width=0.32\columnwidth]{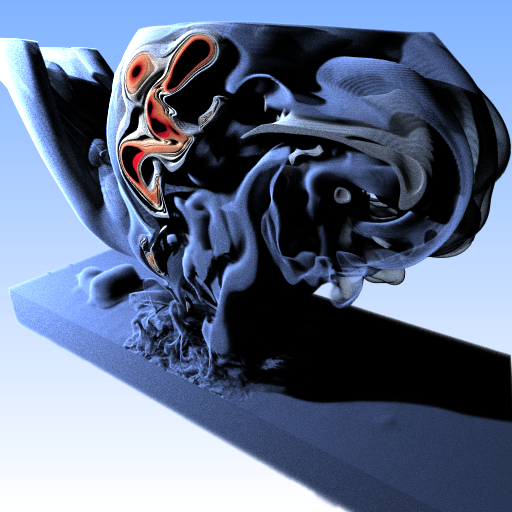}
\includegraphics[width=0.32\columnwidth]{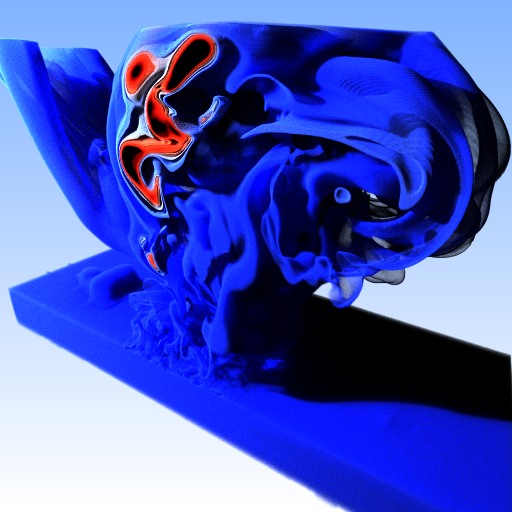}
\vspace{-1em}
\caption{\label{fig:render-modes}%
\added{AMR volumetric path tracing with different quality settings.
Top: exajet vorticity field, bottom: LANL Impact, t=46112.  Left:~absorption
plus emission (lighting model used by sci-vis ray marchers).
Middle:~path traced single scattering with point light source. Right:~full
global illumination using multi scattering.}
\vspace{-2em}
}
\end{figure}
\removed{In this paper we address the problem of computing free-flight distances for
transmittance estimates with Woodcock tracking~[] using AMR
volumes.
For interactivity, we need spatially varying local majorants $\bar
\mu_{i}$ associated with segments of the rays that we compute the distances
for. We extend the \exabrick{} data structure to provide majorants, and to
provide point estimates for the extinction coefficient $\mu(x)$ at locations
$x\in \mathbb{R}^3$.}

\removed{As illustrated in Fig.~2,}
\removed{At the level boundaries, correct local
majorants are defined as the maximum extinction of all the overlapping
\emph{filter domains}. One data structure to achieve this are the ABRs, which
can be traversed to obtain majorants and at the same time can also be used to
sample $\mu(x)$ directly.
Generally, we need a data structure to obtain \emph{some} majorant as
a conservative upper bound representing all the densities in that region,
and the ABRs and their maximum densities are one possible way to achieve that.
Ideally, the \emph{traversal} data structure used to obtain majorants can also
directly be used to sample the point extinction coefficient $\mu(x)$; if the
data structure does not allow that, we need an additional data structure for
sampling.}

\removed{An assumption made by \exabrick{} is that, once an ABR was found, ABR BVH
traversal costs are amortized by taking multiple samples from the spatial
partition, either on the same ray, or on multiple rays sharing a CUDA warp.
This assumption of cost amortization through coherent data accesses suits the
typical access patterns of a sci-vis volume ray marcher as shown in
Fig.~3a.}

\removed{This cost amortization is not so obvious if the workload is incoherent.
The highlighted (third from left) camera path in Fig.~3b
is typical for multiscattering; the path is subject to null collisions
and scatters arbitrarily, and while doing so, crosses level boundaries over
and over. At level boundaries, different cells from different bricks must be
accessed and majorants be recomputed. In contrast, a volume ray marcher would
predictably only cross level boundaries once the ray enters and exits the
spatial partition. In fact, the whole traversal of a ray marcher through the
domain (apart from potential early exits due to the opacity termination
threshold being reached) can be predicted a priori.}

\removed{It is thus all but obvious if a path tracer will successfully amortize the
costs of traversing the ABR BVH by taking enough samples when a leaf was found.
And if not, it is unclear if using the sampling data structure for
traversal---after all the sampling data structure is optimized to form as large
as possible groups of same-level cells---also provides good local majorants.
Local majorants that do not tightly fit the local extinction will generate many
null collisions, which requires the volume to be sampled over and over.
Consequently, we on the one hand have to minimize the traversal costs, and on
the other hand the number of null collisions when a spatial partition was found
via traversal.}

\removed{Taking a closer look at the ABRs generated for a typical AMR data set (cf.
Fig.~2), it is questionable if they can provide good majorants. In
some regions, ABRs often span only a single finest-level AMR cell.
The costs of traversing that ABR with a path tracer can usually not be
amortized over multiple rays in a warp due to divergence. Effectively,
locating and traversing such a single cell ABR is then more costly than using
direct cell location without an acceleration structure.}

\removed{We do not know if the ABRs plus OptiX BVH form a good majorant traversal data
structure, but our observations indicate that the assumptions that may hold for
a volume ray marcher might not necessarily hold for a volumetric path tracer
as well. In the following we thus propose alternative options to accelerate
traversal and sampling, by allowing the two tasks to become \emph{decoupled}
and by using different data structures and traversal methods. We base our
implementation off the original \exabrick{} framework, which is open source and
uses NVIDIA CUDA and OptiX~7.}



\section{Sci-Vis Volumetric Path Tracer} \label{sec:pathtracer}
\added{We implement volumetric path tracing (VPT) with OptiX~7. We support three
visualization modes with different quality settings: \emph{absorption plus
emission}, realized with Woodcock tracking; \emph{single scattering}, where an
additional bounce to a light source chosen at random is performed to compute
volumetric shadows; and \emph{multi scattering} using a global phase function
that allows the light to bounce arbitrarily. The different quality settings can
have a tremendous impact on visual fidelity, as can be seen in
\cref{fig:render-modes}.}

\added{Inside the OptiX ray generation program, we generate cameras ray with random
jitter to compute screen space samples. For the tracking estimator implementation,
the rays traverse an acceleration structure (as previously described); at each
sampling position they compute the extinction coefficient $\mu(x)$ by first
sampling into the volume (using an AMR sample acceleration structure), and then
post-classifying the sample using the RGB$\alpha$ transfer function.}

\added{Given those two operations, which are described and evaluated in detail
below---segment traversal and volume sampling---free flight distances are
computed that are required by all three visualization modes. For the
absorption and emission model, the sample color is proportional to the albedo
from the transfer function at the position where the collision occurred. The
single scattering estimator casts a shadow feeler towards a randomly chose
light source, and the albedo gets weighted by the transmission between sample
position and light source that we compute using ratio
tracking~\cite{novak:2014}. In the case of multi scattering we evaluate a
Henyey Greenstein phase function and perform bounces in a \texttt{while} loop
inside the ray generation program. Based on the throughput, we terminate paths
at random using Russian roulette.}

\added{Since we interpret the RGB component of the RGB$\alpha$ transfer function as
albedo, the path throughput is an RGB tuple as well---and so is the final
screen space sample. We accumulate the screen space samples in a device-side
accumulation buffer using a weighted average. Using an interactive prototypical
viewer, whenever the camera changes, we reset accumulation. The accumulation
buffer content is periodically presented as sRGB color tone mapped and drawn
into an OpenGL pixelbuffer object mapped using \mbox{CUDA/GL} interop.}

\added{We describe how to implement the two core routines for VPT---ray segment
traversal and AMR data sampling---in the following sections. The most obvious
way to implement those operations is by using the existing ABRs (cf.\
\cref{sec:exabrick}). In the following sections, we explain the computation of
ABR majorants and introduce alternative methods for spatial and object
subdivision, which facilitate traversal and sampling of the AMR data structure.}

\section{Acceleration Structures for Traversal and Sampling}
\removed{Traversing and recovering spatially varying local majorants requires the
ray segment $[t_{min},t_{max}]$ that is integrated over to be partitioned
spatially; regardless of how this is achieved, the majorants retrieved along
the ray need to be unambiguous, and the easiest way to achieve this is by means
of a spatial partitioning of the underlying data. We take three different
subdivision schemes into account, an overview of which can be found in Fig.~5.}
\added{Volume acceleration data structures partition the ray into segments
$[t_{0i},t_{1i}]$ that each have their own majorant $\bar \mu_{i}$. Inside the
segments, the rays are sampled. In this section we describe different
strategies to implement those two operations. A restriction of hardware ray
tracing is that when a ray traverses a BVH and the intersection or hit program
is called, tracing rays against another hardware BVH is not permitted. It
is thus not valid to use one BVH to traverse segments, and while integrating
a segment traversing another BVH for sample reconstruction.}

\added{Alternatives to using two different hardware accelerated BVHs are using
software acceleration structures for either of the operations; using the same
acceleration structure for traversal \emph{and} sampling; or halting the
(outer) ray segment traversal for traversal, and then restarting it by setting
the ray's $t_{min}$ parameter to the previous segment's $t_{max}$. Each
strategy we propose implements one of those alternatives.}


\vspace{-0.5em}
\subsection{Ray Segment Traversal Data Structures} \label{sec:traversal}
%

\added{In this subsection we discuss acceleration structures for ray segment
traversal. Sampling data structures are discussed in \cref{sec:sampling}. To
facilitate interactive transfer function updates, all traversal accelerators
(except the non-interactive reference described in \cref{sec:reference}) maintain two lists
for each leaf node or leaf cell: one list with \mbox{min/max} ranges
representing the actual data, and a list of majorants that is updated when the
transfer function changes, by iterating over the list of ranges. The
non-interactive reference stores the list of majorants only and does not
support interactive transfer function updates.}

\subsubsection{\removed{Status Quo: Woodcock on ABRs}\added{Active Brick Region Traversal}} \label{sec:abr-traversal}
\begin{figure}[tb]
\centering
\begin{tikzpicture}
\node[anchor=south west,inner sep=0] (image) at (0,0) {\includegraphics[width=0.99\columnwidth]{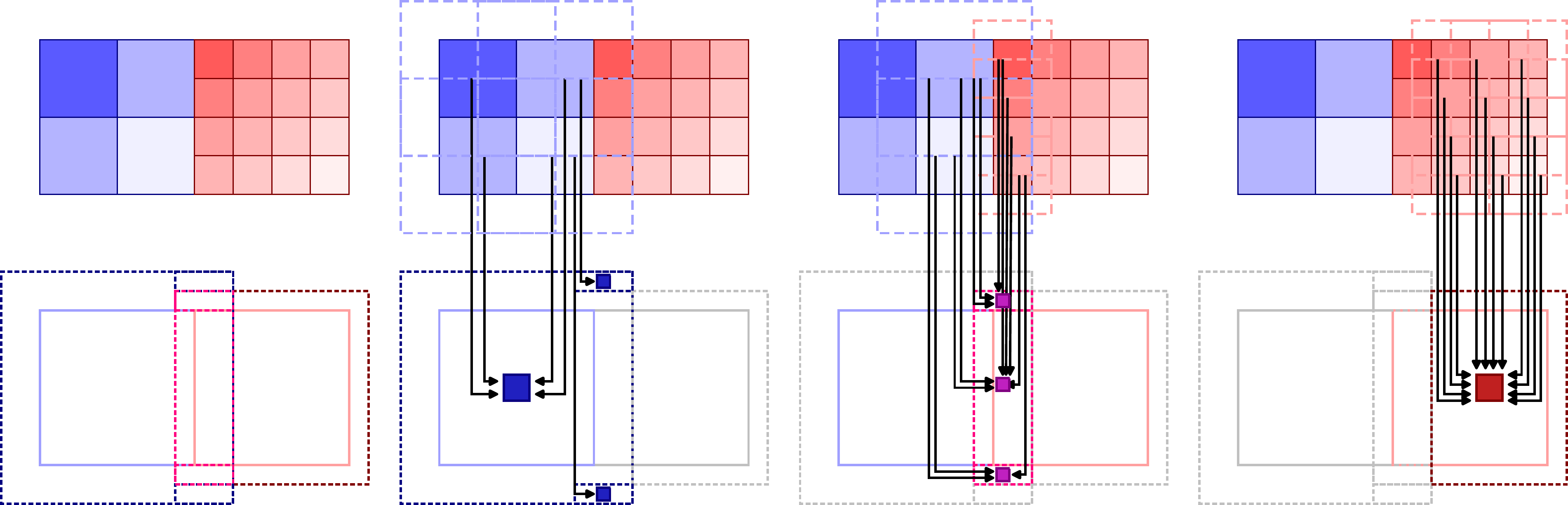}};
\node[anchor=south west,inner sep=0] at (0.9,-0.4) {(a)};
\node[anchor=south west,inner sep=0] at (3.0,-0.4) {(b)};
\node[anchor=south west,inner sep=0] at (5.2,-0.4) {(c)};
\node[anchor=south west,inner sep=0] at (7.3,-0.4) {(d)};
\end{tikzpicture}
\vspace{-2pt}
\caption{\label{fig:abr_min_max}%
\added{Computing min/max \emph{data} ranges for ABRs. (a)~Level~2 (blue) and
level~0 (red) bricks. ABRs for the two bricks are shown at the bottom.
(b)~Min/max's of the ABRs whose tent basis function domains overlap the cells from the level~2 brick
exclusively. Dashed lines show the basis domains. (c)~\mbox{Min/max} computation
for the ABRs that overlap cells from both levels. (d)~\mbox{Min/max}
computation for the level~0 cells only. The \mbox{min/max} ranges are later
used as indices into the alpha transfer function to obtain majorants on-the-fly.}
\vspace{-2em}
}
\end{figure}
One way to extend \exabrick{} to obtain local majorants
\removed{for Woodcock tracking}
is to use the ABRs directly. The ABR extents are precomputed when the data set
is loaded, and so is the list of \mbox{min/max} value ranges. One advantage of
this mode is that it utilizes the same acceleration structure for both segment
provisioning ($\langle t_{0i}, t_{1i}, \bar \mu_{i} \rangle$) and sampling, as
the ABRs also store the overlapping brick IDs and with that all the information
necessary for sample reconstruction (cf.\ \cref{sec:abr-sampling}); additional
BVH traversal is hence not required.
\removed{With a self-written BVH implementation, we could now simply continue to the
next ABR without restarting the traversal; this would require us to use a
traversal routine that visits the primitives whose bounding boxes do not
overlap in front-to-back order. While optimized software traversal methods do
that and in all likelihood this is also implemented by
OptiX, the latter does give no such guarantees. Therefore, we
resort to the safer option of restarting traversal whenever we have processed
an ABR with a ray that has not yet exited the volume.}

\removed{ABRs are constructed once when the data is loaded; each ABR also stores the
min/max range of its data. When the alpha transfer function changes, we
classify the value ranges using the alpha values to interactively obtain new
majorants.}
\added{How min/max ranges for ABRs are computed is shown in
\cref{fig:abr_min_max}. Given the two bricks at the top, we show the ABRs
resulting at the bottom of \cref{fig:abr_min_max}a. \cref{fig:abr_min_max}b--d
illustrates how the cells of the two bricks project to the ABRs based on their
basis function domains.}

\subsubsection{\removed{Traversing the Brick Spatial Decomposition}\added{Direct Brick Traversal}} \label{sec:brick-traversal}
\begin{figure}[tb]
\centering
\begin{tikzpicture}
\node[anchor=south west,inner sep=0] (image) at (0,0) {\includegraphics[width=0.99\columnwidth]{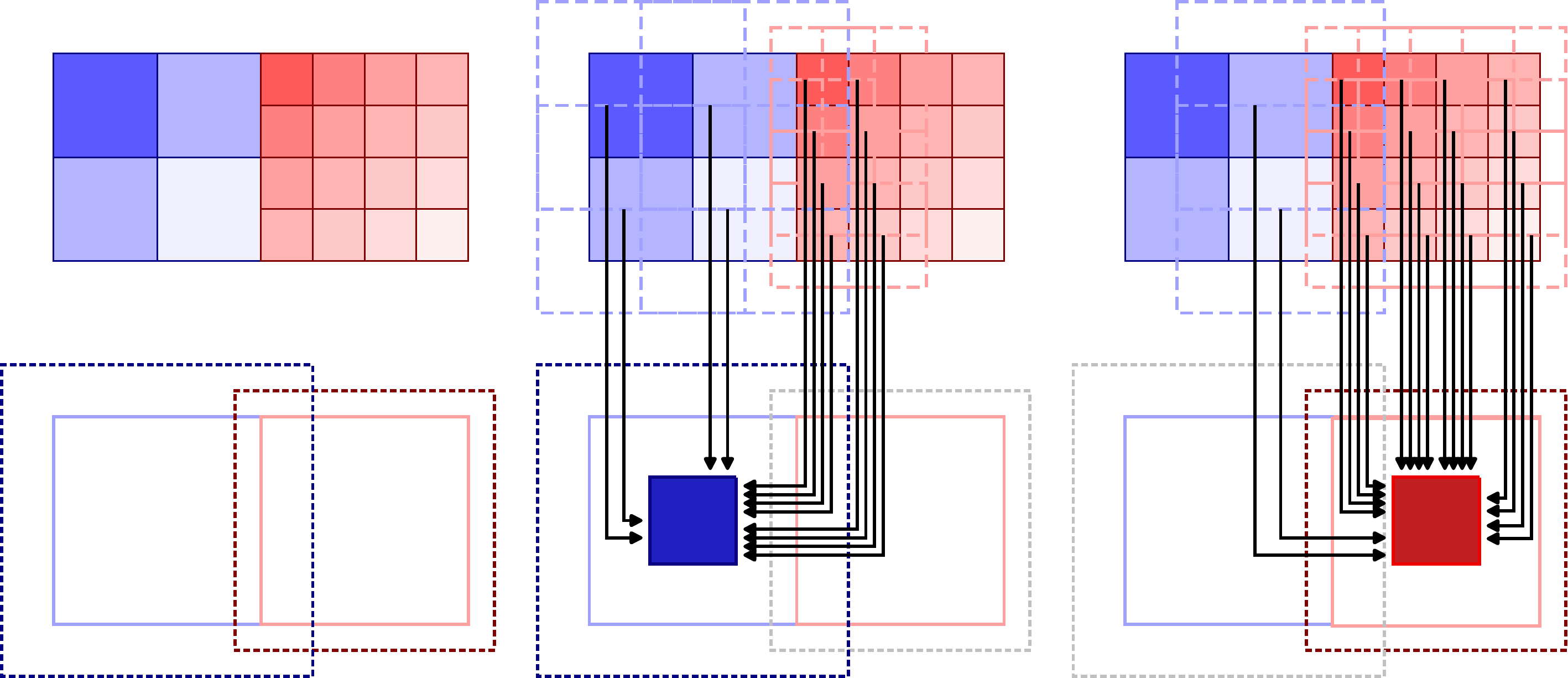}};
\node[anchor=south west,inner sep=0] at (1.25,-0.4) {(a)};
\node[anchor=south west,inner sep=0] at (4.1,-0.4) {(b)};
\node[anchor=south west,inner sep=0] at (6.95,-0.4) {(c)};
\end{tikzpicture}
\vspace{-1em}
\caption{\label{fig:extbrick_min_max}%
\added{Computing \mbox{min/max} data ranges for direct brick traversal.
(a)~Level~2 brick (blue) and level~0 brick (red) at the top, with the respective
basis function domains (dashed outlines) shown at the bottom. (b)~All cells whose
basis domains overlap the brick domains of the level~2 brick contribute their data
value. (c)~Cells that contribute their data value to the data range of the level~0
brick. Compared to ABR data ranges (cf. \cref{fig:abr_min_max}), we obtain fewer
spatial/object regions where cells overlap, but their \emph{data ranges} are much
more conservative, and so are the majorants that are computed from those data
ranges on-the-fly.}
\vspace{-2em}
}
\end{figure}
\removed{The \exabrick{} builder will only save out the leaf nodes of the
kd-tree that is used to cluster same-level cells into bricks. This gives us two
options to traverse the bricks; either with ray tracing hardware, by building
an OptiX BVH, or by saving out the kd-tree during construction, and then traversing
the tree in software. With software traversal, and particularly using a
spatial subdivision such as a kd-tree, we can implement the optimization
detailed in Section~5.1 where the whole volume is traversed in a
single sweep.}
Another data structure that is directly available to us (although we usually
do not maintain a \emph{traversal} data structure for that) are the (Exa-)bricks
themselves.
\added{While the default mode traverses ABRs, and bricks are only indirectly
accessed during ABR traversal, it is also possible to traverse the bricks directly
using an OptiX BVH.}

\removed{When constructing min/max ranges, we have to be careful that the value ranges
also account for overlapping cells from neighboring bricks, as the spatial
partitions traversed by the rays, and hence the majorants, represent the
overlap regions, and not just the brick bounds. Therefore, on construction, we
first compute the reconstruction filter domains of each AMR cell (in our case,
the cell bounds, plus half a cell's width padding at the boundary), and project
\emph{these domains} onto the bricks when computing value ranges. At runtime,
we again maintain that precomputed list of ranges and classify it when the
transfer function changes to obtain majorants.}

\removed{While the spatial partitions are coarser with this approach compared to
traversing ABRs, consequently, the majorants are more conservative than ABR
majorants; again, since the brick data structure is optimized for cell location
and (ideally) spans large regions of space, a single high-frequency cell can
easily ``infect'' a whole brick that is otherwise completely empty.}

\added{We show min/max data range computation for bricks in
\cref{fig:extbrick_min_max}. Every cell that overlaps a brick's domain
contributes its data value to the \mbox{min/max} range of that brick. We observe that
the subdivision with bricks is much coarser than the ABR subdivision (cf.
\cref{fig:abr_min_max}), so that a single overlapping cell can easily
``infect'' the whole brick with its value. A brick that would otherwise have been
empty can now become ``active'' only due to that single cell---where the ABR
subdivision would effectively have split only the region of overlap off and assigned
it its own \mbox{min/max} range.}

\added{Bricks can be traversed in two different ways---either via their bounds (solid lines
at the bottom of \cref{fig:extbrick_min_max}a--c) or
via their domains (dashed lines at the bottom of \cref{fig:extbrick_min_max}a--c). The
primary trade-off is that bounds traversal results in less overlap and potentially
less costly BVH traversal. Domains are, however, later also used for sampling
(cf.\ \cref{sec:brick-sampling}), while the bounds cannot be used for sampling directly; i.e.,
by traversing domains, we can safe a costly cell location operation that
involves a restart and another BVH traversal. Besides, we can safe memory by
storing a single data structure for traversal \emph{and} sampling.}

\subsubsection{Traversal Using a Uniform Grid} \label{sec:grid-traversal}
\begin{figure}[tb]
\centering
\begin{tikzpicture}
\node[anchor=south west,inner sep=0] (image) at (0,0) {\includegraphics[width=0.99\columnwidth]{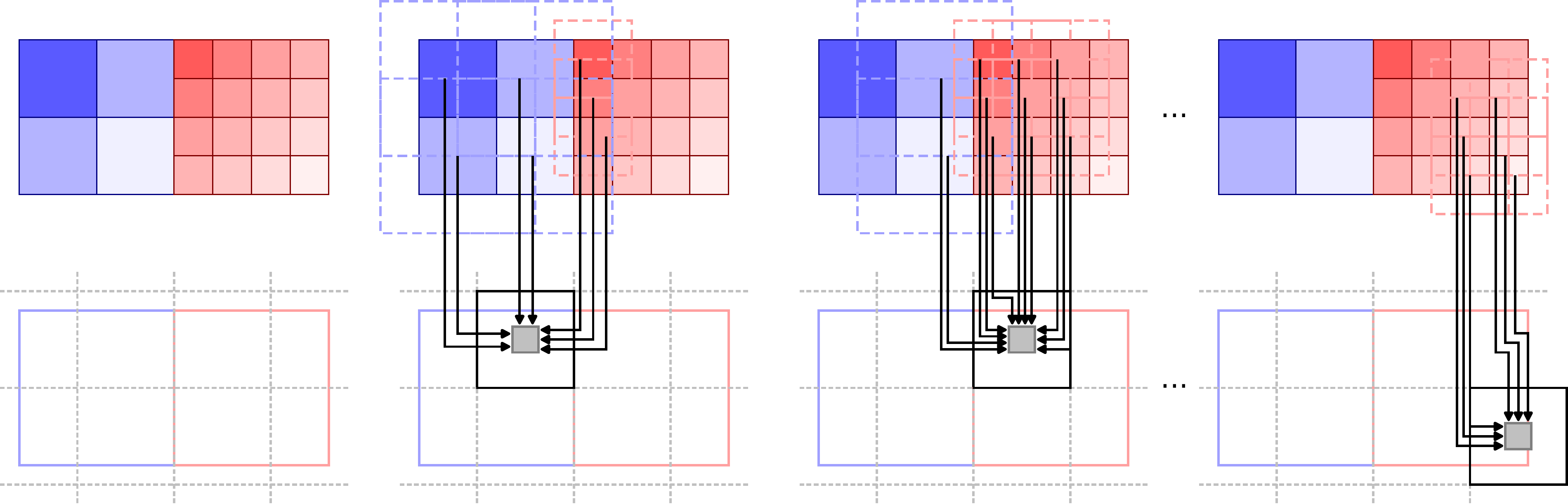}};
\node[anchor=south west,inner sep=0] at (0.9,-0.4) {(a)};
\node[anchor=south west,inner sep=0] at (3.0,-0.4) {(b)};
\node[anchor=south west,inner sep=0] at (5.2,-0.4) {(c)};
\node[anchor=south west,inner sep=0] at (7.3,-0.4) {(d)};
\end{tikzpicture}
\vspace{-2pt}
\caption{\label{fig:grid_min_max}%
\added{Projecting cell values to the min/max data ranges of the macrocells
of a uniform grid. (a)~Level~2 (blue) and level~0 (red) brick (top) and the
uniform grid to construct (bottom). (b-d)~Three macrocells, and how their value
range is affected by overlapping AMR cells and their basis function domains.}
\vspace{-2em}
}
\end{figure}
The third option we consider is using a uniform grid with macrocells;
effectively, this is the same method proposed by Szirmay-Kalos et
al.~\cite{szirmaykalos2011freeps}. Only with AMR, it is debatable if \emph{macro}cell
is a good description because coarse AMR cells can easily be several
integer factors bigger in size than the macrocell itself. It is thus also not
that simple to derive a good grid spacing from just the dimensions of the AMR
grid; as a rule of thumb a good cell spacing is one where a macrocell spans
many finest-level cells (a quantitative evaluation can also be found in
\cref{sec:eval}).

\removed{During construction, we thus have to be careful to project the AMR cells not
only to a single, but to all the macrocells that the AMR cell's filter support
overlaps.}
\added{When constructing the grids and min/max data ranges, we thus have to be careful
to compute the real union of the AMR cells' basis function domains with all
their overlapping macrocells.}
We implement this by running a CUDA kernel over all the AMR cells. Each thread
computes the domain of its cell and splats the value range onto the overlapping
macrocells in a loop.
\added{The resulting mapping is exemplarily shown for three macrocells in
\cref{fig:grid_min_max}.}

\removed{We propose two possible ways to traverse the majorant grid; the first
option uses DDA to traverse the grid as presented by
Szirmay-Kalos~[]. The second option comprises
building an OptiX BVH over the macrocells to obtain majorants
via BVH traversal. An optimization we apply to all traversal methods that use
BVHs is to cull completely empty leaves. An empty leaf is one whose majorant
extinction is $0$; before building the BVH with OptiX, we first run a CUDA
kernel over all macrocells and assemble a list of $N \times M \times K$
bounding boxes for a grid of the same size. Bounding boxes associated
with empty cells are initialized to have zero volume; the OptiX BVH builder that
is passed those bounding boxes to construct a BVH for will simply ignore them.
Again, we also associate min/max ranges with the grid cells so we can later
interactively update the majorants.}
\added{The resulting uniform grid can either be traversed in software using DDA, as
proposed by Szirmay-Kalos~\cite{szirmaykalos2011freeps},
or using OptiX: for that we first cull all the empty macrocells (subject to
the alpha transfer function), and construct axis-aligned box user primitives from
the non-empty macrocells, using the same value range/majorant logic as before.
This mode allows us to traverse the macrocells in hardware, but requires more memory
to accommodate the grid itself (which is dominated by the range and majorant lists),
\emph{plus} the OptiX BVH.}

\vspace{-1em}
\subsubsection{\added{Non-Interactive Traversal Structure}} \label{sec:reference}

\added{An interesting question we seek to answer is how bad such interactive
data structures would perform compared to a data structure that can be fully
pre-computed, given a single, hand-picked alpha function. Interactive transfer
functions add additional challenges, yet we note that even without that,
finding good majorants for AMR data given a single transfer function is far
from trivial. However one decides to build the accelerator, one will either
guide the construction by using the cells, which represent a whole set of
spatial arrangements, or the (hand-picked) transfer function, which represents
a potentially different arrangement, but optimizing the accelerator for both is
generally hard.}

\added{As a compromise that we believe should result in relatively
high-quality accelerators we implemented a kd-tree builder that
\emph{pre-applies} the RGB$\alpha$ transfer function to the volume data, then
computes splits based on the know\-ledge of how the spatial arrangement turns
out \emph{after} transfer function application. This process is guided by a
binned surface area heuristic (SAH) as proposed by Fong
et~al.~\cite{fong:2017}. Split plane placement is no longer guided by the AMR
grid, but by the pre-classified data values.}

\added{We use binning~\cite{Zellmann2021} with seven split candidates per axis,
and a priority queue-based builder to obtain a desired, pre-configurable number
of leaf nodes (for the later benchmarks empirically determined as the one
giving optimal frames/sec.\ per benchmark). We only keep the leaf
nodes and feed them into the OptiX BVH builder so we can traverse them in
hardware. This kd-tree needs to be fully rebuilt per transfer function change,
and depending on the configuration can take on the order of several hours and
longer. As such, this is an idealized benchmark telling us what performance we
\emph{could} achieve if the transfer function was known a priori.}

\added{In addition to the challenges discussed above, we found that even with
prior knowledge of the transfer function, finding optimal majorant splits is
non-trivial. First, for large volumes it is likely that the upper level split
candidates have the same majorants on each side, in which case an arbitrary
split must be chosen (the heuristic by Fong et~al.~\cite{fong:2017} biases the
split position towards the spatial median). Second, pre-classifying the AMR
cells only at the known data points is not enough to give correct majorants;
rather, when interpolating between two adjacent cells, again, any of the alpha
values in the voxel range can be assumed, so that it is again necessary to
conservatively classify with the transfer function inside the the whole value
range and not just the discrete cell values.}

\added{We observed that our biased heuristic will often generate spatial median
splits, and only generates SAH-optimized, yet still quite conservative
majorants near the leaf level. We believe that an optimal kd-tree builder would
at least be guided by the AMR cell sizes, as are the brick and ABR builders;
since interactive transfer function updates are an important objective of ours,
further investigation of this is interesting future work but not within the
scope of our paper.}

\subsection{Sample Reconstruction Data Structures} \label{sec:sampling}
Given a ray segment to integrate, the next operation that requires an
acceleration structure is sample reconstruction. We resort to two different
methods for this purpose. both use an OptiX BVH to perform cell location with
zero-length rays~\cite{morrical2019spaceskip,zellmann2022cise} whose origins
align with the sampling positions. The ultimate goal is to find all the cells
that overlap the sampling position, allowing us to compute the reconstructed
value using tent basis functions.
The basis functions are evaluated on-the-fly within an OptiX \emph{intersection} program.
\removed{The Woodcock rejection sampling loop then becomes the one in Fig.~8.
There are two options to obtain the overlap domains for that.}
\added{We explore two options to obtain the overlap regions for that, namely,
using the ABRs, and using brick domains to compute overlap regions on-the-fly.}

\subsubsection{Option 1: Sample Reconstruction via ABRs} \label{sec:abr-sampling}
The first option we propose is to locate cells with OptiX using the ABR BVH. Note
that the implementation is mostly orthogonal to which \emph{traversal} method
is used, so we can, e.g., use the ABR BVH for sampling, but use a grid or brick
subdivision for traversal. If the \emph{traversal} method also employs ABRs,
then point location traversal can be expedited by directly iterating over the
ABR's brick list. Otherwise a zero-length ray must be traced to locate the
cells necessary to reconstruct the value at $x$\removed{ (Fig.~8)}.

\subsubsection{Option 2: Extended (Exa-)Brick Sampling} \label{sec:brick-sampling}
An alternative to using ABRs is to instead perform cell location over the
bricks that the ABRs are based off. The bricks' bounding boxes form a spatial
decomposition without overlap, so we cannot use them directly. Instead, we
\emph{extend} the brick domains to include the reconstruction filter overlap, by
padding them by a half cell in each direction. Note that this is the same data
structure also used for segment traversal illustrated in
\cref{fig:extbrick_min_max}. Again, if both traversal and sampling use the same
acceleration structure, the segment integrated over can be sampled directly.
When implementing sampling this way, in overlap regions, the
OptiX intersection program will encounter multiple overlapping brick domains,
though in contrast to ABRs, there is no explicit adjacency list so the bricks
are only considered adjacent because they are visited one after another using
the sampling ray. This approach essentially computes the overlap regions
dynamically, in contrast to the pre-computed method used with ABRs.



\section{Evaluation} \label{sec:eval}
\begin{table*}[th]
\setlength{\tabcolsep}{2pt}
\begin{tabular}{c|c|ccc|c}
\toprule
Data Set & Rendering & ABR Majorants & Brick Majorants & Grid Majorants & Preclass.\ Majorants \\
\midrule
\makecell[b]{TAC Molecular Cloud\\\hline \emph{102~M cells}\\``cloud''\\\\\\\\\\} &
\includegraphics[width=0.159\linewidth]{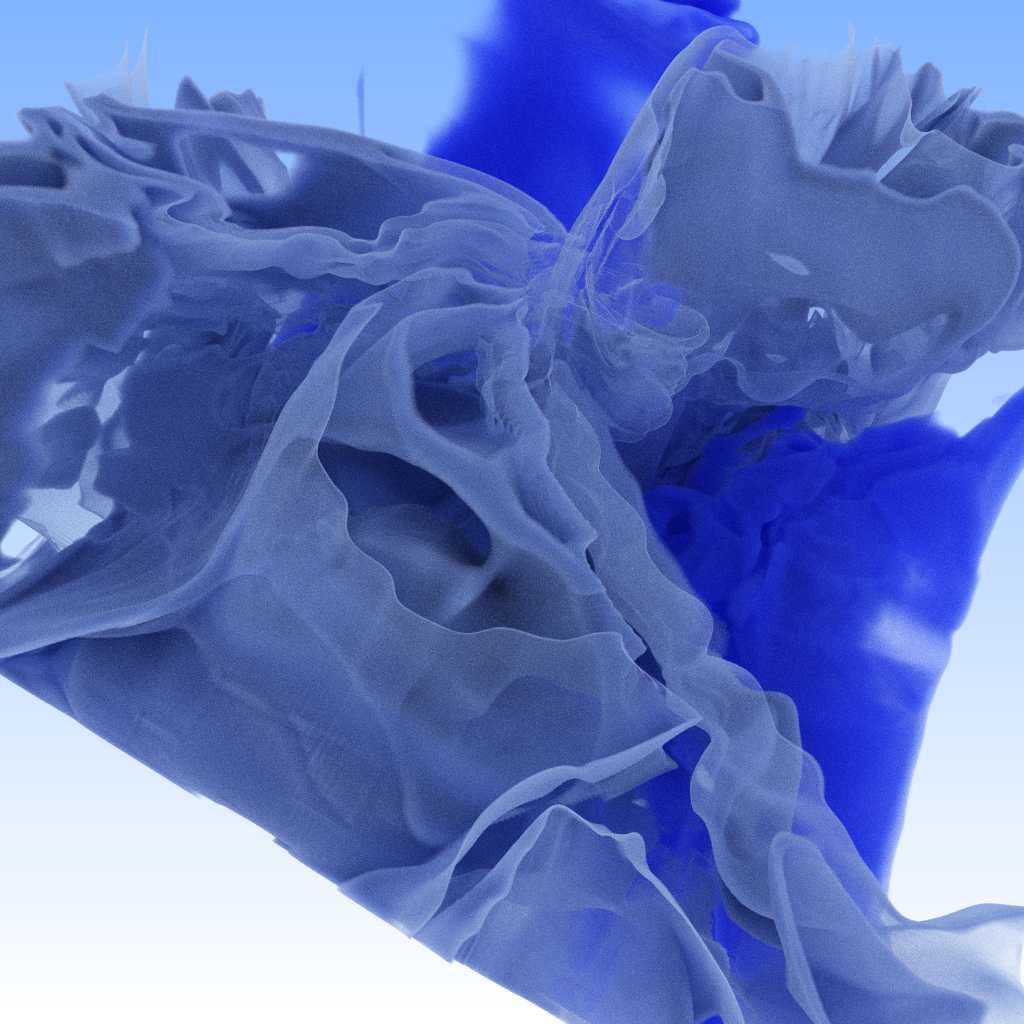}&
\includegraphics[width=0.159\linewidth]{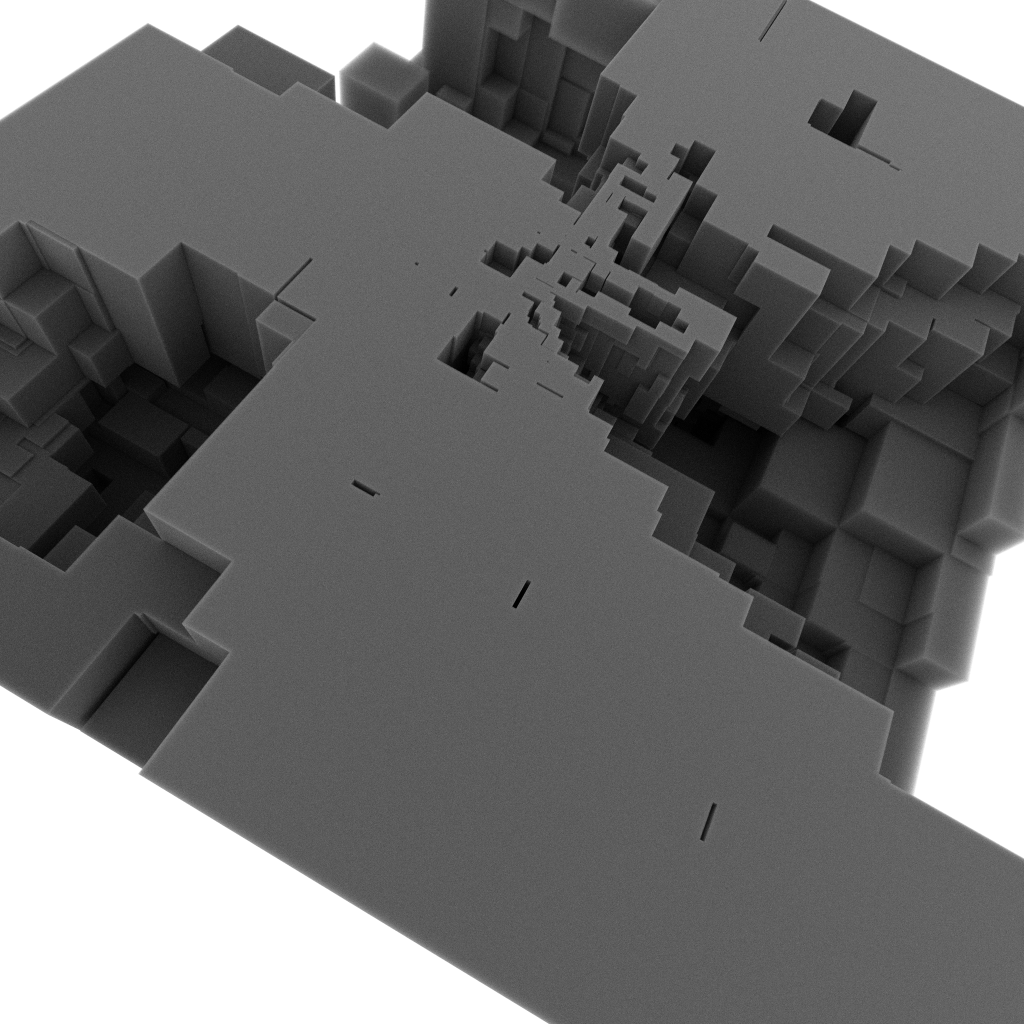}&
\includegraphics[width=0.159\linewidth]{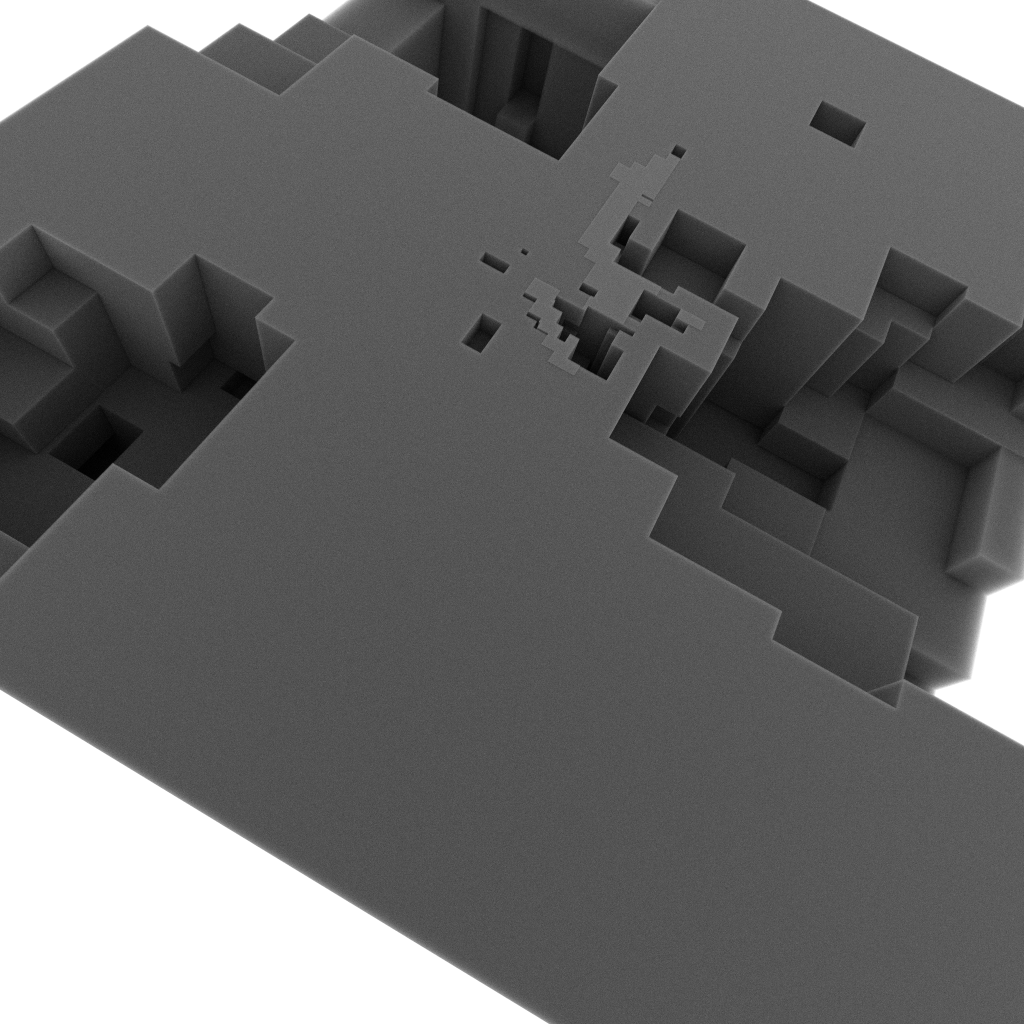}&
\includegraphics[width=0.159\linewidth]{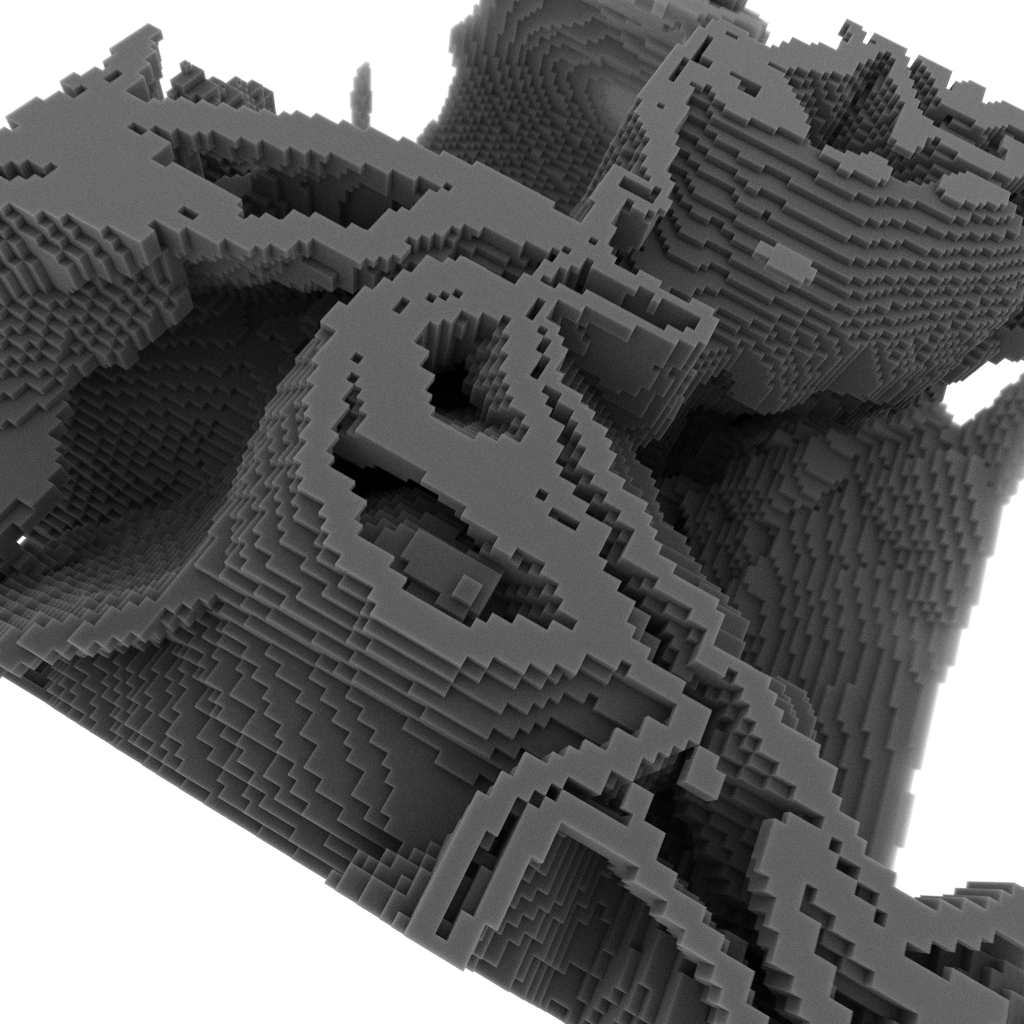}&
\includegraphics[width=0.159\linewidth]{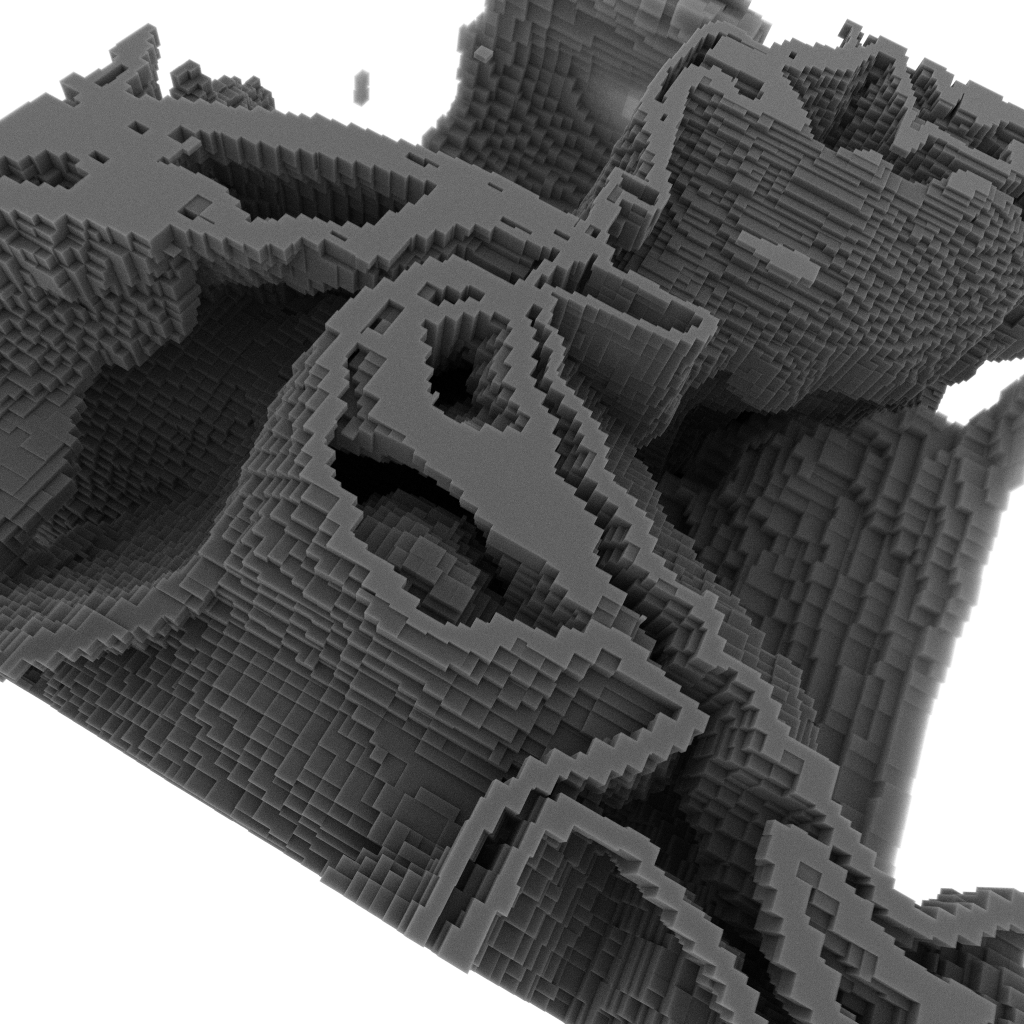}\\

\midrule

\makecell[b]{LANL Impact\\\hline t=20060\\\emph{158~M cells}\\``meteor-20k''\\\\\\\\} &
\includegraphics[width=0.159\linewidth]{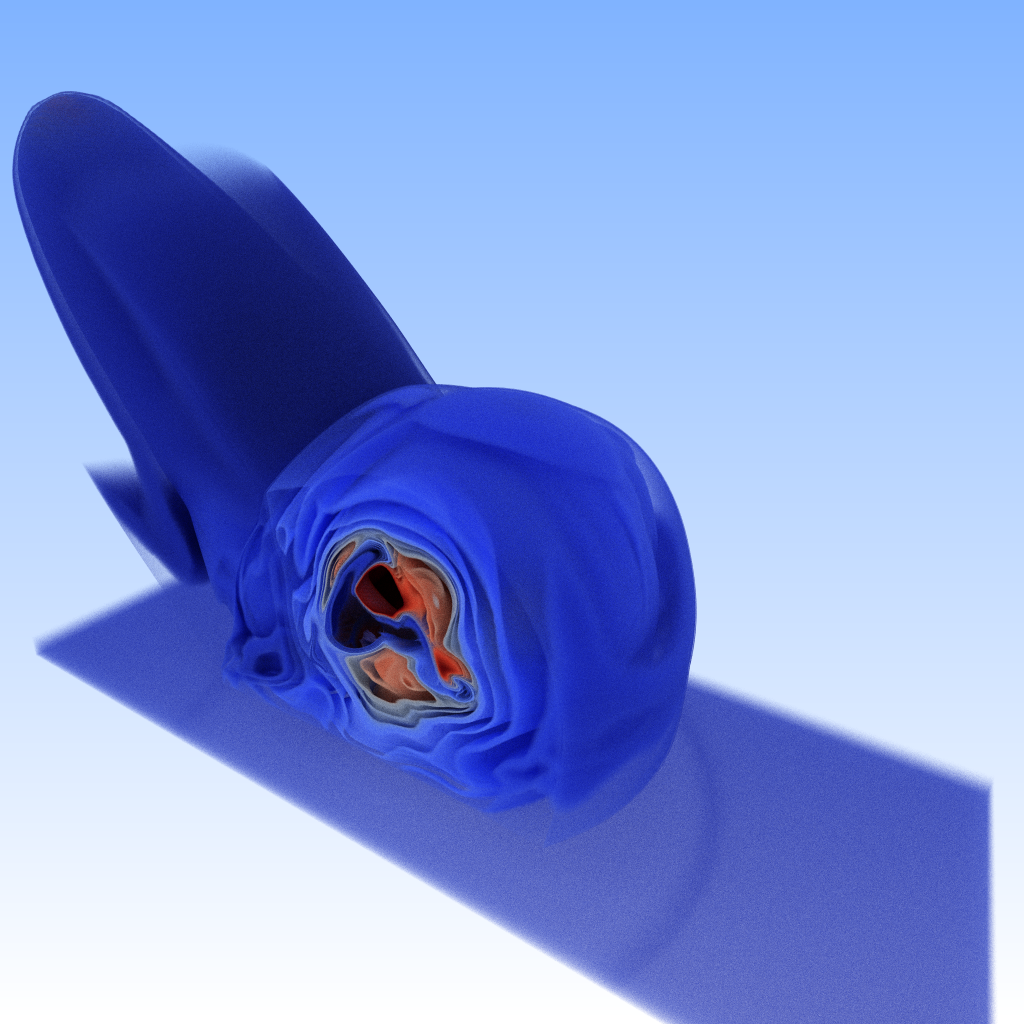}&
\includegraphics[width=0.159\linewidth]{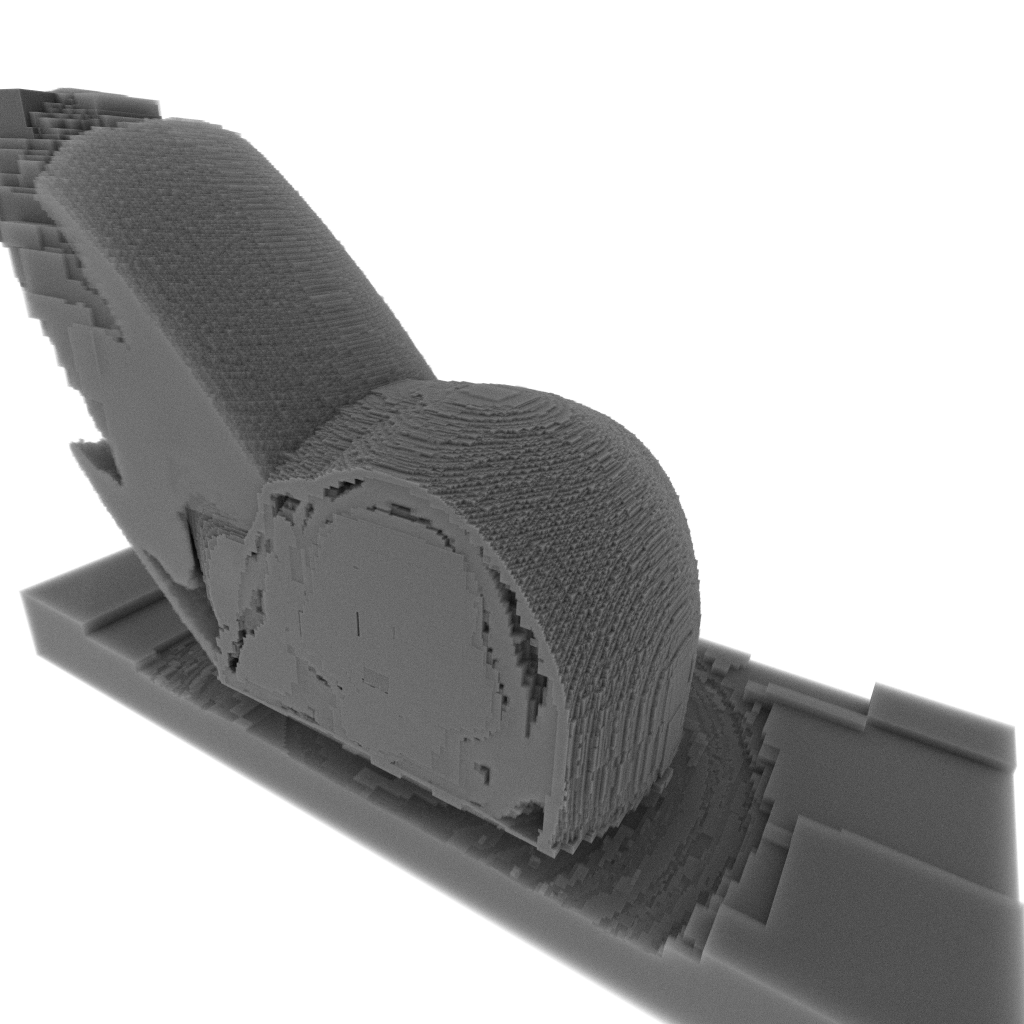}&
\includegraphics[width=0.159\linewidth]{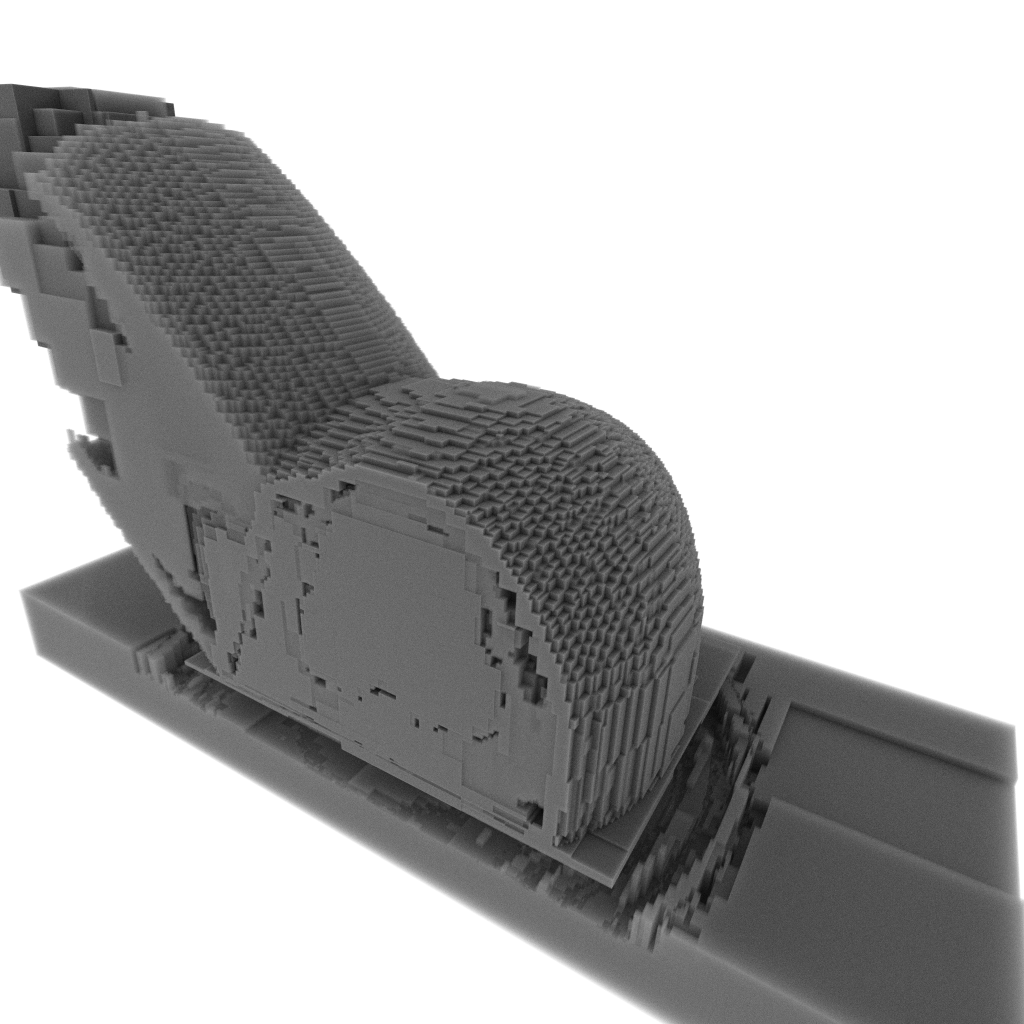}&
\includegraphics[width=0.159\linewidth]{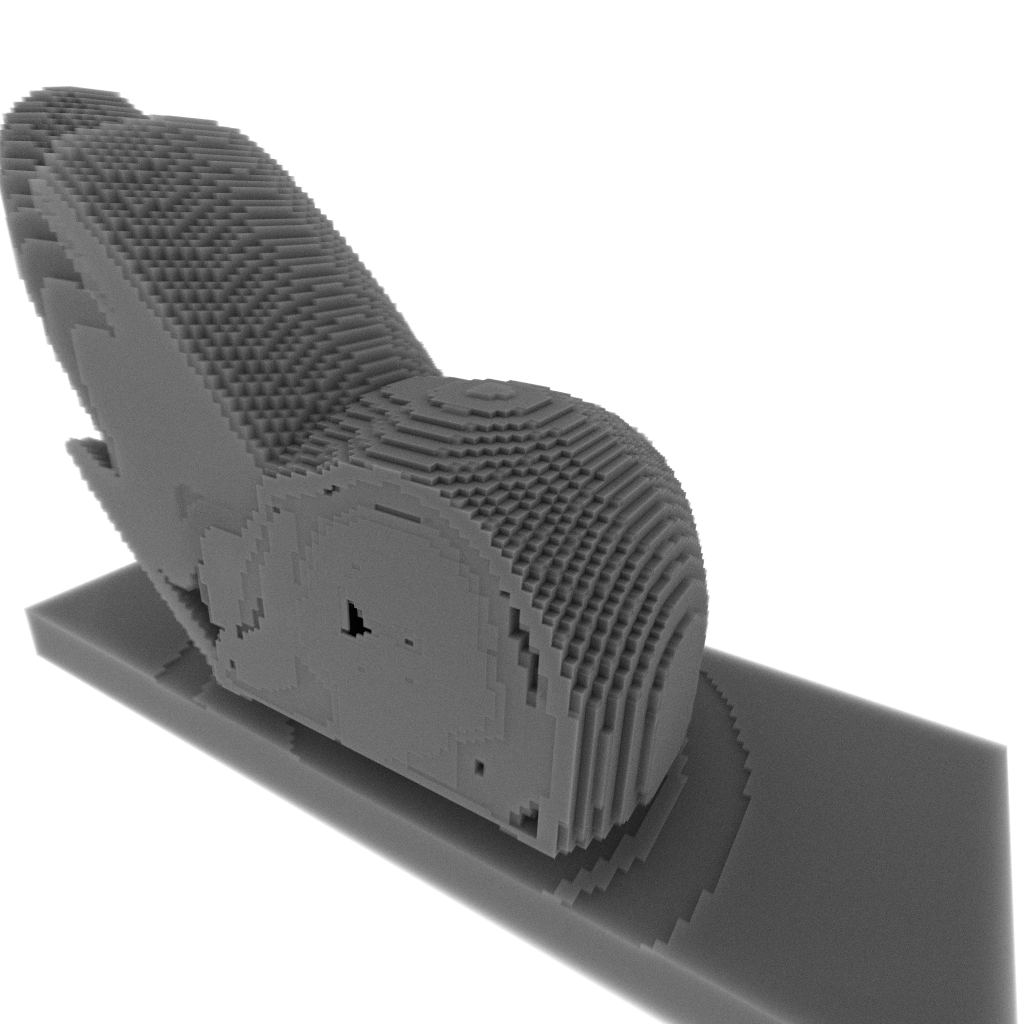}&
\includegraphics[width=0.159\linewidth]{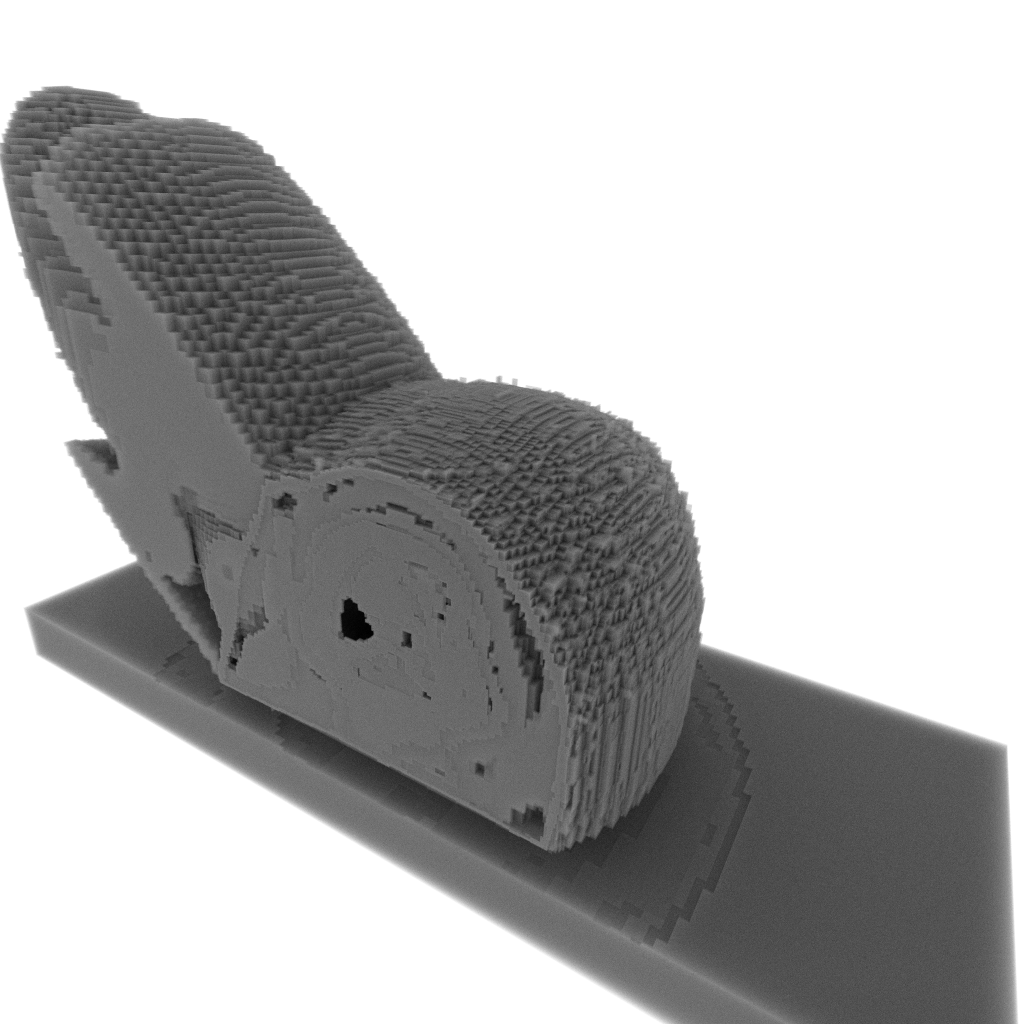}\\

\midrule

\makecell[b]{LANL Impact\\\hline t=46112\\\emph{283~M cells}\\``meteor-46k''\\\\\\\\} &
\includegraphics[width=0.159\linewidth]{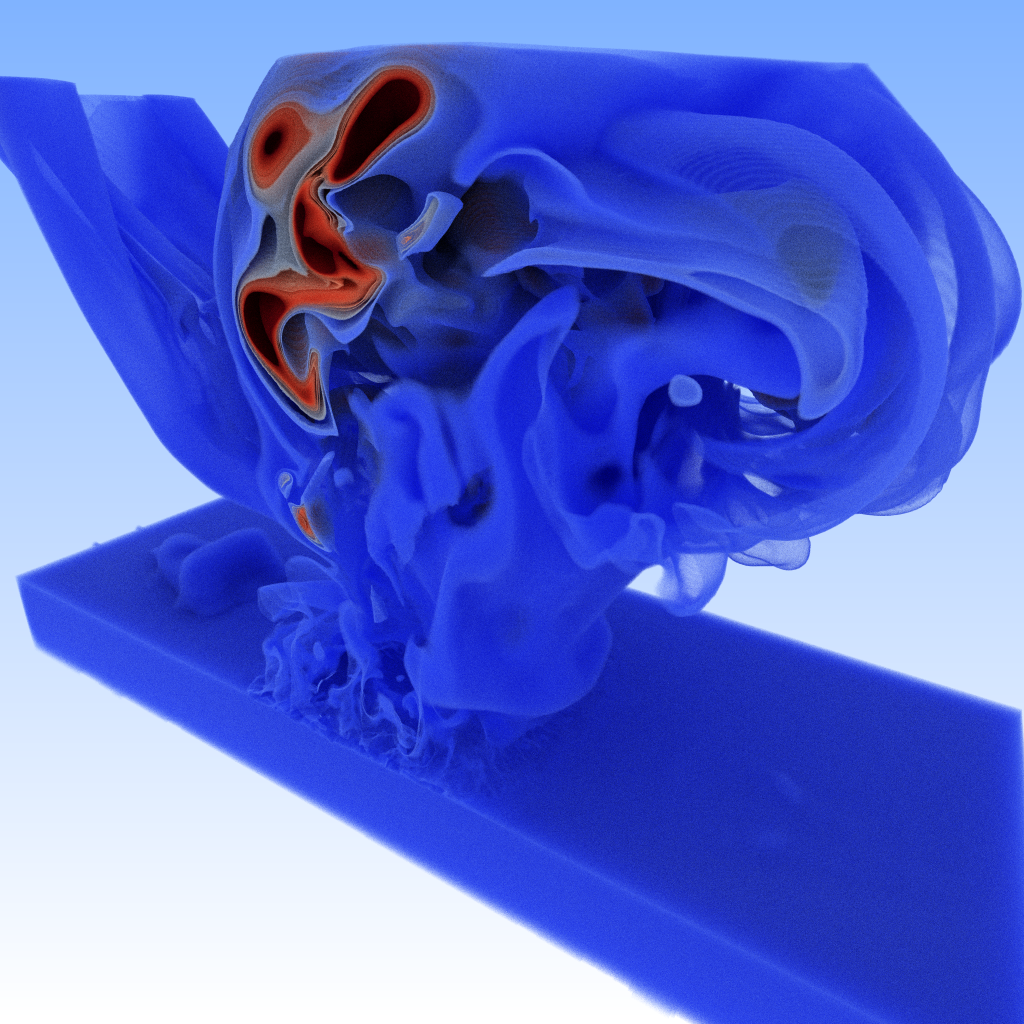}&
\includegraphics[width=0.159\linewidth]{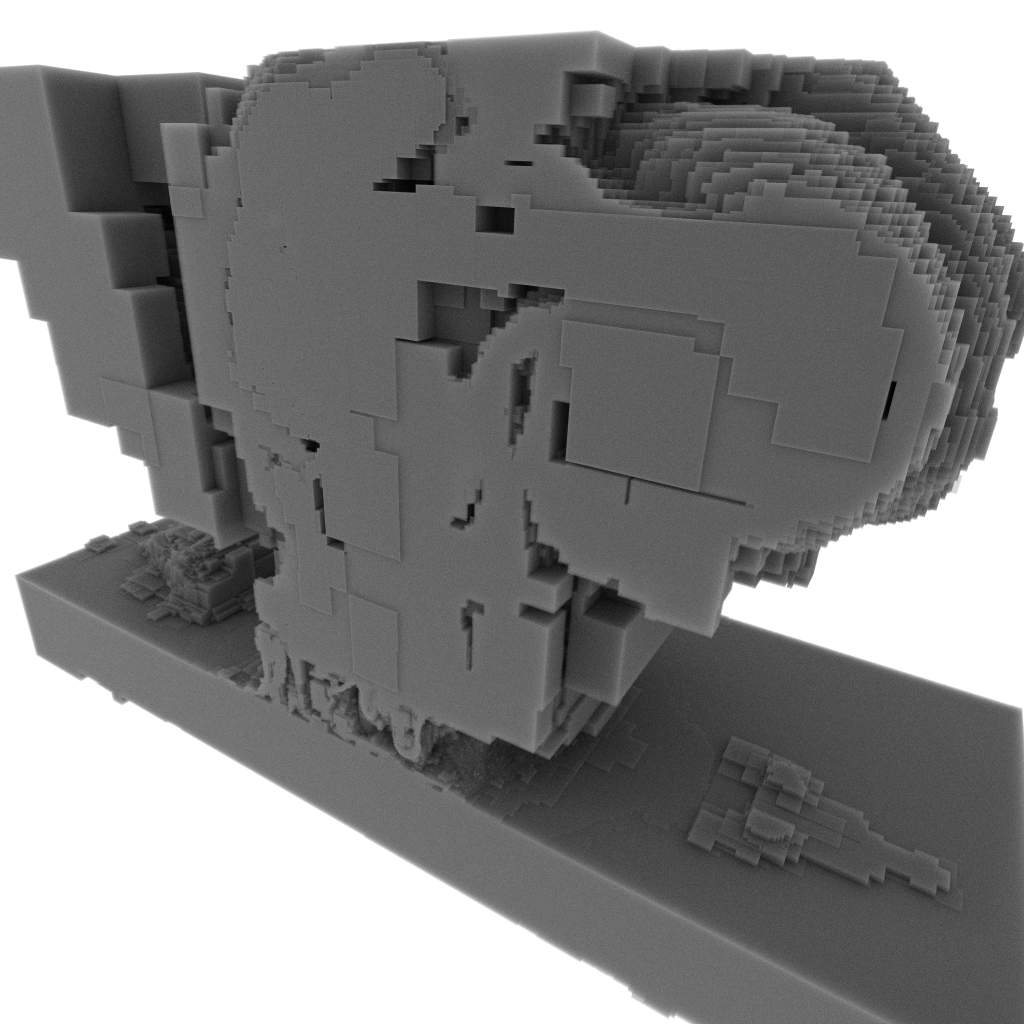}&
\includegraphics[width=0.159\linewidth]{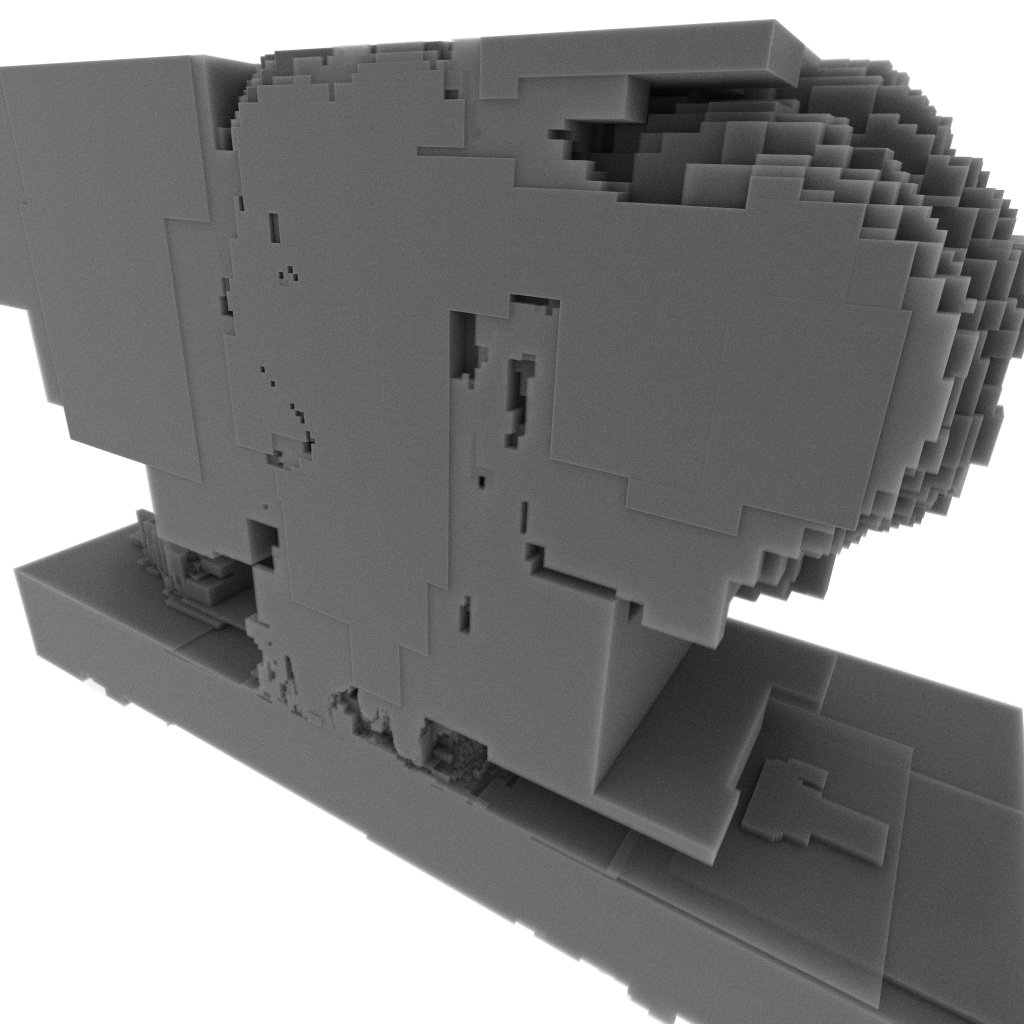}&
\includegraphics[width=0.159\linewidth]{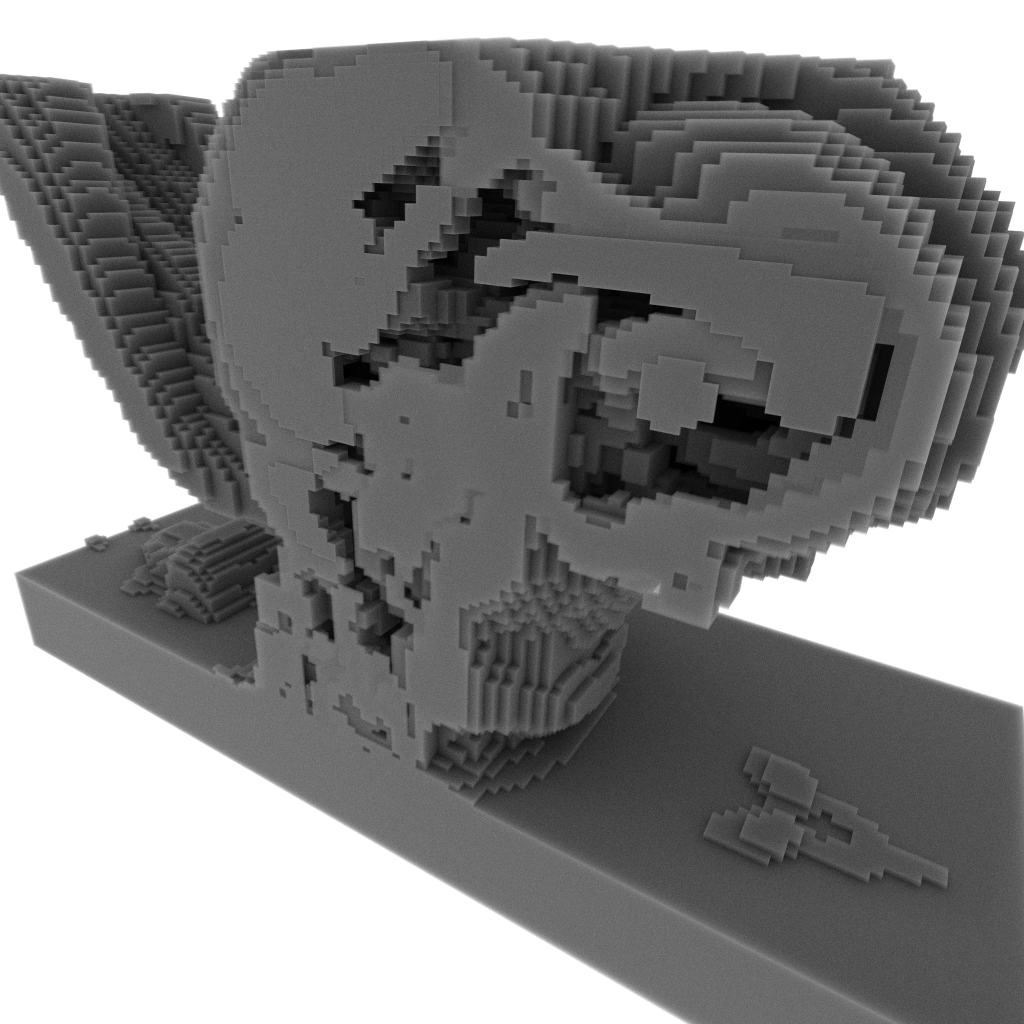}&
\includegraphics[width=0.159\linewidth]{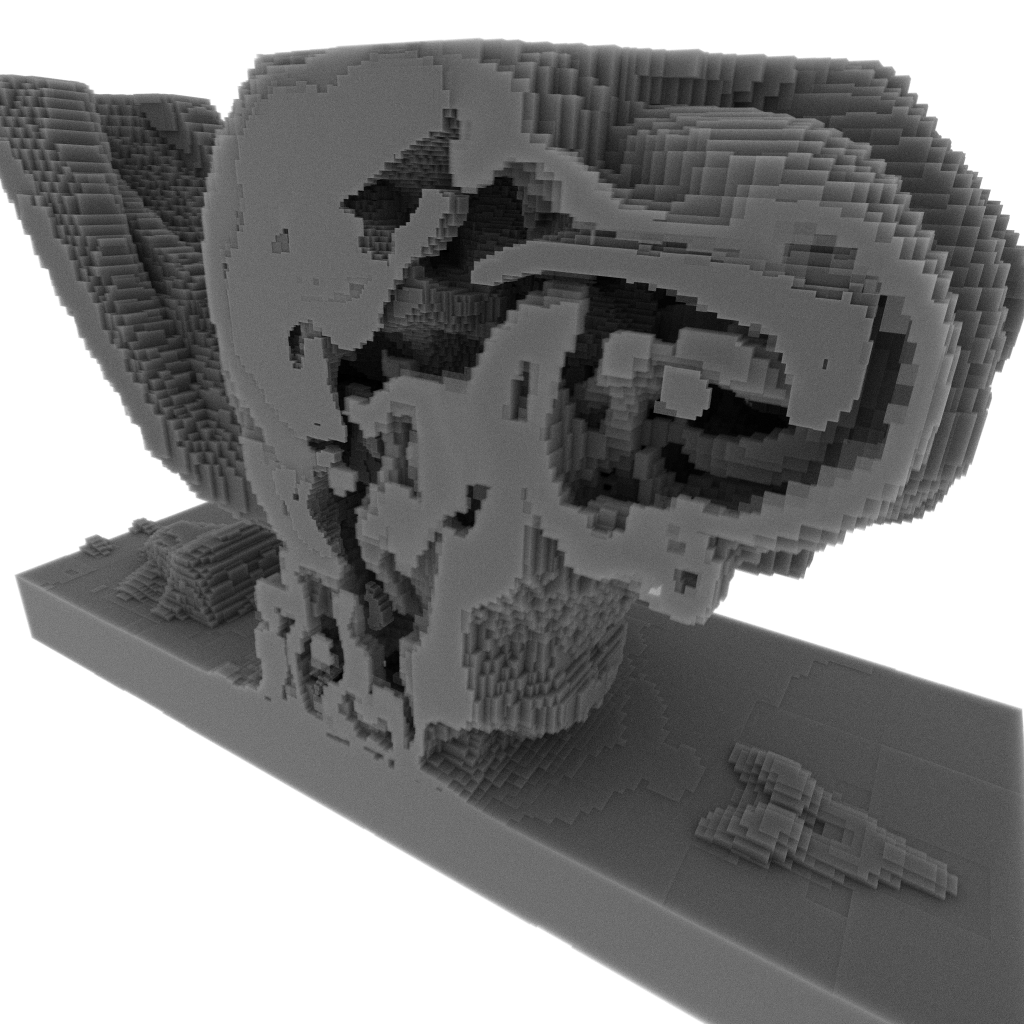}\\

\midrule

\makecell[b]{NASA Landing Gear\\\hline \emph{262~M cells}\\``gear''\\\\\\\\\\} &
\includegraphics[width=0.159\linewidth]{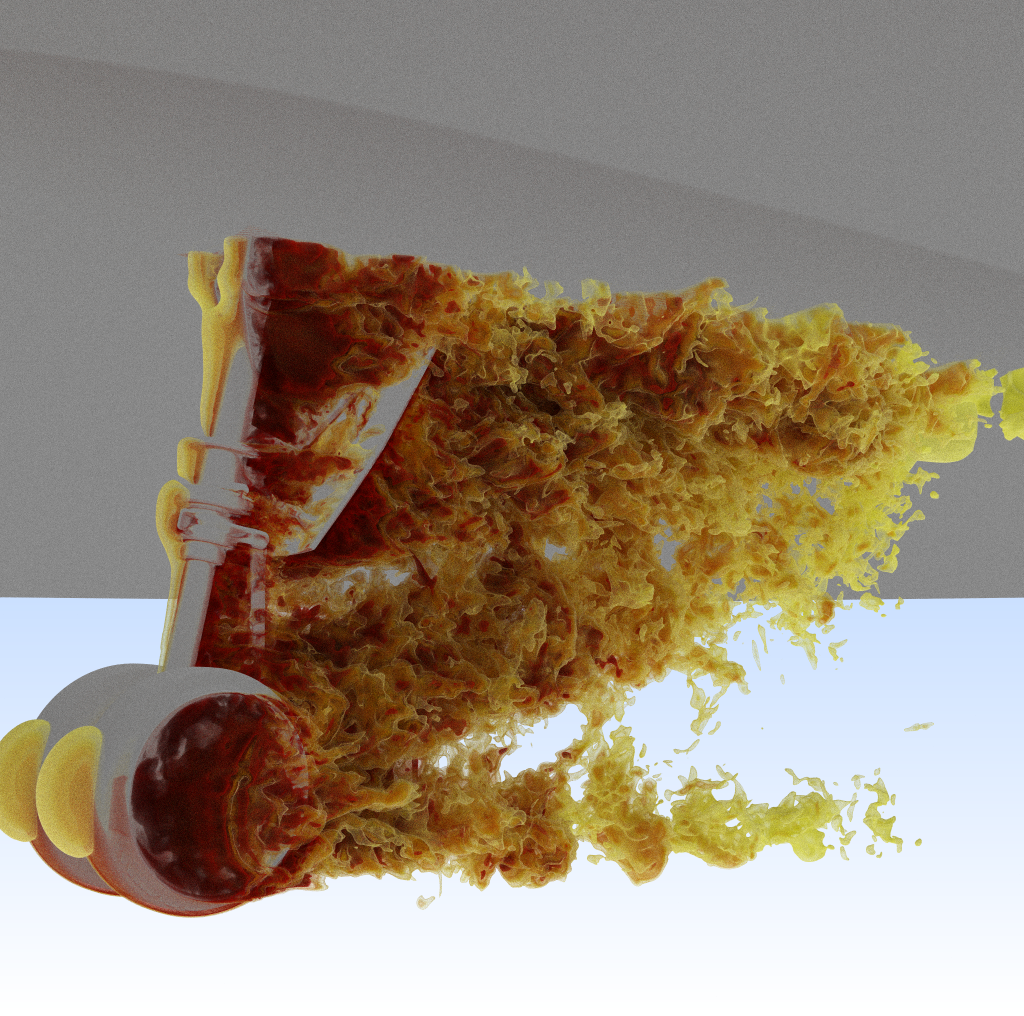}&
\includegraphics[width=0.159\linewidth]{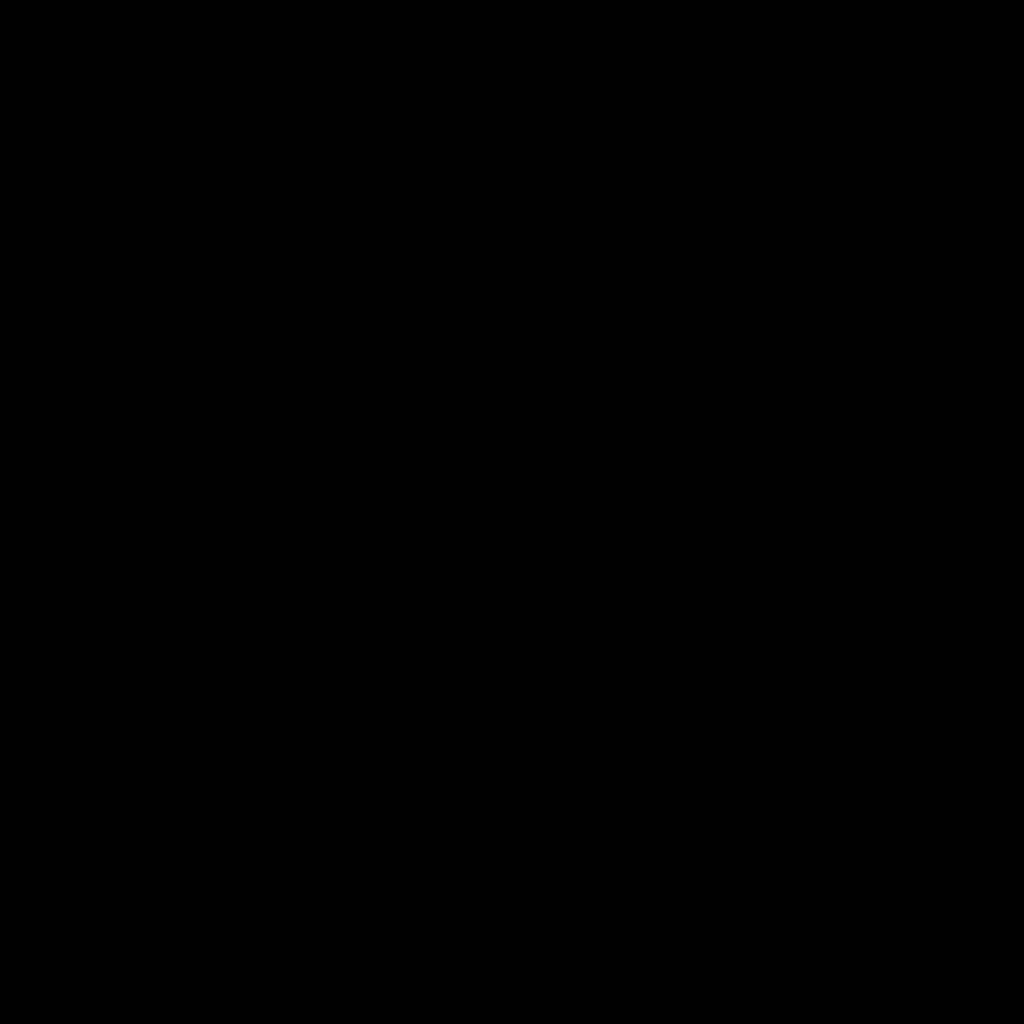}&
\includegraphics[width=0.159\linewidth]{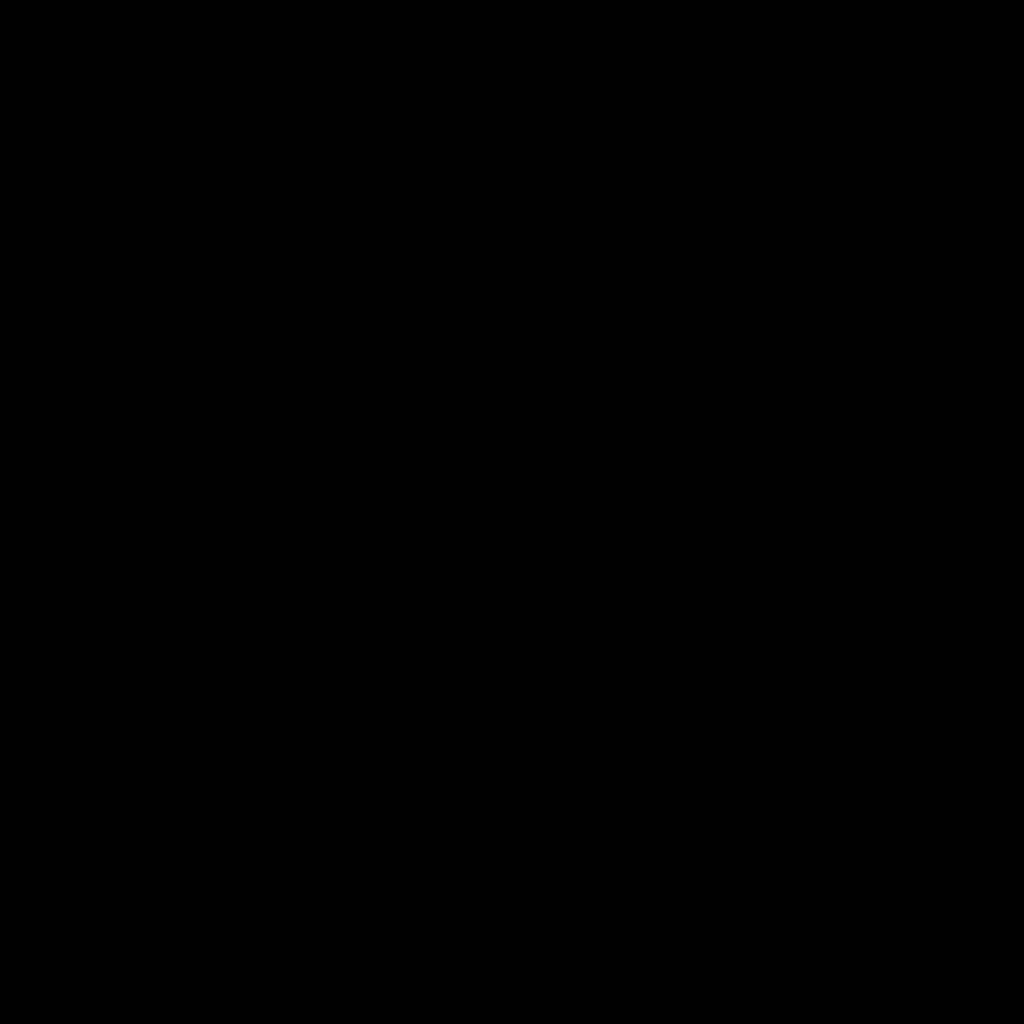}&
\includegraphics[width=0.159\linewidth]{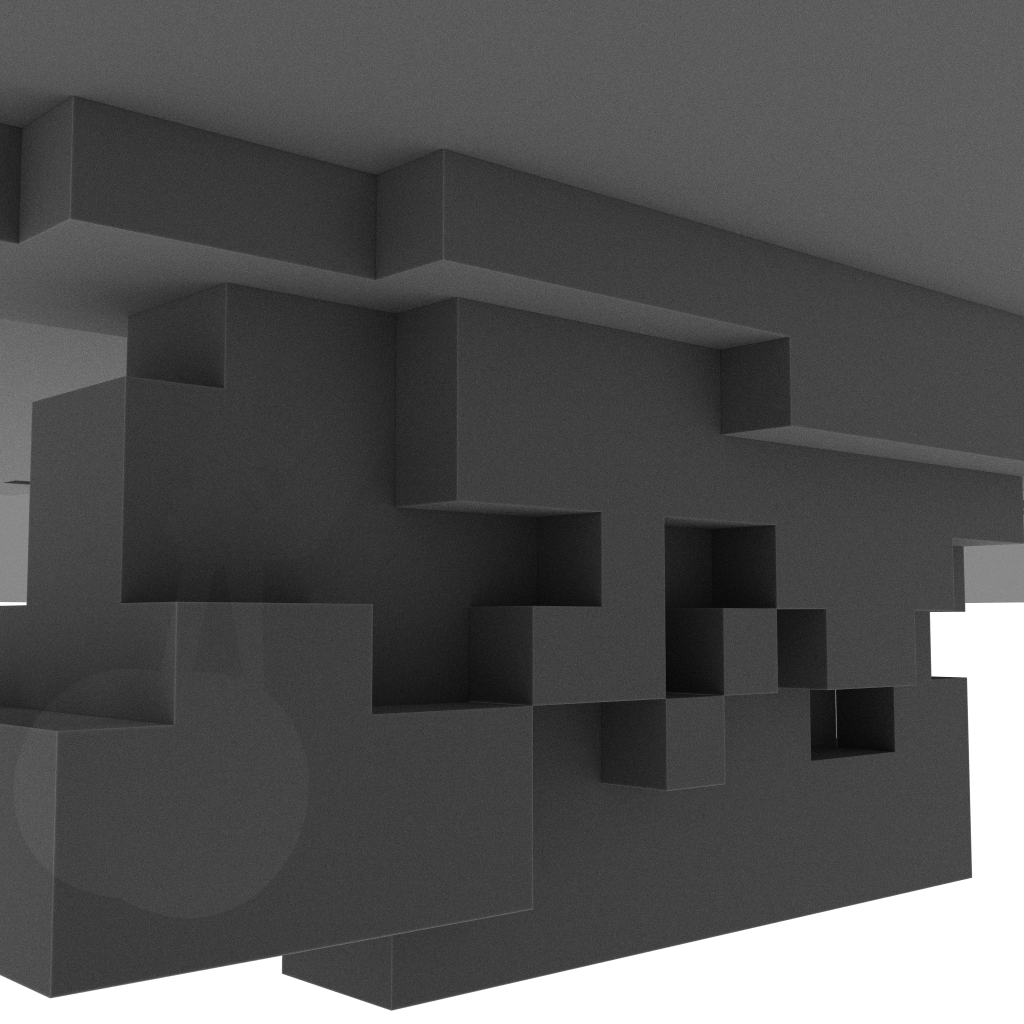}&
\includegraphics[width=0.159\linewidth]{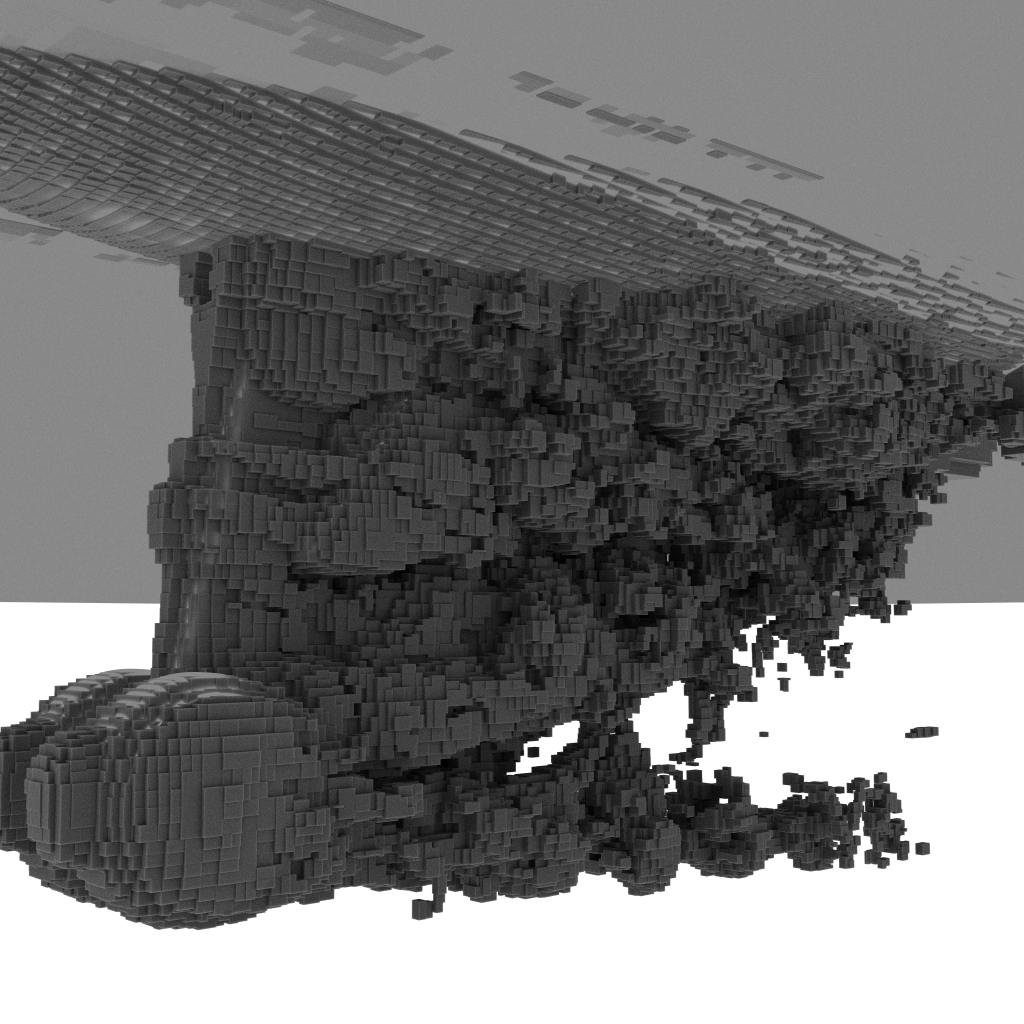}\\

\midrule

\makecell[b]{NASA Exajet\\\hline wing view\\\emph{1.31~B cells}\\``exajet-wing''\\\\\\\\} &
\includegraphics[width=0.159\linewidth]{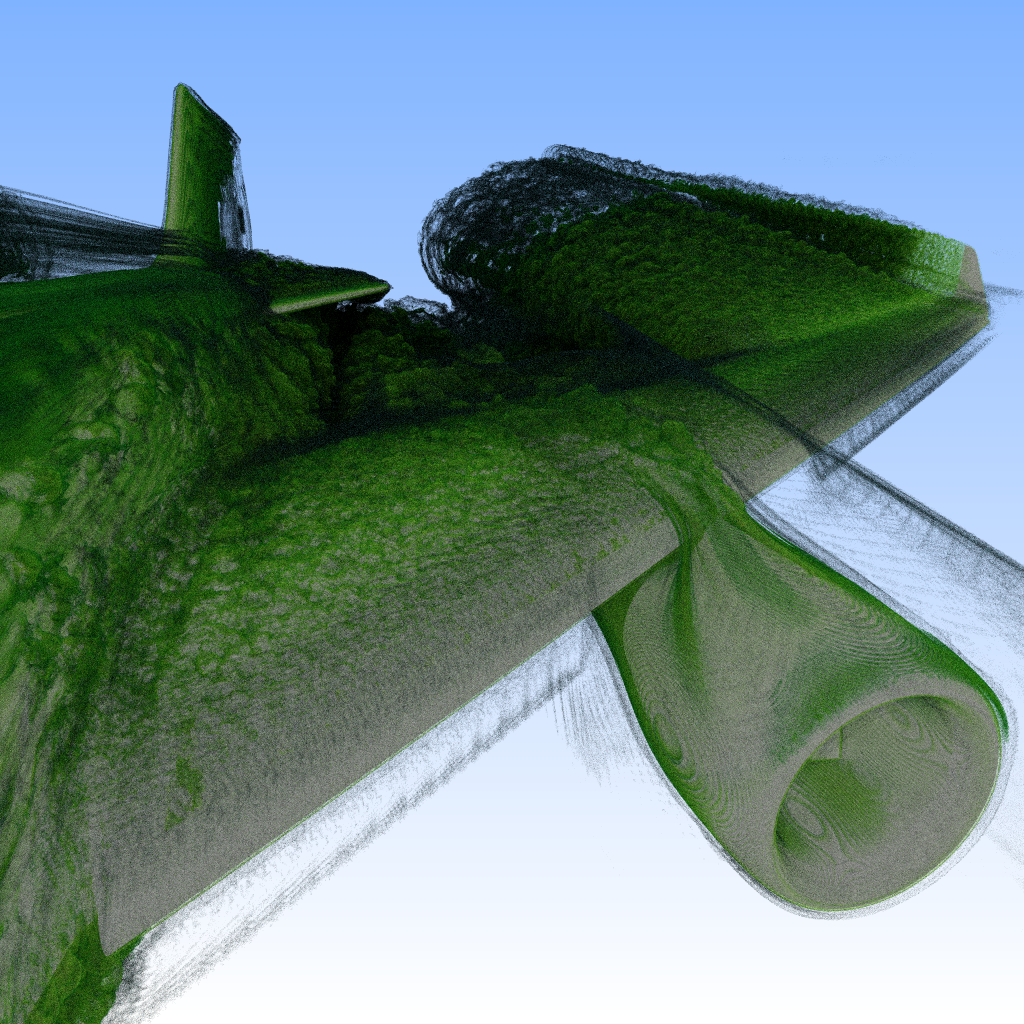}&
\includegraphics[width=0.159\linewidth]{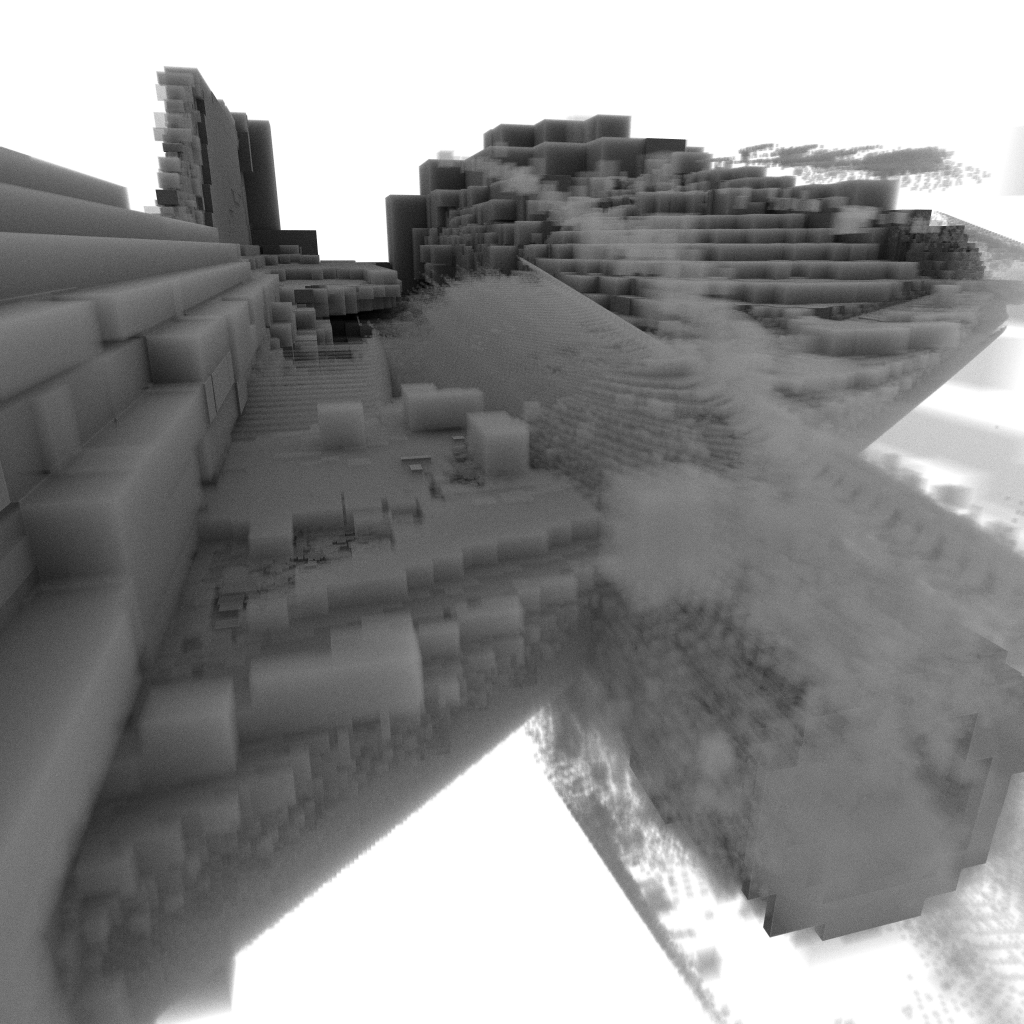}&
\includegraphics[width=0.159\linewidth]{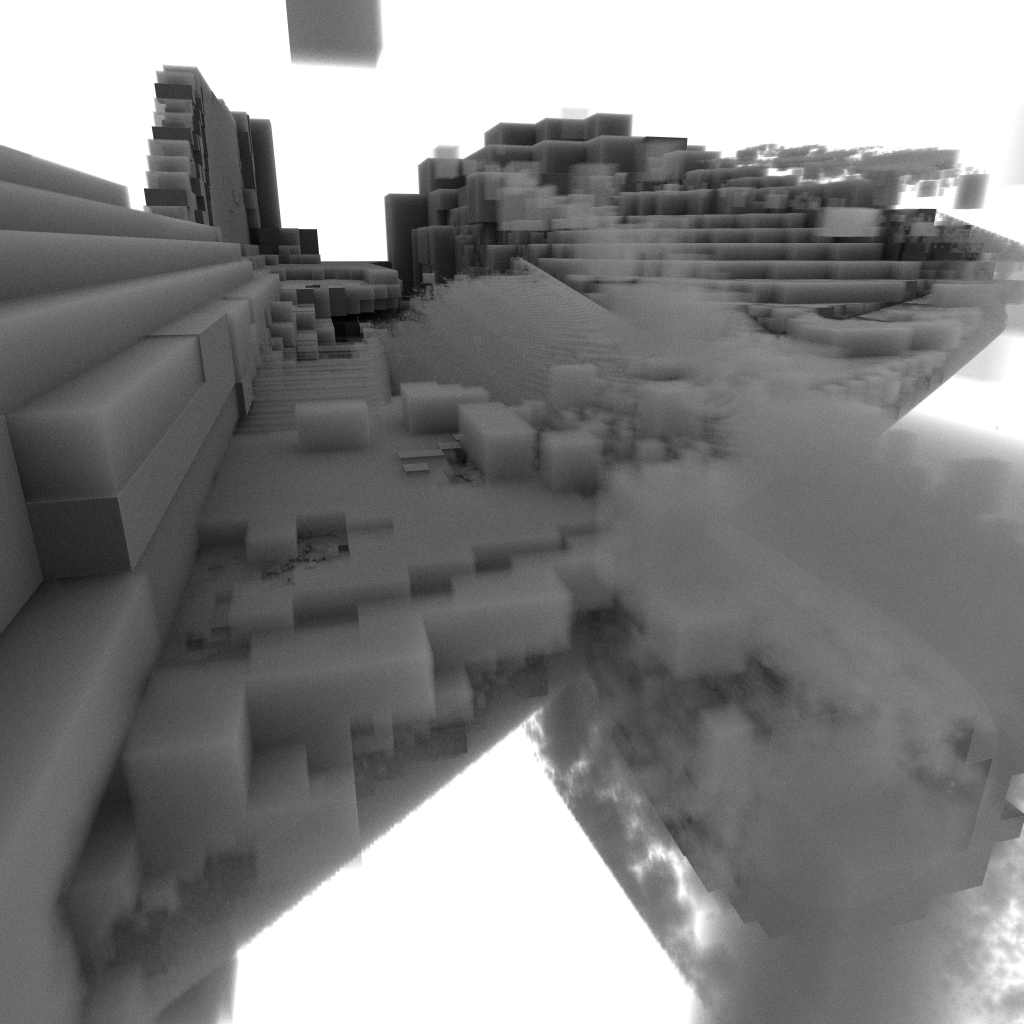}&
\includegraphics[width=0.159\linewidth]{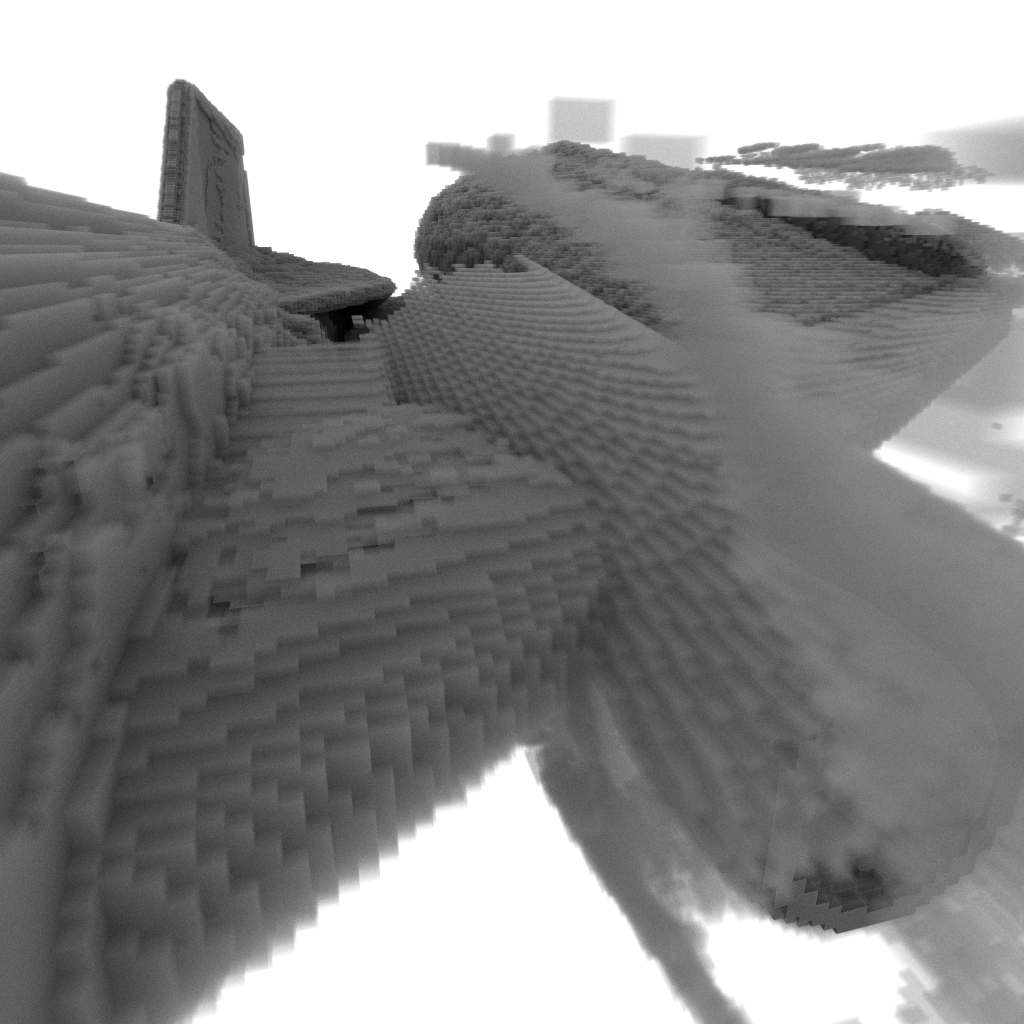}&
\includegraphics[width=0.159\linewidth]{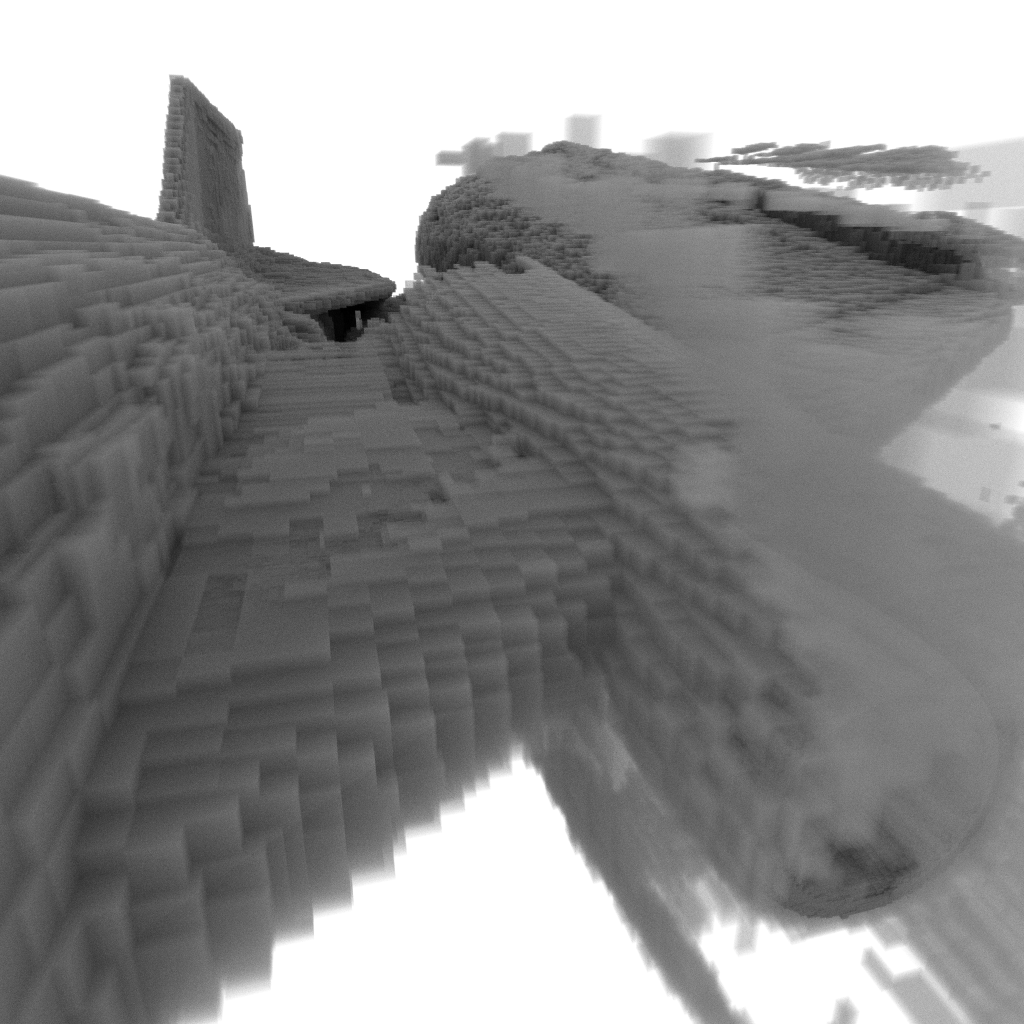}\\

\midrule

\makecell[b]{NASA Exajet\\\hline rear view\\\emph{1.31~B cells}\\``exajet-rear''\\\\\\\\} &
\includegraphics[width=0.159\linewidth]{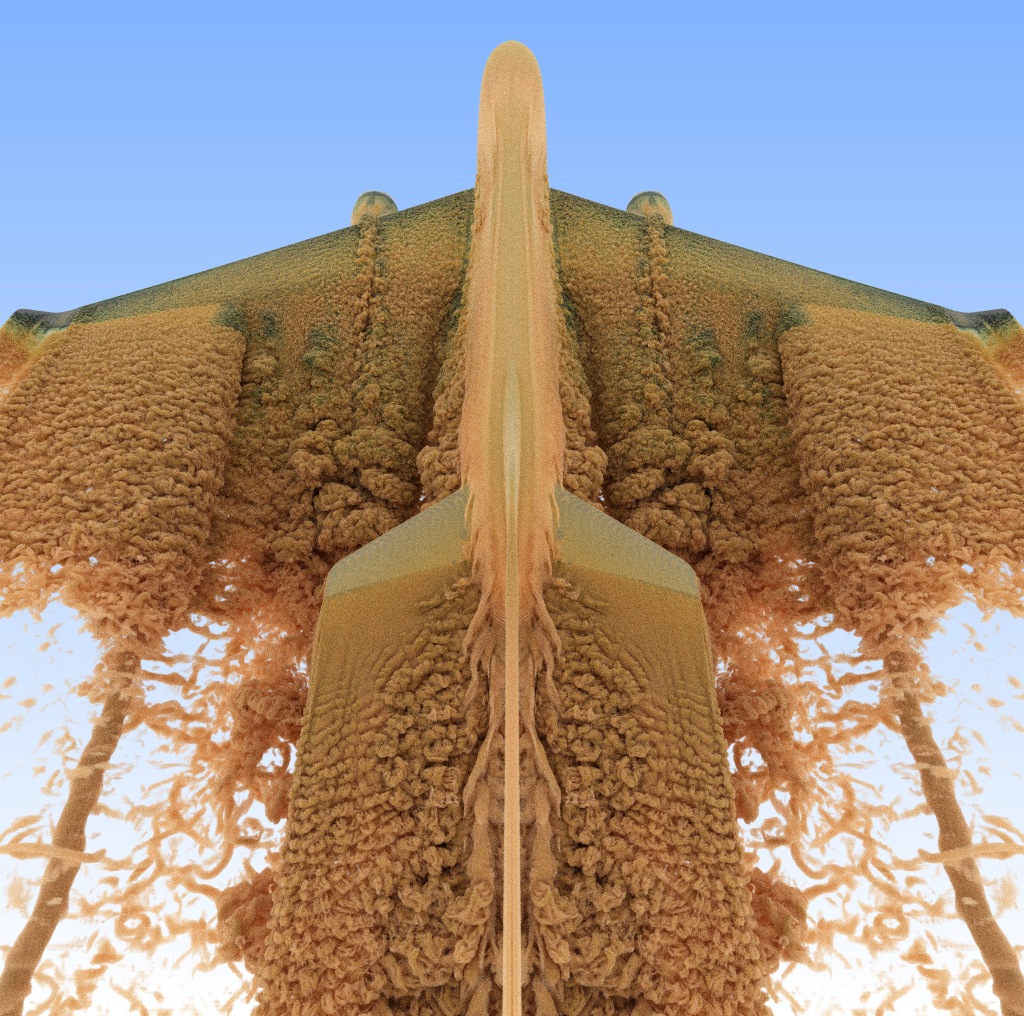}&
\includegraphics[width=0.159\linewidth]{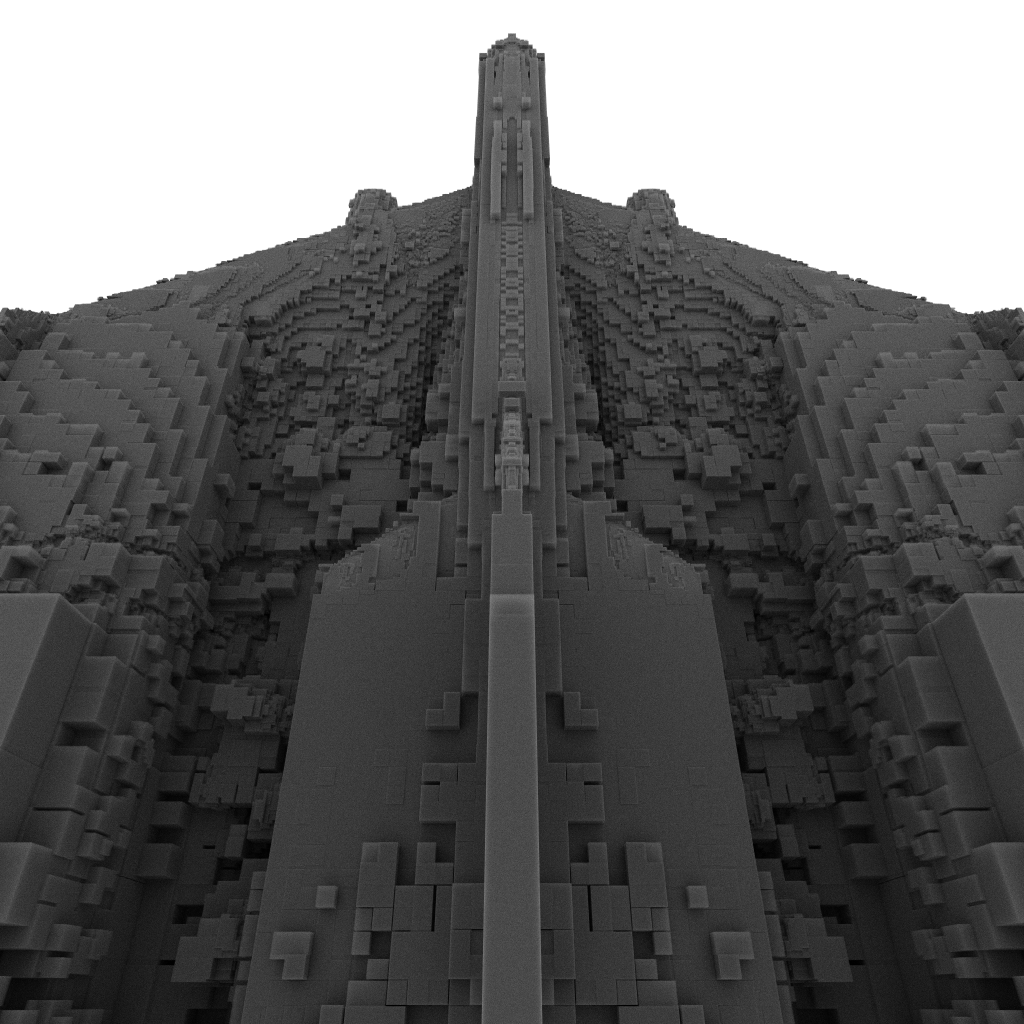}&
\includegraphics[width=0.159\linewidth]{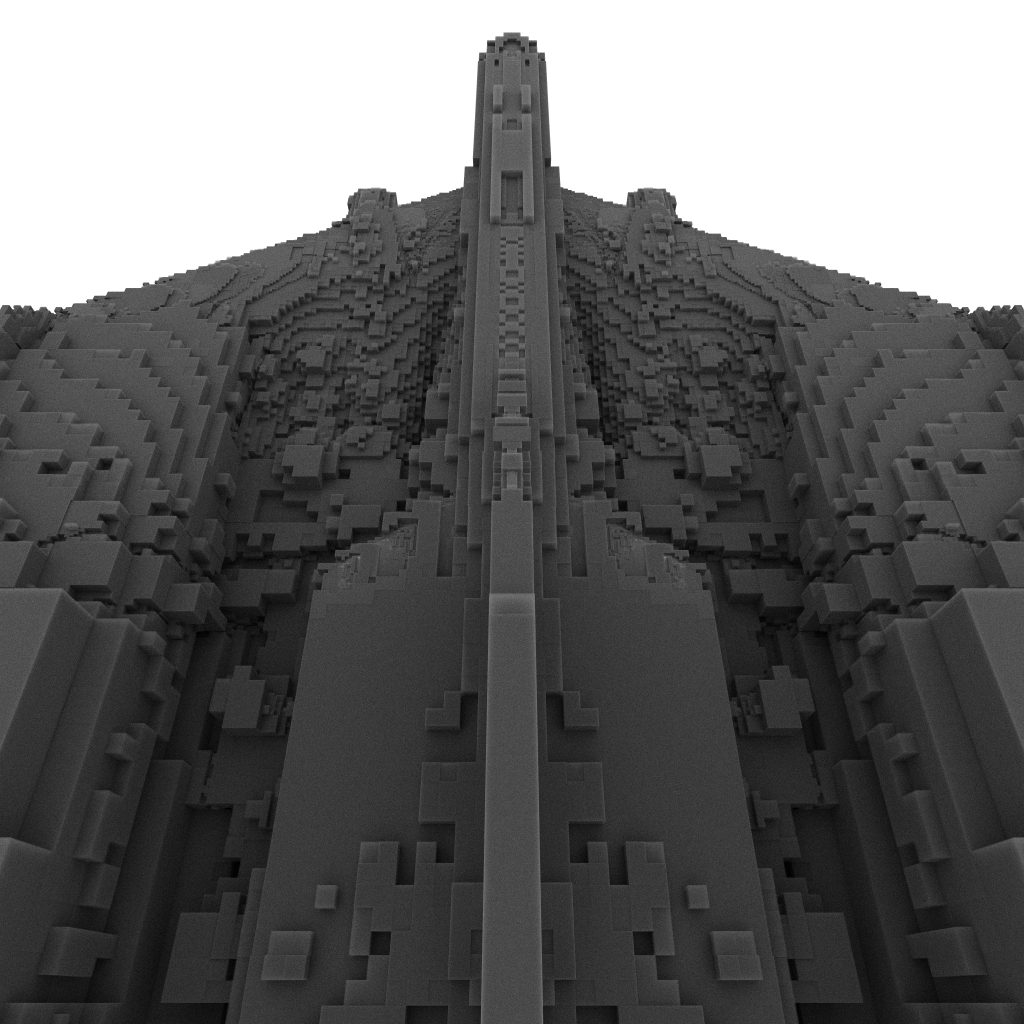}&
\includegraphics[width=0.159\linewidth]{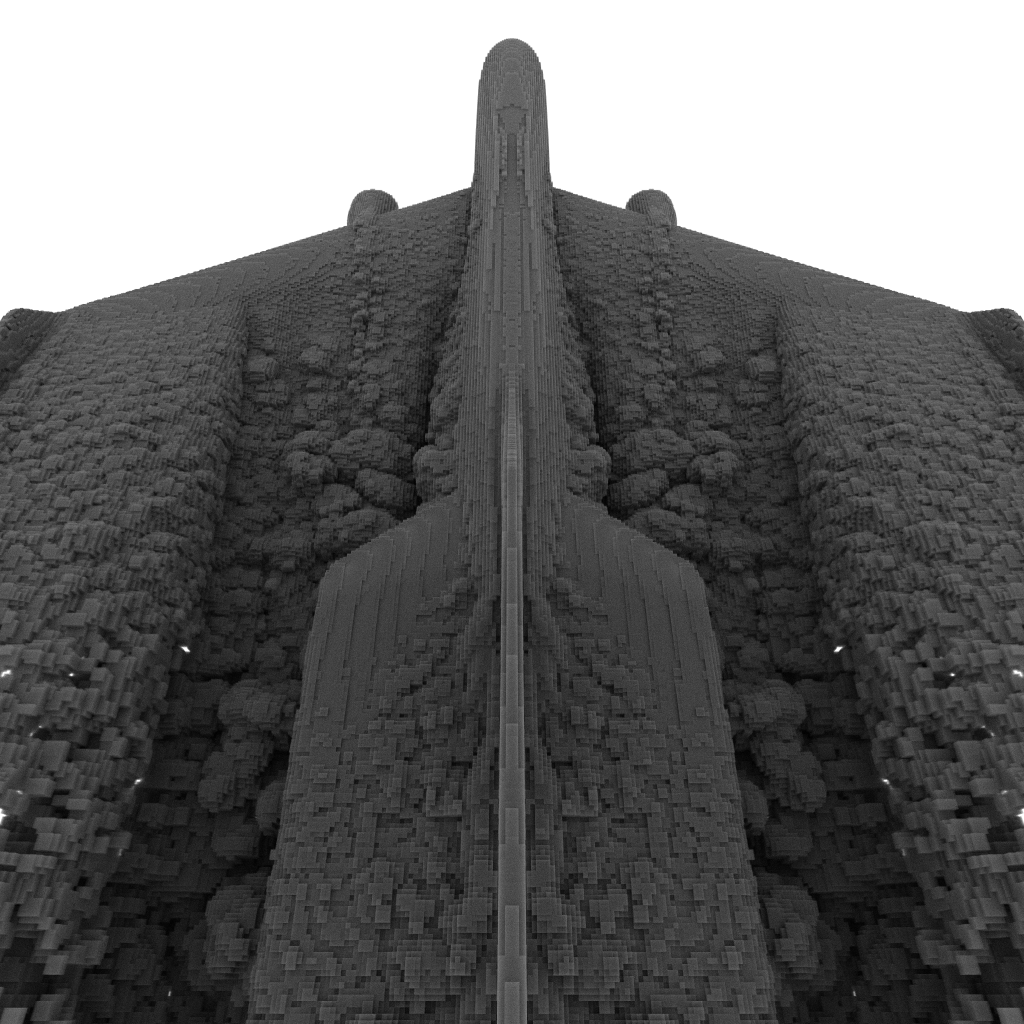}&
\includegraphics[width=0.159\linewidth]{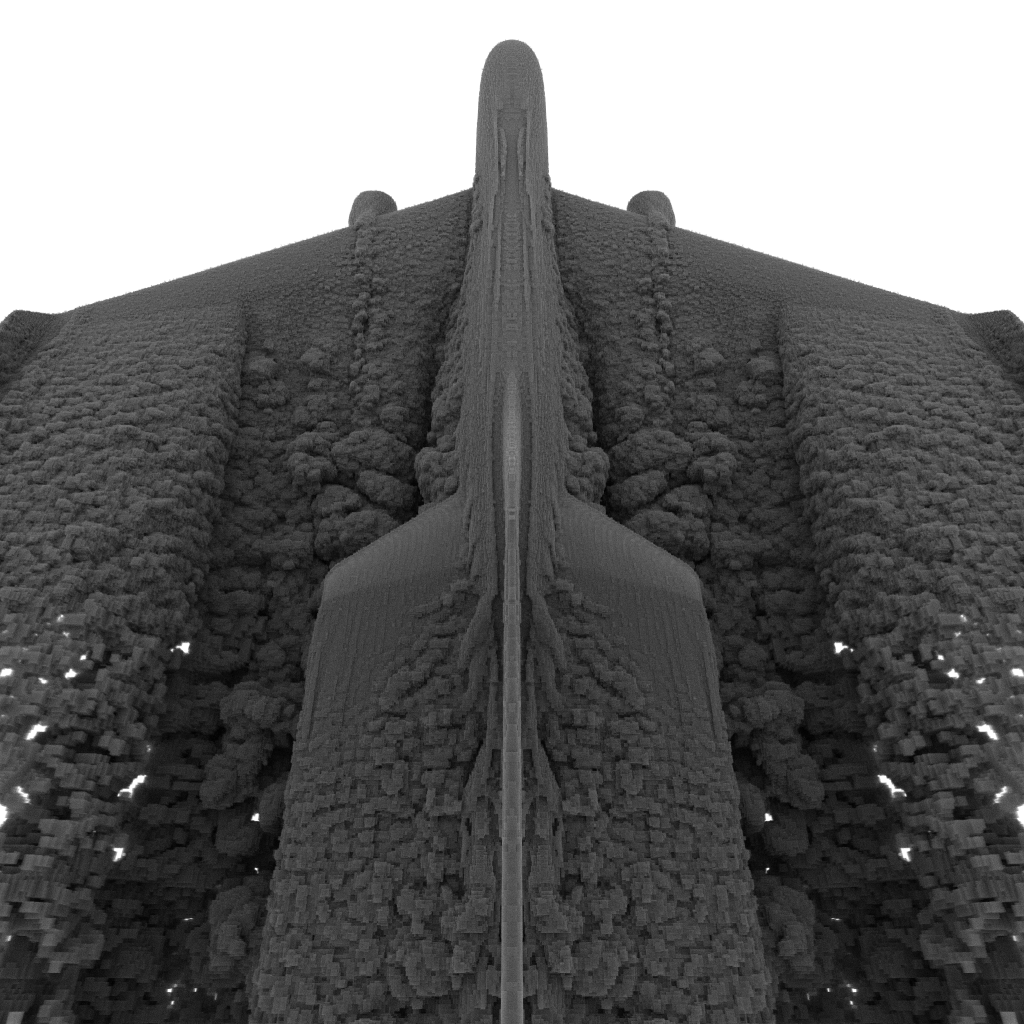}\\

\bottomrule
\end{tabular}
\vspace{-1em}
\caption{\label{fig:majorants3}%
\added{Data sets used for the evaluation, and majorants for the traversal data
structures from \cref{sec:traversal}. Black images indicate that the camera
origin falling inside the density, which happens for the landing gear (also see
\cref{fig:crazy-gear} for a zoomed-out visualization of that data set).}
\vspace{-2.0em}
}
\end{table*}
\begin{figure}[tb]
\includegraphics[width=0.325\columnwidth]{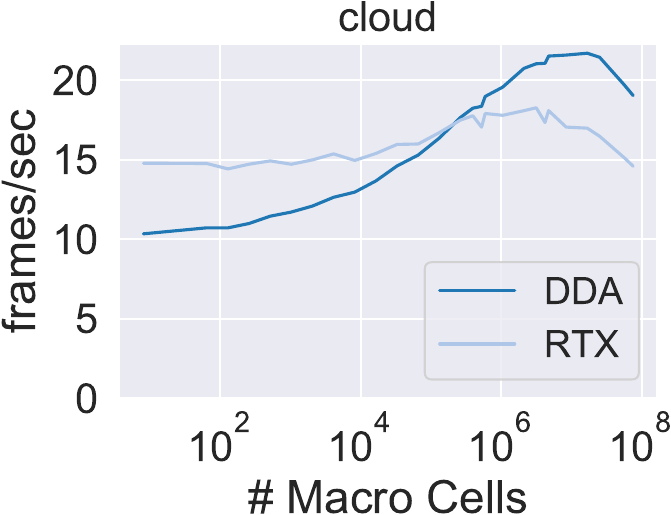}
\includegraphics[width=0.325\columnwidth]{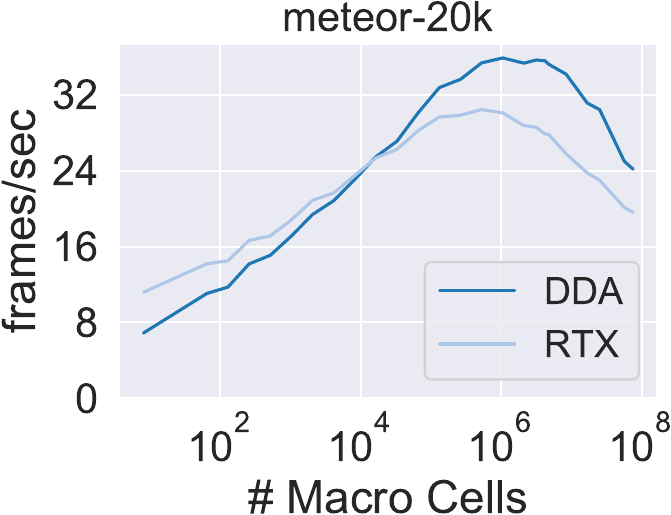}
\includegraphics[width=0.325\columnwidth]{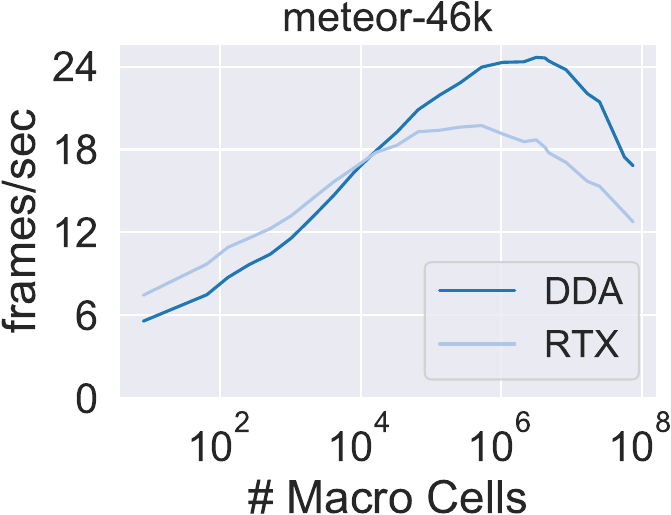}\\
\includegraphics[width=0.325\columnwidth]{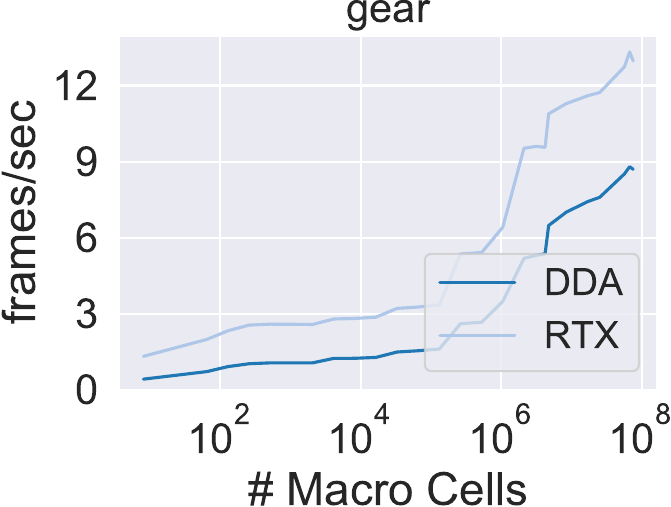}
\includegraphics[width=0.325\columnwidth]{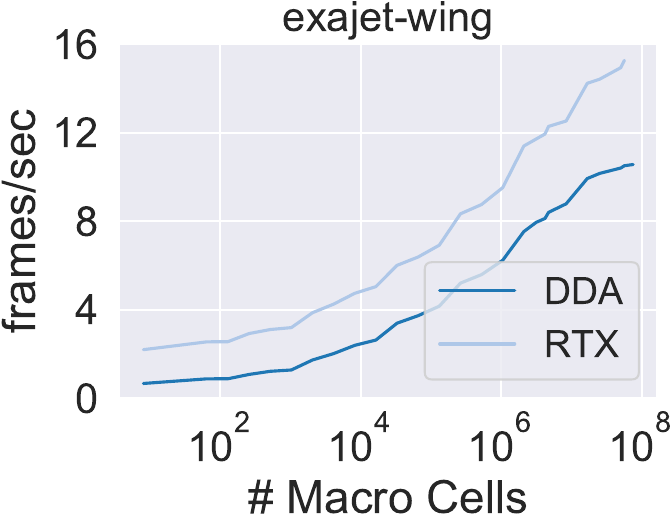}
\includegraphics[width=0.325\columnwidth]{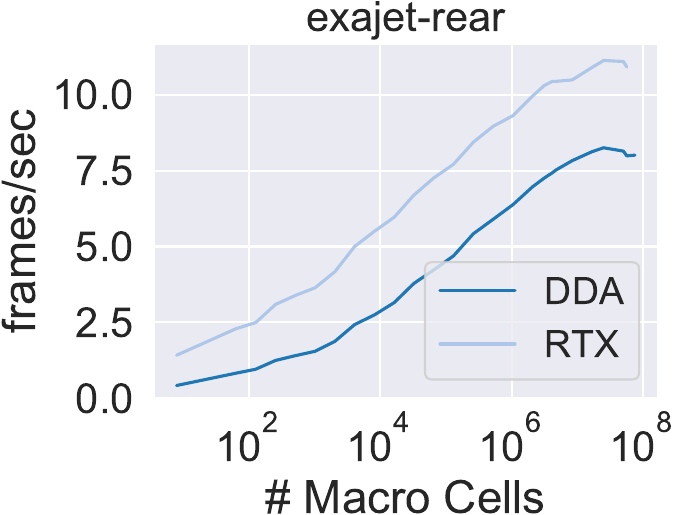}
\vspace{-0.5em}
\caption{\label{fig:grid-eval}
Performance (frames/sec.) for different uniform grid sizes.
Optima are reached at around $128^3$ (cloud),
$128\times128 \times64$ (meteor-20k, meteor-46k),
$1024\times1024\times256$ (gear), and
between $512\times256\times256$ and $512^3$ (exajet) macrocells.
\vspace{-2.5em}
}
\end{figure}
\begin{figure}[tb]
\centering
\stackunder[1pt]{\includegraphics[width=0.32\columnwidth]{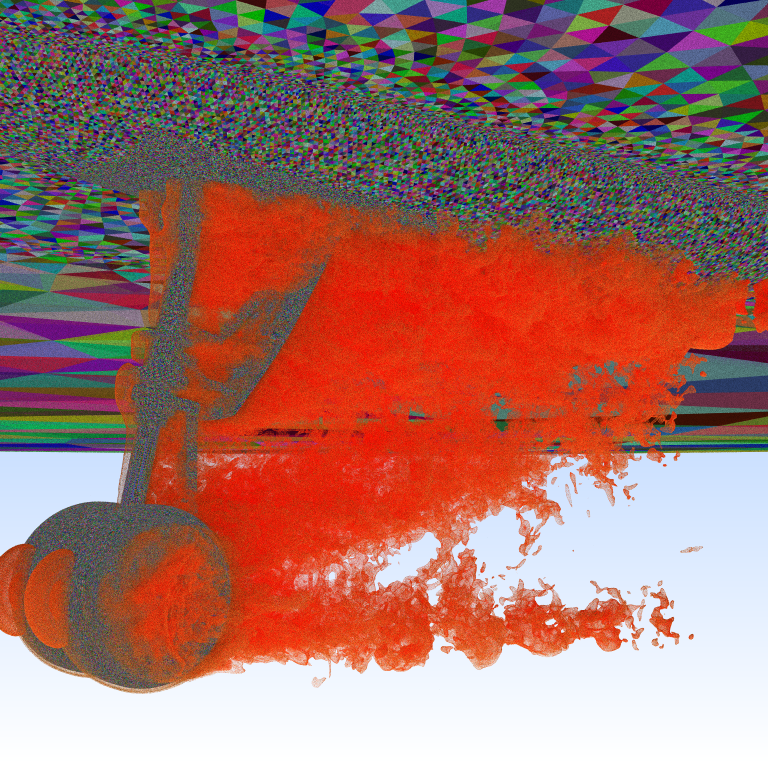}}{(a)}
\stackunder[1pt]{\includegraphics[width=0.32\columnwidth]{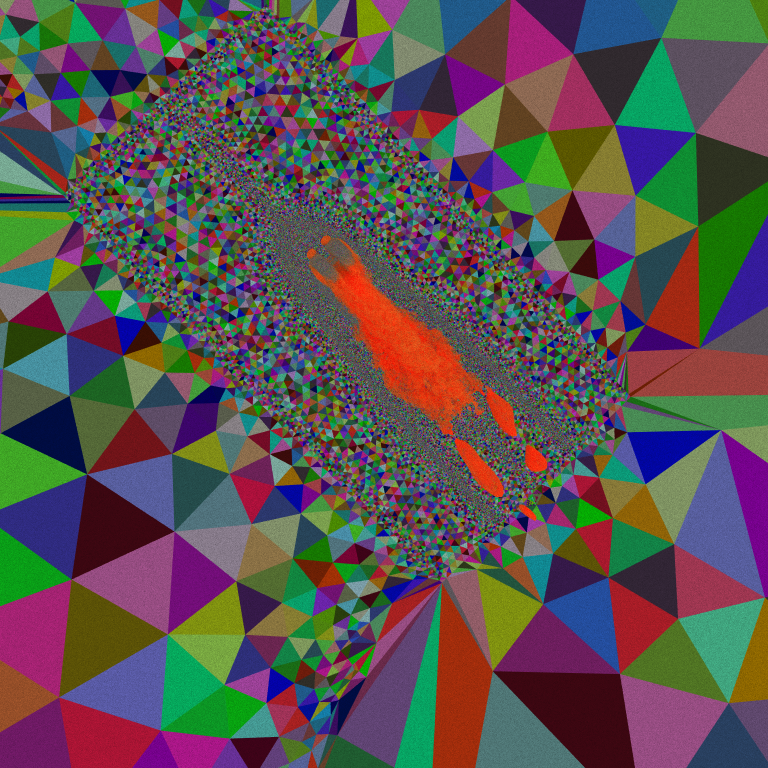}}{(b)}
\stackunder[1pt]{\includegraphics[width=0.32\columnwidth]{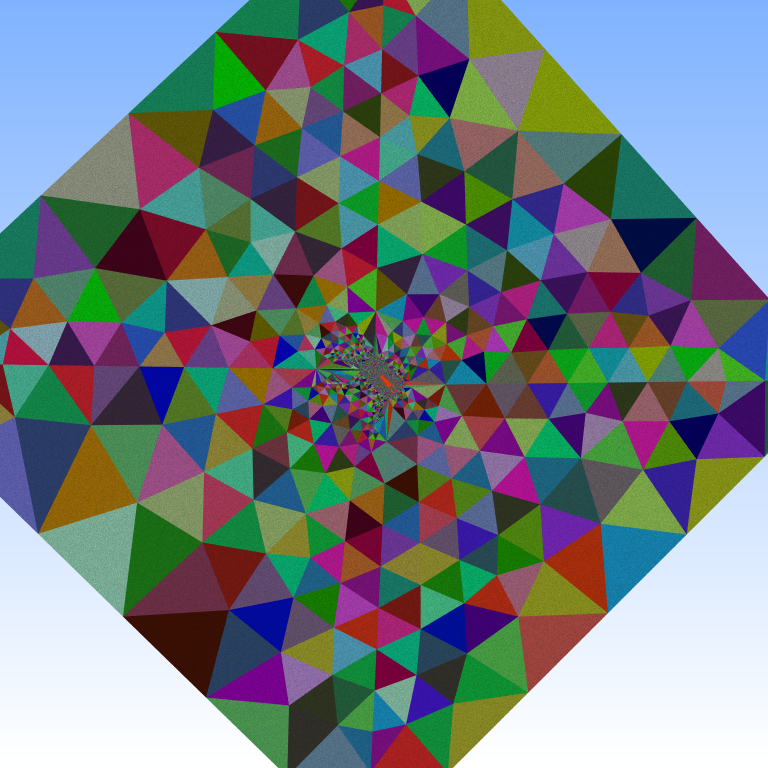}}{(c)}\\
\stackunder[1pt]{\includegraphics[width=0.32\columnwidth]{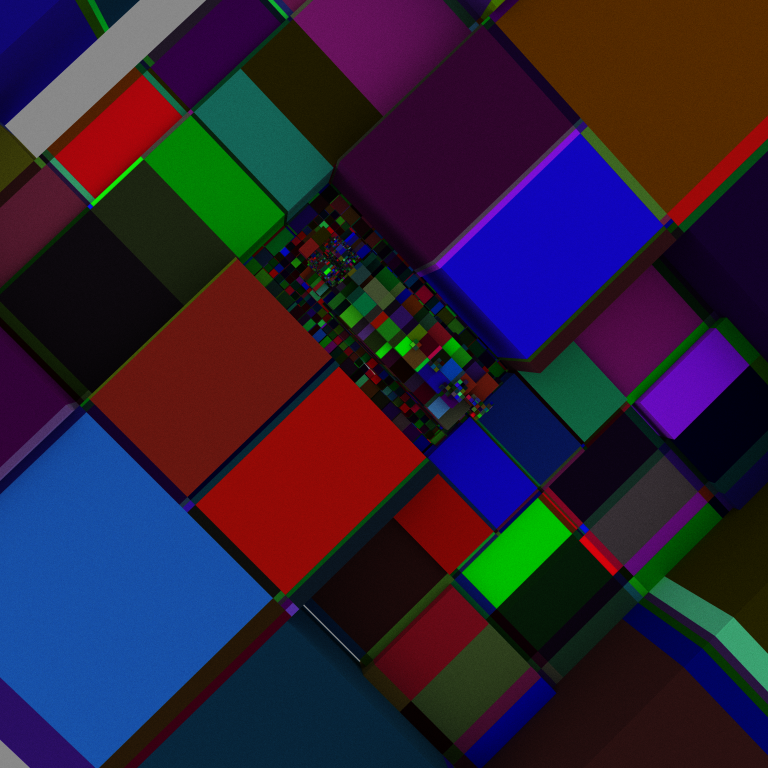}}{(d)}
\stackunder[1pt]{\includegraphics[width=0.32\columnwidth]{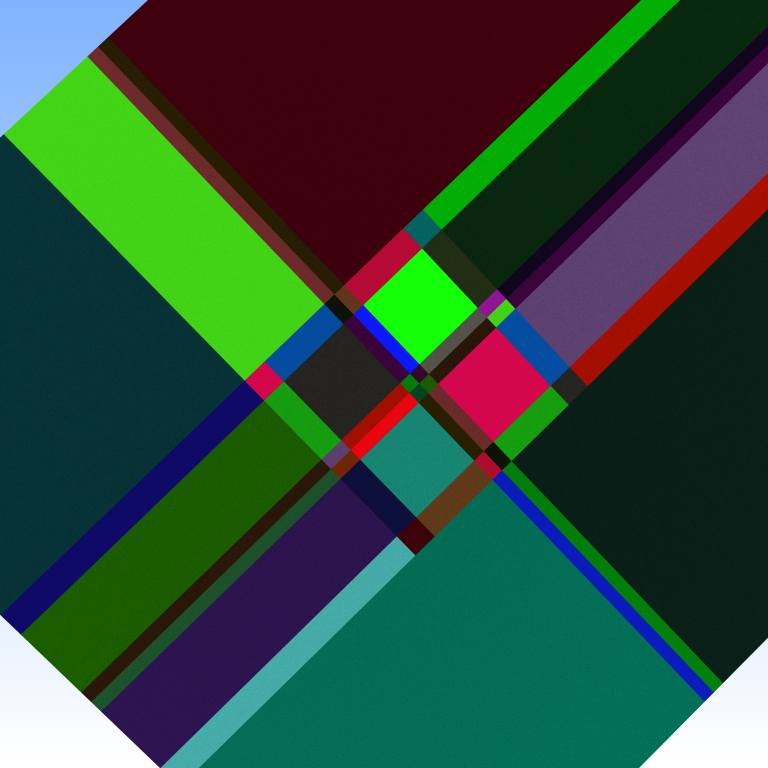}}{(e)}
\caption{\label{fig:crazy-gear}
Landing gear AMR model complexity. (a) The zoomed-in camera position we use in
the paper; triangle mesh colors assigned by \texttt{primID}. (b-c) Different
zooms of the same visualization in (a). (d-e) ABRs for the zoomed-out views in
(b) and (c); the extinction of the whole ABR is set to its majorant extinction,
the RGB albedo is derived from the ABR ID.
\vspace{-2.5em}
}
\end{figure}

\def\first#1{\color{lightgreen}\textbf{#1}}
\def\second#1{\color{midgreen}#1}
\def\third#1{\color{darkgreen}#1}
\def\last#1{\color{darkred}#1}

\begin{table*}[th]
\centering\small{
\setlength{\tabcolsep}{2pt}
\setlength{\extrarowheight}{-10pt}%
\begin{tabular}{c|c|cc|cc|cc|cc|cc|cc|}
\toprule
trav. & samp. & \multicolumn{2}{|c}{cloud} & \multicolumn{2}{|c}{meteor-20k} & \multicolumn{2}{|c}{meteor-46k} & \multicolumn{2}{|c}{gear} & \multicolumn{2}{|c}{exajet-rear} & \multicolumn{2}{|c}{exajet-wing}\\
\midrule
~ & ~ & spiky & foggy & spiky & foggy & spiky & foggy & spiky & foggy & spiky & foggy & spiky & foggy\\
\midrule
ABR & ABR & 19.99 & \third{57.39} & 26.60 & 32.08 & 14.45 & 21.29 & 5.93 & 42.45 & 10.40 & 24.39 & 12.44 & 26.41\\
 & ext.brick & 14.67 & 41.40 & \last{21.12} & \last{28.16} & \last{12.31} & \last{19.68} & 4.41 & 25.57 & 9.45 & 20.62 & 11.30 & 22.65\\
\midrule
grid+DDA & ABR & \third{20.78} & \first{62.13} & \first{35.88} & \first{82.31} & \first{24.28} & \first{66.80} & 8.84 & 27.86 & 8.15 & 14.23 & \last{10.45} & \last{21.29}\\
 & ext.brick & 20.20 & \second{62.10} & \second{34.46} & \second{81.22} & \second{24.01} & \second{66.75} & 7.98 & 27.26 & \last{8.06} & 14.40 & 10.54 & 21.64\\
grid+RTX & ABR & 18.14 & \last{36.59} & 30.07 & 55.88 & 19.13 & 40.53 & \third{13.40} & \third{66.82} & \first{11.12} & 20.84 & 15.01 & 32.14\\
 & ext.brick & 18.37 & 37.68 & 29.47 & 56.50 & 19.20 & 40.87 & 12.30 & 61.14 & \second{11.12} & 21.04 & \third{15.52} & 32.66\\
\midrule
brick.bound & ABR & 15.27 & 57.13 & \third{31.58} & \third{66.26} & 13.87 & \third{42.67} & 3.24 & 21.51 & \third{10.54} & \first{31.44} & 11.68 & \second{34.84}\\
 & ext.brick & \last{14.26} & 55.29 & 30.52 & 66.13 & 13.76 & 42.46 & \last{2.10} & 14.08 & 10.44 & \second{31.31} & 11.83 & \first{35.05}\\
ext.brick & ABR & 15.12 & 55.50 & 27.51 & 52.88 & 12.89 & 35.23 & 3.24 & 21.44 & 10.07 & 28.89 & 11.35 & 32.26\\
 & ext.brick & 14.30 & 54.66 & 27.43 & 53.99 & 13.01 & 35.65 & 2.11 & \last{14.07} & 10.14 & \third{29.30} & 11.63 & \third{33.02}\\
\midrule
preclass. & ABR & \second{21.84} & 48.45 & 30.42 & 40.69 & 20.33 & 34.61 & \first{24.20} & \first{74.37} & 9.87 & \last{11.74} & \second{18.58} & 25.87\\
 & ext.brick & \first{21.99} & 49.75 & 30.18 & 40.96 & \third{20.45} & 34.59 & \second{23.72} & \second{72.94} & 9.92 & 11.84 & \first{19.10} & 26.23\\
\bottomrule
\end{tabular}
} 
\vspace{-0.5em}
\caption{\label{tab:results}%
\removed{Memory consumption and }Performance (frames/sec.) for all the data sets and
different combinations of spatial subdivisions (ABR,grid,bricks)
\removed{traversal implementations (RTX vs. CUDA),}
and sampling modes (BVH over ABRs vs.\ BVH over brick bounds vs.\ BVH over
extended bricks). (Color codes and font faces to highlight the three best and
the worst-performing options, respectively.)
\vspace{-1em}
}
\end{table*}
\begin{table*}[th]
\centering\small{
\setlength{\tabcolsep}{2pt}
\setlength{\extrarowheight}{-10pt}%
\begin{tabular}{c|c|cc|cc|cc|cc|cc|cc|}
\toprule
trav. & samp. & \multicolumn{2}{|c}{cloud} & \multicolumn{2}{|c}{meteor-20k} & \multicolumn{2}{|c}{meteor-46k} & \multicolumn{2}{|c}{gear} & \multicolumn{2}{|c}{exajet-rear} & \multicolumn{2}{|c}{exajet-wing}\\
\midrule
~ & ~ & spiky & foggy & spiky & foggy & spiky & foggy & spiky & foggy & spiky & foggy & spiky & foggy\\
\midrule
ABR & ABR & 2.4 (2.3) & 2.3 (2.3) & 13.1 (6.7) & 14.0 (6.7) & 11.9 (6.6) & 12.3 (6.6) & 3.1 (3.1) & 3.1 (3.1) & 19.1 (12.0) & 20.4 (12.0) & 19.9 (12.1) & 19.1 (12.1)\\
 & ext.brick & 2.4 (2.3) & 2.3 (2.3) & 14.4 (6.7) & 13.7 (6.7) & 12.0 (6.6) & 12.8 (6.6) & 3.2 (3.1) & 3.2 (3.1) & 19.7 (12.0) & 20.1 (12.0) & 20.1 (12.1) & 20.1 (12.1)\\
\midrule
grid+DDA & ABR & 2.3 (2.3) & 2.3 (2.3) & 14.0 (6.7) & 14.0 (6.7) & 13.0 (6.6) & 12.3 (6.6) & 3.9 (3.9) & 3.9 (3.9) & 19.8 (12.7) & 21.1 (12.7) & 20.6 (12.8) & 19.8 (12.8)\\
 & ext.brick & 2.3 (2.3) & 2.3 (2.3) & \first{2.6 (2.6)} & \first{2.6 (2.6)} & \third{3.2 (3.1)} & \third{3.2 (3.1)} & 3.8 (3.8) & 3.9 (3.8) & 9.2 (8.8) & \third{9.0 (8.8)} & 9.1 (8.8) & \third{9.2 (8.8)}\\
grid+RTX & ABR & 2.5 (2.3) & \last{2.6 (2.3)} & 13.1 (6.7) & 14.0 (6.7) & \last{13.0 (6.6)} & \last{13.0 (6.6)} & \last{13.7 (4.4)} & 13.1 (4.4) & \last{20.9 (13.1)} & \last{21.4 (13.1)} & \last{22.2 (13.2)} & \last{20.9 (13.2)}\\
 & ext.brick & \last{2.5 (2.3)} & 2.5 (2.3) & \second{2.7 (2.6)} & 2.8 (2.6) & 3.2 (3.1) & \first{3.1 (3.1)} & 13.6 (4.4) & \last{13.2 (4.4)} & 16.1 (9.2) & 16.8 (9.2) & 16.8 (9.2) & 16.6 (9.2)\\
\midrule
brick.bound & ABR & 2.4 (2.3) & 2.3 (2.3) & 14.4 (6.7) & 14.0 (6.7) & 12.8 (6.6) & 12.8 (6.6) & \third{3.1 (3.1)} & \third{3.1 (3.1)} & 19.2 (12.0) & 19.7 (12.0) & 19.2 (12.1) & 20.5 (12.1)\\
 & ext.brick & \first{2.2 (2.2)} & \first{2.2 (2.2)} & 2.8 (2.6) & \second{2.7 (2.6)} & \first{3.1 (3.1)} & 3.2 (3.1) & \second{3.0 (3.0)} & \second{3.0 (3.0)} & \second{8.5 (8.1)} & \second{8.5 (8.1)} & \second{8.5 (8.1)} & \second{8.5 (8.1)}\\
ext.brick & ABR & 2.3 (2.3) & 2.3 (2.3) & 14.4 (6.7) & \last{14.0 (6.7)} & 12.3 (6.6) & 12.0 (6.6) & 3.2 (3.1) & 3.1 (3.1) & 19.2 (12.0) & 19.7 (12.0) & 19.7 (12.1) & 19.2 (12.1)\\
 & ext.brick & \second{2.2 (2.2)} & \second{2.2 (2.2)} & \third{2.8 (2.6)} & \third{2.7 (2.6)} & \second{3.2 (3.1)} & 3.2 (3.1) & \first{3.0 (3.0)} & \first{3.0 (3.0)} & \first{8.1 (8.1)} & \first{8.1 (8.1)} & \first{8.5 (8.1)} & \first{8.4 (8.1)}\\
\midrule
preclass. & ABR & 2.3 (2.3) & 2.3 (2.3) & \last{14.5 (6.8)} & 13.2 (6.8) & 12.8 (6.7) & 12.4 (6.7) & 3.4 (3.3) & 3.4 (3.3) & 20.0 (12.8) & 21.2 (12.8) & 20.8 (12.9) & 20.6 (12.9)\\
 & ext.brick & \third{2.3 (2.3)} & \third{2.3 (2.3)} & 2.8 (2.7) & 2.8 (2.7) & 3.2 (3.2) & \second{3.2 (3.2)} & 3.6 (3.3) & 3.8 (3.3) & \third{8.9 (8.8)} & 9.2 (8.8) & \third{9.1 (8.8)} & 9.2 (8.8)\\
\bottomrule
\end{tabular}
}
\vspace{-1em}
\caption{\label{tab:results-mem}%
GPU memory consumption in~GB \removed{ and performance (frames/sec.)}for all the data sets and
different combinations of spatial subdivisions (ABR,grid,bricks), traversal
implementations (RTX vs. CUDA), and sampling modes (BVH over ABRs vs. BVH over
extended bricks). \added{Results are sorted by peak memory consumption; we report
the total memory consumption (lower number, usually achieved after acceleration structure
construction) in parentheses.} (Color codes and font faces to highlight the three best and
the worst-performing options, respectively.)
\vspace{-2.5em}
}
\end{table*}
\added{For the evaluation we concentrate on runtime and memory performance of
the different combinations of acceleration structures.
Our goal is to compare the performance of two segment traversal structures: one
adaptable to transfer function changes (sections \cref{sec:abr-traversal},
\cref{sec:brick-traversal}, \cref{sec:grid-traversal}), and one optimized for a
single transfer function (\cref{sec:reference}).
Another question we explore is if spatial acceleration
structures (grids, kd-trees) have an advantage over object structures
(BVHs)---or the other way around, when being used for ray segment traversal; we
are also interested in if, for this kind of volume data it is beneficial to use
the same acceleration structures for traversal and sampling. And finally, we
want to explore if a universal combination of acceleration structures exists,
or if the choice is predominantly dependent on the data set and transfer
function.}

We compare the various combinations of sampling methods (ABR BVH vs. extended
brick BVH), spatial subdivion/majorant traversal data structures (extended
bricks, brick bounds, macrocell grid), and implementation using ray tracing
hardware (RTX BVH) vs.\
\added{software DDA,}
\removed{CUDA software implementation (kd-tree traversal, DDA)}against the baseline
of using ABRs for both traversal and sampling. We use the data sets from
\cref{fig:majorants3}, which also shows majorants for the acceleration
structures.

\added{Our target system for the performance evaluation is a GPU workstation
with an NVIDIA GeForce RTX~4090 GPU (24~GB GDDR memory) and an Intel i7-6800K CPU
and 128~GB main memory on the host.}\removed{Our target system for the
performance evaluation is a GPU workstation with an NVIDIA A6000 GPU (48~GB
GDDR memory) and 64~GB main memory.} Our evaluation concentrates on memory
consumption and rendering performance.
\added{We compute full global illumination with multi-scattering (isotropic
phase function, russian roulette path termination, ambient light surrounding
the whole scene).}\removed{The first benchmark (``DL'') focuses on
interactivity and computes direct lighting (camera ray, plus one shadow ray
towards a point light for next event estimation). The second benchmark (``MS'')
computes full global illumination with multi-scattering (isotropic phase
function, russian roulette path termination, ambient light surrounding the
whole scene). We use Woodcock tracking to compute free-flight distances and
ratio tracking when computing transmittance along shadow rays.}
\added{We use two transfer functions---one that generates a high-opacity visualization
(cf.\ first column in \cref{fig:majorants3}, ``spiky''), and a second that results
in a more translucent appearance (``foggy'').}
For the grid benchmark, we first empirically determined optimal grid
dimensions, the result of which is presented in \cref{fig:grid-eval}. Based on
that we chose $128^3$ (cloud), $128\times128 \times64$ (meteor-20k,
meteor-46k), $512\times512\times256$ (gear), and $512\times256\times384$
(exajet) for the remaining experiments. The performance results in
frames/sec.\ can be found in \cref{tab:results}.
\added{We also report peak GPU memory consumption and the total memory
consumption (i.e., during rendering, when acceleration structures were built
and temporary memory freed) in \cref{tab:results-mem}.}
\removed{We also report the average number of spatial partitions
traversed vs. the number of samples taken per partition in Fig.~9.}

\added{Our results allow for multiple observations. One such observation is that data
structures with overlap (ABRs, bricks, brick domains) can suffer from the
aforementioned problem of single cells infecting large regions with their
contribution. This can also be seen in \cref{fig:majorants3}, where the
majorant regions for ABRs or bricks sometimes appear excessively coarse.
This observation suggests that spatial data structures are to be preferred
over object data structures as this problem is easier to control. However, it is
well-known that grids suffer from the ``teapot in a stadium'' problem. To illustrate
this we include the landing gear data set whose cells are contained within
a large boundary of coarser cells that do not contribute important features
(cf.\ \cref{fig:crazy-gear}). Apart from this extreme case, we however note
that the proposed traversal and sampling methods come close to and often even
outperform the majorant of a laborously pre-computed kd-tree.}\removed{Our
results indicate that grids are generally superior to the other traversal
methods. The number of samples taken overall is significantly reduced most of
the times; at the same time, the number of traversal steps increases, which can
however usually be amortized by the decrease in samples per partition. For the
larger data sets (gear, exajet) we observe that traversing the grid with a BVH
gives significantly better results; this is because DDA has to traverse
numerous empty cells; with the BVHs, we can apply the optimization that removes
empty regions, which we found to have a significant impact. In the case of the
grid, this may be obvious; generally, this optimization is important for high
rendering performance, because traversing ABRs, bricks, or grid cells whose
majorants are 0 causes unnecessary context switches from the ray tracing cores
that traverse the BVH on the chip, to the shader cores that execute the
(software) intersection program.}

\removed{An extreme case where ABR and brick traversal resulted in extraordinarily bad
performance is the landing gear data set; we analyzed the spatial arrangement
of this data set more closely and present some of our findings in
Fig.~??. This data set suffers severely from the ``teapot in a
stadium'' problem; the finely tessellated landing gear is surrounded by huge
coarse-level AMR cells and ABRs that get ``infected'' by the much smaller
finest-level cells at the data set's center (cf. Sec.~??),
which spill their majorants into the regions spanned by the coarser ABRs. It
is those extreme cases that are typical for AMR data, and it is this type of
data set that also leads us to generally recommend decoupling traversal and
sampling, and to use a grid data structure for traversal that is independent of
the sampling data structure.}

We also observed a fixed cost associated with using software traversal with DDA,
which we suspect is due to register pressure (OptiX however does
not allow us to exactly determine per-kernel register usage). This becomes
obvious in the comparison between DDA and RTX grid traversal for small grid sizes 
(e.g., $\le 10^3$ macrocells, cf. \cref{fig:grid-eval}).
\removed{which corroborates our observation that uniform grids with RTX
traversal are superior to the other data structures in terms of performance.}
We however also note that the memory consumption for this option can become
quite high; in the special case of a transfer function without empty space, we
have to build a full BVH over the grid, which limits the number of grid cells
we can use effectively.

\vspace{-0.5em}
\section{Discussion and Conclusion} \label{sec:discussion} \label{sec:conclusion}
We have shown how to take a visualization framework targeted at large-scale AMR
data and modify it to support interactive volumetric path tracing. The benefits
of this are increased visual fidelity and improved time to first image.
Volumetric path tracing seamlessly integrates complex lighting models. It also
supports traditional models if the user desires not to focus on realism. Even
in these cases, we still find path tracing preferable, as it enables smarter
sample placement compared to ray marching, resulting in fewer samples needed
for volume-rendered images. We contribute the first framework plus
acceleration structures optimized for volume path tracing native AMR data on
GPUs with ray tracing hardware.

The base operation performed by a volumetric path tracer is free flight
distance tracking, which requires routines to obtain extinction and majorant
estimates for local regions. The local majorant extinctions
to traverse the volume with the minimum number of samples usually have an
irregular spatial arrangement. AMR hierarchies already focus their less
uniform structures in regions where the entropy of the data is high and
consequently the extinction varies a lot. One might therefore intuitively
assume that the AMR hierarchy would be the best possible majorant traversal
data structure. However, our experiments show that the task of finding
majorants for this kind of data is more complicated.

We find it hard to make a general recommendation as of the best combination of
acceleration structures for traversal and sampling, yet we observe a tendency
for object order data structures to do a bad job in this case because their
overlap allows small regions from neighboring structures to ``infect'' other
structures that are otherwise homogeneous or even empty. Notably,
both combinations that use the same acceleration structure for
traversal and sampling generally do not outperform the other alternatives.

\removed{On the other hand, the same data structures (bricks, ABRs, etc.) \emph{ideally}
cover many same-level cells, regardless of the cell \emph{values} that we
derive the extinction from. Very large regions of space with varying
extinctions are the exact opposite of what one would consider an optimal
majorant data structure.}

In fact, grids often present a viable alternative; in this case the outer loop
traversal over ray segments is implemented in software and the inner loop uses
hardware ray tracing. We find that using hardware ray tracing to traverse the
grid's macrocells pays off for large grids, though this is quite
memory intensive, restricting the maximum grid size one can use.


The solutions we focus on allow us to update RGB$\alpha$ transfer functions
interactively, even for our largest data sets.
\added{The alternative to this is using pre-classified data structures, which
due to the nature of how extinction is computed via alpha transfer function lookups
often even results in inferior acceleration structures, and can take hours to
build even when using optimizations such as priority queues and binning.}
\removed{This
can however only be accomplished by storing auxiliary data: the grids we use,
e.g., do not only store a majorant extinction per cell, but also per-cell
min/max value ranges to quickly obtain those majorants by applying the alpha
transfer function on-the-fly in a GPU kernel. For the $512 \times 512 \times
256$ grid we use for the landing gear, the storage requirements thus triple
from 1~GB (majorants only) to 3~GB (majorants plus value ranges).}
\removed{More severe
optimizations to this end (e.g., refitting instead of rebuilding BVHs as we
currently do) are likely to come at even higher memory costs. The design
decisions we made to this end are different than, e.g., those we would have
made when designing a production renderer with different objectives regarding
interactivity.}

\added{Another observation we made is that transfer functions that have high
frequencies---as is often the case when extracting ISO-surface like
features---can result in extreme performance degredations. Adjusting the slope
of ``alpha peaks'' can lead to extremely ``spiky'' transfer functions, causing
high sample counts and significant frame rate reductions.}

\added{Another point important to discuss is that our solution accounts for the
(pre-tabulated) albedo by multiplying it to the result of the evaluated phase
function. Because of that we cannot use residual tracking
methods~\cite{Kutz2017SDT,novak:2014} as the control variates cannot vary in
space, and reconstructing the correct albedo for a control variate sample is
impossible. To this end, the solution employed in production rendering would be
to have individual majorants plus control volumes for each R,G, and B channel
of the volume. In our case, this would limit memory availability even further,
so we opt to use larger grids, or deeper hierarchies with more leaves, in favor
of residual tracking algorithms, which we believe gives us better overall
performance. A more thorough investigation of this would however be interesting
future work.}

\added{In conclusion, we have presented how to extend a sci-vis volume renderer with
non-trivial sampling structures to support volumetric path tracing. Volumetric
path tracing can significantly improve visual fidelity, but even with standard
sci-vis lighting models has several advantages, such as better sample
placement---and resulting from that faster time to image---as well as
unbiasedness and fewer sampling artifacts. We have evaluated multiple
alternatives to implement this by proposing extensions to the \exabrick{} data
structure; our experiments show that computing majorant extinctions for this
kind of data is not trivial---not even in the case where a single hand-picked
transfer function is used, and even less so when transfer functions are allowed
to change interactively. Our results indicate a slight performance advantage
for spatial over object data structures. Overall we have presented numerous
different strategies to accelerate traversal and sampling that allow for
interactive transfer function updates and demonstrated that these do not
perform significantly worse---and sometimes even outperform---the offline
reference.}



\section*{Acknowledgments}
\ifsubmission
Intentionally omitted for review.
\else
This work was supported by the Deutsche Forschungsgemeinschaft (DFG, German
Research Foundation)---grant no.~456842964. We also express our gratitude to
NVIDIA, who kindly provided us with hardware we used for the evaluation.
\fi

\bibliographystyle{eg-alpha-doi}  
\bibliography{main}

\end{document}